\documentclass[12pt]{article}

\usepackage{amsmath,amsfonts,amssymb}
\usepackage{graphicx,psfrag,epsf,comment,float,lipsum,titlefoot}
\usepackage{enumerate}
\usepackage{enumitem}
\usepackage{bbm}
\usepackage{mathtools}

\usepackage[numbers]{natbib}
\usepackage{authblk}

\usepackage[linktoc=all,colorlinks=true,linkcolor=black,citecolor=blue,urlcolor=blue,bookmarks=false,hypertexnames=true]{hyperref}
\usepackage{url}

\usepackage{xcolor}
\usepackage{algorithm2e}
\usepackage{mathtools}

\usepackage{subcaption}
\usepackage[font=footnotesize,labelfont=bf]{caption} 

\usepackage{tabularx}
\usepackage{array}

\usepackage{booktabs}
\usepackage{longtable}
\usepackage{colortbl}

\usepackage[resetlabels]{multibib}
\newcites{appendix}{Appendix References}

\usepackage[subtle]{savetrees}

\usepackage{fontawesome}

\usepackage{pifont}
\newcommand{\cmark}{\ding{51}}%
\newcommand{\xmark}{\ding{55}}%

\newcommand{\blind}{1}

\newcommand{\bv}{\mathbf{v}}
\newcommand{\bw}{\mathbf{w}}
\newcommand{\bx}{\mathbf{x}}

\newcommand{\bz}{\mathbf{z}}
\newcommand{\bV}{\mathbf{V}}
\newcommand{\bU}{\mathbf{U}}
\newcommand{\bS}{\mathbf{S}}

\newcommand{\bbeta}{\boldsymbol{\beta}}
\newcommand{\bepsilon}{\boldsymbol{\epsilon}}

\newcommand{\bchi}{\boldsymbol{\chi}}

\newcommand{\bmu}{\boldsymbol{\mu}}
\newcommand{\bnu}{\boldsymbol{\nu}}
\newcommand{\balpha}{\boldsymbol{\alpha}}

\newcommand{\bX}{\mathbf{X}}
\newcommand{\bY}{\mathbf{Y}}
\newcommand{\bZ}{\mathbf{Z}}

\newcommand{\R}{\mathbb{R}}
\renewcommand{\P}{\mathbb{P}}
\newcommand{\E}{\mathbb{E}}
\newcommand{\indicator}{\mathbf{1}}

\DeclareMathOperator*{\argmin}{arg\,min}

\newcommand{\method}{Synthetic Combinations}

\newcommand{\data}[1][n]{\mathcal{D}_{#1}}

\newcommand{\partition}{\mathfrak{p}}

\newcommand{\cell}{\mathfrak{t}}

\newcommand{\node}{\mathfrak{t}}

\definecolor{cyan}{cmyk}{1, 0.2, 0, 0}

\newcommand\independent{\protect\mathpalette{\protect\independenT}{\perp}}
\def\independenT#1#2{\mathrel{\rlap{$#1#2$}\mkern2mu{#1#2}}}

\newtheorem{theorem}{Theorem}[section]
\newtheorem{lemma}[theorem]{Lemma}
\newtheorem{proposition}[theorem]{Proposition}
\newtheorem{corollary}[theorem]{Corollary}

\newtheorem{definition}[theorem]{Definition}
\newtheorem{assumption}[theorem]{Assumption}
\newtheorem{remark}[theorem]{Remark}
\newtheorem{proof}[theorem]{Proof}
\DeclarePairedDelimiter{\braces}{\lbrace}{\rbrace}
\begin{document}

\def\spacingset#1{\renewcommand{\baselinestretch}%
{#1}\small\normalsize} \spacingset{1}


\if1\blind
{\title{\vspace{-10mm}\bf \method: A Causal Inference Framework for Combinatorial Interventions}
  \author{
    \normalsize Abhineet Agarwal \vspace{-1em}\\
    \normalsize Department of Statistics, University of California, Berkeley\\
    \vspace{0.5em}
    Anish Agarwal\footnote{Part of this work was completed while Anish was a post-doc at Amazon, Core AI.}\\
    \normalsize Department of Industrial Engineering and Operations Research, Columbia University\\
   \vspace{0.5em}
    Suhas Vijaykumar\\
    \normalsize Amazon, Core AI
    }
  \maketitle
} \fi

\if0\blind
{
  \bigskip
  \bigskip
  \bigskip
  \begin{center}
    {\LARGE\bf \method: A Causal Inference Framework for Combinatorial Interventions}
\end{center}
  \medskip
} \fi

\begin{abstract}
\label{sec:abstract}
Consider a setting where there are $N$ heterogeneous units and $p$ interventions.
Our goal is to learn unit-specific potential outcomes for any combination of these $p$ interventions, i.e., $N \times 2^p$ causal parameters. 
Choosing a combination of interventions is a problem that naturally arises in a variety of applications such as factorial design experiments, recommendation engines, combination therapies in medicine, conjoint analysis, etc.  
Running $N \times 2^p$ experiments to estimate the various parameters is likely expensive and/or infeasible as $N$ and $p$ grow.
Further, with observational data there is likely confounding, i.e., whether or not a unit is seen under a combination is correlated with its potential outcome under that combination.
To address these challenges, we propose a novel latent factor model that imposes structure across units (i.e., the matrix of potential outcomes is approximately rank $r$), and combinations of interventions (i.e., the coefficients in the Fourier expansion of the potential outcomes is approximately $s$ sparse). 
We establish identification for all $N \times 2^p$ parameters despite unobserved confounding.
We propose an estimation procedure, Synthetic Combinations, and establish it is finite-sample consistent and asymptotically normal under precise conditions on the observation pattern.
Our results imply consistent estimation given $\text{poly}(r) \times \left( N + s^2p\right)$ observations, while previous methods have sample complexity scaling as $\min(N \times s^2p, \ \ \text{poly(r)} \times (N + 2^p))$.
We use Synthetic Combinations to propose a data-efficient experimental design. 
Empirically, Synthetic Combinations outperforms competing approaches on a real-world dataset on movie recommendations. 
Lastly, we extend our analysis to do causal inference where the intervention is a permutation over $p$ items (e.g., rankings). 

\end{abstract}
\noindent%
{\it Keywords:} Matrix Completion, Fourier Analysis of Boolean Functions, Latent Factor Models, Synthetic Controls, Learning to Rank
\vfill

\newpage
\spacingset{1.75}

\section{Introduction}
Modern-day decision-makers, in settings from e-commerce to public policy to medicine, often encounter settings where they have to pick a combination of interventions and ideally would like to do so in a highly personalized manner.
Examples include recommending a curated basket of items to customers on a commerce platform, deciding on a combination of therapies for a medical patient, enacting a collection of socio-economic policies for a specific geographic location, feature selection in a machine learning model, doing a conjoint analysis in surveys, etc.
Despite the ubiquity of this setting, it comes with significant empirical challenges: with $p$ interventions and $N$ units, a decision maker must evaluate $N \times 2^p$ potential combinations in order to confirm the optimal personalized policy.
With large $N$ and even with relatively small $p$ (due to its exponential dependence), it becomes infeasible to run that many experiments.
Of course, in observational data there is the additional challenge of potential unobserved confounding.
Current methods tackle this problem by following one of two approaches: (i) they impose structure on how combinations of interventions interact, or (ii) they assume latent similarity in potential outcomes across units. 
However, as we discuss in detail below, these approaches require a large number of observations to estimate all $N \times 2^p$ potential outcomes because they do not exploit structure across both units and combinations.
Hence the question naturally arises: {\em how can one effectively share information across both units and combinations of interventions?}

\vspace{2mm}
\noindent
{\bf Contributions.} 
Our contributions may be summarized as follows.

\textbf{(1)} For a given unit $n \in [N]$, we represent its potential outcomes over the $2^p$  combinations as a Boolean function from $\{-1, 1\}^p$ to $\mathbb{R}$, expressed in the Fourier basis.
To impose structure across combinations, we assume that for a unit $n$, the Fourier coefficients $\balpha_n \in \mathbb{R}^{2^p}$ induced by this basis are sparse, i.e., have at most $s$ non-zero entries.
To impose structure across units, we assume that this matrix of Fourier coefficients across units $\mathcal{A} = [\balpha_n]_{n \in [N]} \in \mathbb{R}^{N \times 2^p}$ has rank $r$.
This simultaneous sparsity and low-rank assumption is indeed what allows one to share information across both units and combinations.

\textbf{(2)} We establish identification for the $N \times 2^p$ potential outcomes of interest, which requires that any confounding is mediated by the (unobserved) matrix of Fourier coefficients $\mathcal{A}$.

\textbf{(3)} We design a two-step algorithm ``\method'' and prove it consistently estimates the various causal parameters, despite potential unobserved confounding. 
The first step of \method, termed ``horizontal regression'', learns the structure across combinations of interventions via the Lasso. 
The second step, termed ``vertical regression'', learns the structure across units via principal component regression (PCR).

\begin{table}[t!]
\scriptsize
    \centering
    \begin{tabular}{|c|c|c|c|}
     \hline
    Learning Algorithm & Exploits combinatorial structure ($\lVert \balpha_n \rVert_0 = s$) & Exploits inter-unit structure $\text{rank}(\mathcal{A}) = r$) & Sample Complexity \\
    \hline 
    Lasso     & \cmark & \xmark & $O(N \times s^2p)$ \\
    \hline 
    Matrix Completion     & \xmark & \cmark & $O(\text{poly}(r) \times (N + 2^p)$ \\
    \hline 
    \method   & \cmark & \cmark & $O(\text{poly}(r) \times (N + s^2p)$ \\
    \hline
    \end{tabular}
    \caption{Comparison of sample complexity of different methods to estimate all $N \times 2^p$ causal parameters. \method~combines the strengths of both the Lasso and matrix completion methods by exploiting structure across both combinations and units to estimate all potential outcomes with significantly fewer observations. }
    \label{tab:sample_complexity_summary}
\end{table}

\textbf{(4)} Our results imply that \method~is able to consistently estimate unit-specific potential outcomes given a total of $\text{poly}(r) \times \left( N + s^2p\right)$ observations (ignoring logarithmic factors).
This improves over previous methods that do not exploit structure across both units and combinations, which have sample complexity scaling as $\min(N \times s^2p, \ \ \text{poly}(r) \times (N + 2^p))$.
A summary of the sample complexities required for different methods can be found in Table \ref{tab:sample_complexity_summary}. 
We provide an example showing how replacing Lasso with ``classification and regression trees'' (CART) leads to a sample complexity of $\text{poly}(r) \times \left( N + s^2\right)$ under additional regularity assumptions on the Fourier coefficients.
A key technical challenge in our theoretical analysis is studying how the error induced in the first step of \method~percolates through to the second step. 
To tackle it, we reduce this problem to that of high-dimensional error-in-variables regression with linear model misspecification, and do a novel analysis of this statistical setting.
Under additional conditions, we establish asymptotic normality of the estimator.
%

%
\textbf{(5)} We show how \method~can be used to inform experiment design for combinatorial interventions.
In particular, using ideas from compressed sensing, we propose and analyze an experimental design mechanism that ensures the key assumptions required for consistency of \method~are satisfied. 

\textbf{(6)} To empirically evaluate \method, we apply it to a real-world dataset for user ratings on sets of movies \cite{sharma2019sets}, and find it outperforms both Lasso and matrix completion methods.
Further, we show two of our key modeling assumptions---low-rank and sparse Fourier coefficients---hold in this dataset.
We perform additional numerical simulations that corroborate our theoretical findings and show the robustness of \method~to unobserved confounding.

\textbf{(7)}
We discuss how to extend \method~to estimate counterfactual outcomes when the intervention is a permutation over $p$ items, i.e., rankings. 
Learning to rank is a widely studied problem and has many applications such as search engines and matching markets.

\subsection{Related Work}\label{sec:related_work}

\noindent
{\bf Learning over combinations.}
To place structure on the space of combinations, we use tools from the theory of learning (sparse) Boolean functions, in particular, the Fourier transform. 
Sparsity of the Fourier transform as a complexity measure was proposed by \citet{karpovsky1976finite}, and was used by \cite{brandman1990spectral} and others to characterize and design learning algorithms for low-depth trees, low-degree polynomials, and small circuits. 
Learning Boolean functions is now a central topic in learning theory, and is closely related to many important questions in ML more broadly; see e.g.~\cite{mossel2003learning} for discussion of the $k$-Junta problem and its relation to relevant feature selection. 
We refer to \citet[Chapter 3]{o2008some} for further background on this area. 
In this paper, our focus is on general-purpose statistical procedures for learning sparse Boolean functions with noise. 
We build on the work of \cite{negahban2012learning} on variants of the Lasso procedure, and the works of \cite{breiman2017classification,syrgkanis2020estimation,klusowski2020sparse,klusowski2021universal} and others on CART. 
In particular, we highlight the works of \cite{negahban2012learning} and \cite{syrgkanis2020estimation} which respectively show that Lasso and CART can be used to efficiently learn sparse Boolean functions.
Recent work \cite{bravohermsdorff2023intervention} has also explored the use of graphical models to learn across combinations. 

\noindent {\bf Matrix completion.}  
Previous works have shown that imputing counterfactual outcomes with latent factor structure and potential unobserved confounding can be equivalently expressed as low-rank matrix completion with missing not at random data \citep{bai2019matrix, athey2021matrix, agarwal2020synthetic, agarwal2021causalmatrix}.
The observation that low-rank matrices may typically be recovered from a small fraction of the entries by nuclear-norm minimization has had a major impact on modern statistics \cite{candes2010power,recht2011simpler,candes2012exact}. 
In the noisy setting, proposed estimators have generally proceeded by minimizing risk subject to a nuclear-norm penalty, such as in the SoftImpute algorithm of \citep{mazumder2010spectral}, or minimizing risk subject to a rank constraint as in the hard singular-value thresholding (HSVT) algorithms analyzed by \cite{keshavan2009matrix,keshavan2010matrix,gavish2014optimal, chatterjee2015matrix}.
We refer the reader to \cite{ieee_matrix_completion_overview, nguyen2019low} for a comprehensive overview of this vast literature. 

\noindent
{\bf Causal latent factor models.}
There is a rich literature on how to learn personalized treatment effects for heterogeneous units. 
This problem is of particular importance in the social sciences and in medicine, where experimental data is limited, and has led to several promising approaches including instrumental variables, difference-in-differences, regression discontinuity, and others; see \cite{angrist1996identification,imbens2015causal} for an overview.
Of particular interest is the synthetic control method  \cite{abadie2010synthetic}, which exploits an underlying factor structure to effectively impute outcomes under control for treated units.
Building on this, recent works \citep{agarwal2020synthetic, agarwal2021causalmatrix} have shown how to use latent factor representations to efficiently share information across heterogeneous units and treatments.
This paper generalizes these previous works to where a treatment is a combination of interventions and {\em most treatments have no units that receive it}.
To do so, we impose latent structure across combinations of interventions. 

In doing so, we aim to bridge such latent factor models with highly practical settings where multiple interventions are delivered simultaneously, such as slate recommendations, medicine (e.g. basket trials, combination therapies), conjoint analysis in surveys \cite{egami2019causal,haimmueller2014causal,goplerud2023estimating}, and factorial design experiments (e.g., multivariate tests in digital experimentation) \citep{duflo2007using,dasgupta2015causal,wu2011experiments,zhao2022regression} .
For example, structure across combinations is often assumed in the analysis of factorial design experiments, where potential outcomes are generally assumed to only depend on main effects and pairwise interactions between interventions \citep{bertrand2004emily,delaCuesta2021ImprovingTE,eriksson2014employers,george2005statistics}.
We discuss factorial design experiments in detail in Section \ref{sec:combinatorial_inference_applications}.
%


\section{Setup and Model}\label{sec:model}
In this section, we first describe requisite notation, background on the Fourier expansion of real-valued functions over Booleans, and how it relates to potential outcomes over combinations. 
Next, we introduce the key modeling assumptions and data generating process (DGP), along with the target causal parameter of interest.
We also provide a brief summary of how the formalism can be extended to potential outcomes over permutations in Section \ref{subsec:rankings_summary}, and a more detailed discussion in Appendix \ref{sec:permutations}.

\subsection{Notation}\label{sec:notation}

\noindent
{\bf Representing combinations as binary vectors}.
Let $[p] = \{1,\ldots p\}$ denote the set of $p$ interventions. 
Denote by $\Pi$ the power set of $[p]$, the set of all possible combinations of $p$ interventions, where we note $|\Pi| = 2^p$.
Then, any given combination $\pi \in \Pi$ induces the following binary representation $\bv(\pi) \in \{-1,1\}^p$ defined as follows: $\bv(\pi)_{i} = 2~\mathbbm{1}\{i \in \pi\} -1.$

\vspace{2mm}
\noindent
{\bf Fourier expansion of Boolean functions.}
Let $\mathcal{F}_{\text{bool}} = \{f : \{-1,1\}^p \to \mathbb{R}\}$ be the set of all real-valued functions defined on the hypercube $\{-1,1\}^p$. 
Then $\mathcal{F}_{\text{bool}}$ forms a Hilbert space defined by the following inner product: for any $f,g \in \mathcal{F}_{\text{bool}}$, $\langle f, g \rangle_{B} = \frac{1}{2^p}\sum_{\bx \in \{-1,1\}^p}f(\bx)g(\bx).$
This inner product induces the norm $\langle f, f \rangle_{B} \coloneqq \lVert f \rVert_{B}^2 = \frac{1}{2^p}\sum_{\bx \in \{-1,1\}^p}f^2(\bx).$
We construct an orthonormal basis for $\mathcal{F}_{\text{bool}}$ as follows: for each subset $S \subset [p]$, define a basis function $\chi_{S}(\bx) = \prod_{i \in S} x_i$ where $x_i$ is the $i^\text{th}$ coefficient of $\bx \in  \{-1,1\}^p$. 
One can verify that for any $S \subset [p]$ that $\lVert \chi_{S} \rVert_B = 1$, and that $\langle \chi_S, \chi_{S'} \rangle_B = 0$ for any $S' \neq S$. 
Since $|\{\chi_S: S \subset [p] \}| = 2^p$, the functions $\chi_S$ are an orthonormal basis of $\mathcal{F}_{\text{bool}}$.
We will refer to $\chi_S$ as the Fourier character for the subset $S$.

Hence, any $f \in \mathcal{F}_{\text{bool}}$ can be expressed via the following Fourier decomposition: $f(\bx) = \sum_{S \subset[p]}\alpha_{S}\chi_{S}(\bx),$
where the Fourier coefficient $\alpha_S$ is given by computing $ \alpha_S = \langle f, \chi_S \rangle_B $.
For a function $f$, we refer to $\balpha_f = [\alpha_S]_{S \in [p]} \in \mathbb{R}^{2^p}$ as the vector of Fourier coefficients associated with it. 
For a given binary vector $\bx \in \{-1, 1\}^{p}$, we refer to $\bchi(x) = [\chi_{S}(\bx)]_{S \in [p]} \in \{-1, 1\}^{2^p}$ as the vector of Fourier character outputs associated with it.
Hence any function $f : \{-1,1\}^p \to \mathbb{R}$ can be re-expressed as follows:
$f(\bx) =  \langle \balpha_f, \bchi(x) \rangle.$
For $\pi \in \Pi$, abbreviate $\chi_S(\bv(\pi))$ and $\bchi(\bv(\pi))$ as $\chi^{\pi}_S$ and $\bchi^\pi$ respectively.

\vspace{2mm}
\noindent
{\bf Observed and potential outcomes.}
Let $Y^{(\pi)}_n \in \mathbb{R}$ denote the {\em potential outcome} for unit $n$ under combination $\pi$ and $Y_{n\pi} \in \{\mathbb{R} \cup \star \}$ as the {\em observed outcome}, where $\star$ indicates a missing value, i.e., the outcome associated with the unit-combination pair $(n, \pi)$ was not observed. 
Let $\bY = [Y_{n\pi}] \in \{\mathbb{R} \cup \star \}^{N \times 2^p}.$
Let $\mathcal{D} \subset [N] \times [2^p],$ refer to the subset of observed unit-combination pairs, i.e., 
\begin{equation}\label{eq:SUTVA}
Y_{n\pi} = 
\begin{cases}
Y^{(\pi)}_n, & \text{if $(n, \pi) \in \mathcal{D}$}\\
\star, & \text{otherwise}.
\end{cases}
\end{equation}
Note that \eqref{eq:SUTVA} implies stable unit treatment value assignment (SUTVA)  holds.
Let $\Pi_S \subseteq \Pi$ denote a subset of combinations. 
For a given unit $n$, let $\bY_{\Pi_S,n} = [Y_{n\pi_i} : \pi_i \in \Pi_S] \in \{\mathbb{R} \cup \star \}^{|\Pi_S|}$ represent the vector of observed outcomes for all $\pi \in \Pi_S$. 
Similarly, let $\bY_n^{(\Pi_S)} = [Y^{(\pi_i)}_{n} : \pi_i \in \Pi_S] \in \mathbb{R}^{|\Pi_S|}$ represent the vector of potential outcomes. Denote $\bchi(\Pi_S) = [\bchi^{\pi_i} : \pi_i \in \Pi_S] \in \{-1, 1\}^{|\Pi^S| \times 2^p }$.

\subsection{Model and Target Causal Parameter}\label{sec:causal_model}
Define $Y_n^{(\cdot)} : \pi \to \mathbb{R}$ as a real-valued function over the hypercube $\{-1,1\}^p$ associated with unit $n$.
It takes as input a combination $\pi$, converts it to a $p$-dimensional binary vector $\bv(\pi)$, and outputs the real-valued potential outcome $Y_n^{(\pi)}$.
Given the discussion in Section \ref{sec:notation} on the Fourier expansion of Boolean functions, it follows that $Y_n^{(\pi)}$ always has the representation $\langle \balpha_n, \bchi^\pi \rangle$ for some $\balpha_n \in \mathbb{R}^{2^p}$. 
Thus, without any loss of generality, the $\balpha_n$ are unit-specific latent variables, i.e., the Fourier coefficients, encoding the treatment response function. 
Below, we state our key assumption on these induced Fourier coefficients.

\begin{assumption} [Potential Outcome Model]
\label{ass:observation_model} 
For any unit-combination pair $(n,\pi)$,  assume the potential outcome has the following representation,
\begin{equation}\label{eq:observed_outcome}
Y^{(\pi)}_{n} = \langle \balpha_{n}, \bchi^\pi \rangle + \epsilon_n^{\pi},
\end{equation}
where $\balpha_n \in \mathbb{R}^{2^p}$ and $ \bchi^\pi \in \{-1, 1\}^{2^p}$ are the Fourier coefficients and characters, respectively.
We assume the following properties: (a) low-rank: the matrix $\mathcal{A} = [\balpha_{n}]_{n \in [N]}$ has rank $r \in [\min\{N, 2^p\}]$; (b) sparsity: $\balpha_{n}$ is $s$-sparse (i.e. $\lVert \balpha_n \rVert_0 \leq s$, where $s \in [2^p]$) for every unit $n \in [N]$; (c) $\epsilon^{\pi}_n$ is a residual term specific to $(n,\pi)$ which satisfies $\E[\epsilon_n^{\pi} ~ | ~ \mathcal{A}] = 0$. 

\end{assumption}
The assumption is then that each $\balpha_n$ is $s$-sparse and $\mathcal{A}$ is rank-$r$; $\epsilon_n^{\pi}$ is the residual from this sparse and low-rank approximation, and it serves as the source of uncertainty in our model. 

Given $\E[\epsilon_n^{\pi} ~ | ~ \mathcal{A}] = 0$ and that $\mathcal{A}$ is rank-$r$, it implies the matrix $\E[\bY_N^{(\Pi)}] = [\E[\bY_n^{(\Pi)}]: n \in [N]] \in \mathbb{R}^{2^p \times N}$ is also rank $r$, where the expectation is defined with respect to $\epsilon_n^{\pi}$.
This is because $\E[\bY_N^{(\Pi)}]$ can be written as $\E[\bY_N^{(\Pi)}] = \bchi(\Pi) \mathcal{A}^T$, and since $\bchi(\Pi)$ is an invertible matrix, $\text{rank}(\E[\bY_N^{(\Pi)}]) = \text{rank}(\mathcal{A})$.
%
The low-rank property places \emph{structure across units}; that is, there is sufficient similarity across units so that $\E[\bY_n^{(\Pi)}]$ for any unit $n$ can be written as a linear combination of $r$ other rows of $\E[\bY_N^{(\Pi)}]$. 
This is a standard assumption used to encode latent similarity across units in matrix completion and its related applications (e.g., recommendation engines).

Sparsity establishes \emph{unit-specific} structure; that is, the potential outcomes for a given user only depend on a small subset of the Fourier characters $\{\chi_S: S \subset [p] \}$. 
This subset of Fourier characters can be different across units.
As discussed in Section \ref{sec:related_work}, sparsity is commonly employed when studying the learnability of Boolean functions. 
In the context of recommendation engines, sparsity is implied if the ratings for a set of goods only depend on a small number of combinations of items within that set---if say only $k < p$ items mattered, then $s \le 2^k$.
Sparsity is also often assumed implicitly in factorial design experiments, where analysts typically only include pairwise interaction effects between interventions and ignore higher-order interactions \citep{george2005statistics}---here $s \le p^2$.
These various models of imposing combinatorial structure in greater detail in Section \ref{sec:combinatorial_inference_applications}. 

Next, we present an assumption that formalizes the dependence between the missingness pattern induced by the treatment assignments $\mathcal{D}$ and the potential outcomes $Y^{(\pi)}_n$, i.e., the type of confounding we can handle, and provide an interpretation of the induced data generating process (DGP) for potential outcomes.

\begin{assumption}[Selection on Fourier coefficients]
\label{ass:selection_on_fourier} 
For all $n \in [N]$, $\pi \in \Pi$, $Y^{(\pi)}_n \independent \mathcal{D} \mid \mathcal{A}$.
\end{assumption}
\vspace{-2mm}
\noindent \textbf{Data Generating Process.} 
Given Assumptions \ref{ass:observation_model} and \ref{ass:selection_on_fourier}, one sufficient DGP that is line with the assumptions made can be summarized as follows: (i) unit-specific latent Fourier coefficients $\mathcal{A}$ are either deterministic or sampled from an unknown distribution; we will condition on this quantity throughout.
(ii) Given $\mathcal{A}$, we sample mean-zero random variables $\epsilon_n^{\pi}$, and generate potential outcomes according to our model $Y_n^{(\pi)} = \langle \balpha_n, \chi^{\pi} \rangle + \epsilon_n^{\pi}$.
(iii) $\mathcal{D}$ is allowed to depend on unit-specific latent Fourier coefficients $\mathcal{A}$ (i.e., $\mathcal{D}=f(\mathcal{A})$).
We define all expectations w.r.t.~noise, $\epsilon_n^{\pi}$.
This DGP introduces unobserved confounding since $Y_n^{(\pi)} \not\independent \mathcal{D}$. 
However, this DGP does imply that Assumption \ref{ass:selection_on_fourier} holds, i.e., conditional on the Fourier coefficients $\mathcal{A}$, the potential outcomes are independent of the treatment assignments $\mathcal{D}$.
This conditional independence condition can be thought of as ``selection on latent Fourier coefficients”, which is analogous to the widely made assumption of “selection on observables.” The latter requires that potential outcomes are independent of treatment assignments conditional on {\em observed} covariates---we reemphasize that $\mathcal{A}$ is unobserved. 

\vspace{2mm}
\noindent 
\textbf{Target parameter.} 
For any unit-combination pair $(n,\pi)$, we aim to estimate $\E[Y^{(\pi)}_{n} \mid \mathcal{A}]$, where the expectation is w.r.t.~$\epsilon_n^{\pi}$, and we condition on the set of Fourier coefficients $\mathcal{A}$.
\subsection{Extension to Permutations}
\label{subsec:rankings_summary}
We provide a brief summary of how the formalism established above for combinations can be extended to permuations (i.e., rankings), and provide a detailed discussion in Appendix \ref{sec:permutations}.

\noindent \textbf{Binary Representation of Permutations.} 
Let $\tau: [p] \rightarrow [p]$ denote a permutation on a set of $p$ items such that $\tau(i)$ denotes the rank of item $i \in [p]$.  
There are a total of $p!$ different permutations, and denote the set of permutations by $\mathbb{S}_p$.
Every permutation $\tau \in \mathbb{S}_p$ induces a binary representation $\mathbf{v}(\tau) \in \{-1,1\}^{\binom{p}{2}}$, which can be constructed as follows.
For an item $i$, define  $\mathbf{v}^{i}(\tau) \in \{-1,1\}^{p-i}$ as follows: $\mathbf{v}_j^{i}(\tau)= \mathbbm{1}\{ \tau(i) > \tau(j)\} - \mathbbm{1}\{ \tau(i) < \tau(j)\}$ for items $1 \leq i  <  j \leq p$.
That is, each coordinate $\mathbf{v}_j^{i}(\tau)$ indicates whether items $i$ and $j$ have been swapped for items $j > i$. 
Then, $\mathbf{v}(\tau) = [\mathbf{v}^{i}(\tau): i \in [p]] \in \{-1,1\}^{\binom{p}{2}}$.
For example, with $p = 4$, the permutation $\tau([1,2,3,4]) = [1,3,4,2]$ has the binary representation $\mathbf{v}(\tau) = (-1,-1,-1,1,1,-1)$.

\noindent \textbf{Fourier Expansion of Functions of Permutations.} Since a permutation $\tau$ can be expressed as a binary vector, any function $f: \mathbb{S}_p \rightarrow \mathbb{R}$ can be thought of as a Boolean function. 
Then, given the discussion on Fourier expansions of Boolean functions in Section \ref{sec:notation}, any function $f: \mathbb{S}_p \rightarrow \mathbb{R}$ admits the following Fourier decomposition:
$f(\tau) = \sum_{S \subset [\binom{p}{2}]} \alpha_S \chi_{S}(\mathbf{v}(\tau)) \coloneqq \langle  \balpha_f, \bchi^{\tau} \rangle$, where $\balpha_f = [\alpha_S]_{S \in [\binom{p}{2}]} \in \mathbb{R}^{\binom{p}{2}}$, and $\bchi^{\tau} = [\chi_S(\mathbf{v}(\tau)]_{S \in [\binom{p}{2}]} \in \{-1,1\}^{\binom{p}{2}}$ for $\tau \in \mathbb{S}_p$. 

\noindent \textbf{Model and DGP.} We propose a similar model to the one discussed for combination, that is, model the potential outcome for a unit-permutation pair $(n,\tau)$ as
$Y_n^{(\tau)} = \langle \balpha_n, \bchi^{\tau} \rangle + \epsilon_n^{\tau}$, where we assume sparsity ($||\balpha||_0 = s \leq 2^{\binom{p}{2}}$), low-rank structure ($\text{rank}(\mathcal{A}) = r \in \{\min\{N,2^{\binom{p}{2}}\}\}$), and $\E[\epsilon_n^{\tau} ~|~ \mathcal{A}] = 0$. 
Additionally, one sufficient DGP for permutations is one that is analogous to what is stated above for combinations.

Expressing potential outcomes over permutations $Y_n^{(\tau)}$ as Boolean functions allows us to easily adapt the proposed estimator, and our theoretical results to rankings. 
For simplicity, we focus on combinations for the rest of the paper and provide a detailed discussion of our results for permutations in Appendix \ref{sec:permutations}.

\section{Combinatorial Inference Applications}
\label{sec:combinatorial_inference_applications}
In this section, we discuss how classical models for learning functions over Booleans relates to the proposed potential outcome model. 
In particular, we discuss two well-studied models for functions over combinations: low-degree boolean polynomials and $k$-Juntas.
We also discuss applications such as factorial design experiments and recommendation systems. 

\noindent \textbf{Low-degree boolean Polynomials.} 
A special instance of sparse boolean functions is a low-degree polynomial which is defined as follows.
For a positive integer $d \leq p$,  $f : \{-1,1\}^p \rightarrow \mathbb{R}$ is a $d$-degree polynomial if its Fourier transform satisfies the following for any input $\bx \in \{-1,1\}^p$
\begin{equation}
\label{eq:low_degree_fourier_transform}
 f(\bx) = \sum_{S \subset |p|, |S| \leq d} \alpha_S \chi_S (\bx).
\end{equation}
\noindent In this setting, $s  \leq \sum^d_{i=0} \binom{p}{i} \approx p^d$. That is, degree $d$-polynomials impose sparsity on the potential outcome by limiting the degree of interaction between interventions. Next, we discuss how potential outcomes are typically modeled as low-degree polynomials in factorial design experiments. 

\noindent
\emph{Factorial Design Experiments.} Factorial design experiments consist of $p$ treatments where each treatment arm can take on a discrete set of values, and units are assigned different combinations of these treatments. 
In the special case that each treatment arm only takes on two possible values, this experimental design mechanism is referred to as a $2^p$ factorial experiment. 
Factorial design experiments are widely employed in the social sciences, agricultural, and industrial applications \citep{duflo2007using,dasgupta2015causal,wu2011experiments}, amongst others. 
A common strategy to determine treatment effects in this setting is to assume that the potential outcome only depends on main effects and pairwise interactions, with higher-order interactions being negligible \citep{bertrand2004emily,eriksson2014employers,george2005statistics}. 
By only considering pairwise interactions, it is easy to see that $d = 2$, i.e., $s = p^2$.
The reader can refer to \cite{zhao2022regression} for a detailed discussion of various estimation strategies and their validity in different settings. 

The model and algorithm proposed (see Assumption \ref{ass:observation_model}) capture these various modeling choices commonly used in factorial design experiments by imposing sparsity on $\balpha_n$ and learning which of the coefficients are non-zero in a data-driven manner.
That is, the \method~algorithm can adapt to low-degree polynomials without pre-specifying the degree, $d$, i.e., it automatically adapts to the inherent level of interactions in the data.
However, the additional assumption made  compared to the literature is that there is structure across units, i.e., the matrix $\mathcal{A} = [\balpha_n]_{n \in [N]}$ is low-rank.
Further, we require that units are observed under multiple combinations, which is the case in recommendation engines where users are exposed to multiple combinations of goods or in crossover design studies in which units receive a sequence of treatments \cite{vonesh1996linear,jones2003design}. 

\noindent \textbf{$k$-Juntas.} Another special case of sparse Boolean functions are $k$-Juntas which only depend on $k < p$ input variables. More formally, a function $f : \{-1,1\}^p \to \mathbb{R}$ is a $k$-junta if there exists a set $K \subset [p]$ with $|K| = k$ such that the fourier expansion of $f$ can be represented as follows
\begin{equation}
\label{eq:k_junta}
    f(\bx) = \sum_{S \subset K} \alpha_{S} \chi_{S}(\bx)
\end{equation}
Therefore, in the setting of the $k$-Junta, the sparsity index $s \leq 2^k$. In contrast to low-degree polynomials where the function depends on all $p$ variables but limits the degree of interaction between variables, $k$-Juntas only depend on $k$ variables but allow for arbitrary interactions amongst them.  We discuss how two important applications can be modeled as a $k$-Junta.

\noindent 
\emph{Recommendation Systems.} Recommendation platforms such as \emph{Netflix} are often interested in recommending a combination of movies that maximizes a user's engagement with a platform. 
Here, units $n$ can be individual users, and $\pi$ represents combinations of different movies.
The potential outcomes are the engagement levels for a unit $n$ when presented with a combination of movies $\pi$, with $\balpha_n$ representing the latent preferences for that user. 
In this setting, a user's engagement with the platform may only depend on a small subset of movies. For example, a user who is only interested in fantasy will only remain on the platform if they are recommended movies such as \emph{Harry Potter}.
Under this behavioral model, the potential outcomes (i.e., engagement levels) can be modeled as a $k$-Junta with the non-zero coefficients of $\balpha_n$ representing the combinations of the $k$ movies that affect engagement level for a user $n$. 
The potential outcome observational model (Assumption \ref{ass:observation_model}) captures this form of sparsity while also reflecting the low-rank structure commonly assumed when studying recommendation systems.
Once again, \method~can adapt to $k-$Juntas without pre-specifying the subset of features $K$, i.e., it automatically learns the important features for a given user. 
This additional structure in $k-$Juntas is well-captured by the CART estimator, which leads to tighter finite-sample bounds, as detailed in Section \ref{subsec:CART_finite_sample}.

\noindent 
\emph{Knock-down Experiments in Genomics.} A key task in genomics is to identify which set of genes are responsible for a phenotype (i.e., physical trait of interest such as blood pressure) in a given patient. 
To do so, geneticists use knock-down experiments which measure the difference in the phenotype after eliminating the effect of a set of genes in an individual. 
To encode this process in the language of combinatorial causal inference, we can think of units as different individual patients or sub-populations of patients, an intervention as knocking a particular gene out,  and $\pi$ as a combination of genes that are knocked out. 
The potential outcome $Y_n^{(\pi)}$ is the expression of the phenotype for a unit $n$ when the combination of genes $\pi$ are eliminated via knock-down experiments, and the coefficients of $\balpha_n$ represent the effect of different combination of genes on the phenotype for unit $n$. 
A typical assumption in genomics is that phenotypes only depend on a small set of genes and their interactions. 
In this setting, one can model this form of sparsity by thinking of the potential outcome function as a $k$-junta, as well as capturing the similarity between the effect of genes on different patients via our low-rank assumption. 
%

%
%
%
%
%
%
%

\section{Identification of Potential Outcomes}
\label{sec:identification}
\vspace{-2mm}
We show $\E[Y^{(\pi)}_{n} \mid \mathcal{A}]$ can be written as a function of observed outcomes, i.e., we establish identification of our target causal parameter.
As discussed earlier, our model allows for \emph{unobserved confounding}: whether or not a unit is seen under a combination may be correlated with its potential outcome under that combination due to unobserved factors, as long as certain conditions are met.  
We introduce necessary notation and assumption required for our result. 
For a unit $n \in [N]$, denote the subset of combinations we observe them under as $\Pi_n \subseteq \Pi$. 
%
%
For $\pi \in \Pi$, let $\tilde{\bchi}^{\pi}_n \in \mathbb{R}^{2^p}$ denote the restricted Fourier characteristic vector where we zero out all coordinates of $\bchi^{\pi} \in \mathbb{R}^{2^p}$ that correspond to the coefficients of $\balpha_n$ which are zero. 
For example, if $\balpha_n = (1,1,0,0,\ldots 0)$ and $\bchi^{\pi} = (1,1,\ldots 1)$, then  $\tilde{\bchi}^{\pi}_n = (1,1,0, \ldots 0)$. 
We then make the following assumption.


\begin{assumption}[Donor Units]
\label{ass:donor_set_identification} Assume there exists a set of ``donor units'' $\mathcal{I} \subset [N]$, such that the following two conditions hold: 
\vspace{-2mm}
\begin{enumerate}[leftmargin=*]
 \item [(a)] Horizontal span inclusion: For any donor unit $u \in \mathcal{I}$ and combination $\pi \in \Pi$, suppose  $\tilde{\bchi}_u^{\pi} \in  span(\tilde{\bchi}_u^{\pi_i}: \pi_i \in \Pi_u)$. That is, there exists $\bbeta_{\Pi_u}^{\pi} \in \mathbb{R}^{|\Pi_{u}|}$ such that $\tilde{\bchi}_u^{\pi} = \sum_{\pi_{i} \in  \Pi_{u}} \beta_{\pi_i}^{\pi} \tilde{\bchi}_u^{\pi_i}.$
\vspace{-2mm}
 \item [(b)]  Linear span inclusion: For any unit $n  \in [N] \setminus \mathcal{I}$, suppose $\balpha_{n} \in span(\balpha_{u} : u \in \mathcal{I})$. That is, there exists $\bw^{n}$ such that $\balpha_{n} = \sum_{u \in \mathcal{I}}w_{u}^{n}\balpha_{u}$
\end{enumerate}
\end{assumption}
\vspace{-1.5mm}
Horizontal span inclusion requires that the set of observed combinations for any ``donor'' unit is ``diverse'' enough that the projection of the Fourier characteristic for a target intervention is in the span of characteristics of observed interventions for that unit. 
Linear span inclusion requires that the donor unit set is diverse enough such that the Fourier coefficient of any ``non-donor'' unit is in the span of the Fourier coefficients of the donor set. 

\noindent \textbf{Motivating example.} 
The following serves as motivation for existence of a donor set $\mathcal{I}$ satisfying Assumption \ref{ass:donor_set_identification}. 
Suppose Assumption \ref{ass:observation_model} holds and that  each unit belongs to one of $r$ \emph{types}.
That is, for all $n \in [N]$,
$\balpha_n \in \{\balpha(1),\balpha(2),\ldots,\balpha(r)\} \subset \mathbb{R}^{2^p},$ where each $\balpha(i)$ is $s$-sparse. 
Having only $r$ distinct sets of Fourier coefficients implies the low-rank property in Assumption \ref{ass:observation_model} (a). 
Suppose the donor set $\mathcal{I} \subset [N]$ is chosen by sampling a subset of units independently and uniformly at random with size satisfying $\Omega(r\log(r/\gamma))$. 
Similarly, sample $\Pi_\mathcal{I} \subset \Pi$ combinations independently and uniformly at random with size satisfying $\Omega(s\log(r|\mathcal{I}|/\gamma))$, and assign all donor units this set of combinations. 
Then, the following result shows this sampling scheme ensures both horizontal and linear span inclusion are satisfied with high probability as long as there are $\Omega(N)$ units of each type.

\begin{proposition}
\label{prop:motivating_example}
Let Assumptions \ref{ass:observation_model} and Assumptions \ref{ass:selection_on_fourier} hold. Moreover, suppose that for some $c > 0$, there are at least $cN/r$ units of each type. Then, under the proposed sampling scheme described above, Assumption \ref{ass:donor_set_identification} is satisfied with probability at least $1 - \gamma$.  
\end{proposition}
Proposition \ref{prop:motivating_example} shows that (ignoring logarithmic factors) sampling $r$ units and assigning these units $s$ combinations at random is sufficient to induce a valid donor set. 
In Section \ref{sec:experimental_design}, we show that Assumption \ref{ass:donor_set_identification} holds even when we relax our assumption that there are only $r$ types of units. 
Specifically, it is established that a randomly selected donor set of size $\omega(r\log(rs))$ will satisfy linear span inclusion as long as the $r$ non-zero singular values of the matrix $\mathcal{A}$ are of similar magnitude (see Assumption \ref{ass:balanced_spectrum} below).
Note that the motivating example discussed above does not induce unobserved confounding since the treatment mechanism is completely randomized (i.e., $\mathcal{D}$ does not depend on $\mathcal{A}$).
Hence, next we present a simple but representative example that both induces unobserved confounding and satisfies Assumption \ref{ass:donor_set_identification}.

\noindent \textbf{Natural model of unobserved confounding.} For any unit $n$, suppose  $\balpha_n \in \{\balpha(1) = (1, 1, 4, 0 \ldots, 0), \ \balpha(2) =(1, 1, 5, 0 \ldots, 0), \ \balpha(3) = (0.5, 0.5, 2, 0 \ldots, 0), \ \balpha(4) = (0.5, 0.5, 2.5, 0 \ldots, 0)\}$.
One can verify $\text{rank}(\mathcal{A}) = 2 $.
Define the treatment assignment to be such that combination $\pi$ for unit $n$ is observed only if $| \langle \balpha_n, \bchi^{\pi} \rangle | \geq 2$ (i.e., $| \mathbb{E}[Y^{(\pi)}_{n}] | \ge 2$).
Missingness patterns where only outcomes with large absolute values are observed is common in applications such as recommendation engines, where one is only likely to observe ratings for combinations that users either strongly like or dislike.
For units of type $\{\balpha(3),\balpha(4)\}$, we verify in Appendix \ref{subsec:proofs_natural_model} that they do not satisfy horizontal span inclusion (i.e., Assumption \ref{ass:donor_set_identification} (a) does not hold). 
For units of type $\{\balpha(1),\balpha(2)\}$, combinations $\{\pi_1,\pi_2,\pi_3,\pi_4\}$ with the following binary representations $\{v(\pi_1) = (1,1,1,1,\ldots,1), v(\pi_2) = (1,-1,-1,1,\ldots,1)\},v(\pi_3) = (1,-1,1,1,\ldots,1),v(\pi_4) = (1,1,-1,1,\ldots,1)$ are observed. 
These combinations have associated restricted Fourier characteristics $\Tilde{\bchi}^\pi$ as follows: $\{\Tilde{\bchi}^{\pi_1} = (1,1,1,0,\ldots,0), \Tilde{\bchi}^{\pi_4} = (1,-1,-1,0,\ldots,0)\}, \Tilde{\bchi}^{\pi_3} = (1,-1,1,0,\ldots,0),\Tilde{\bchi}^{\pi_2} = (1,1,-1,0,\ldots,0)$. 
One can verify that these observed combinations ensure that horizontal span inclusion holds for units with type $\{\balpha(1),\balpha(2)\}$.
Since $\text{rank}\{\balpha(1),\balpha(2)\} =2$, vertical span inclusion also holds.
It is straightforward to generalize this example to setting with $r$ types.

\vspace{-3mm}
\subsection{Identification Result}
\vspace{-2mm}
\noindent Given these assumptions, we now present our identification theorem. 
\label{subsec:identification_theorem}

\begin{theorem}
\label{thm:identification}
Let Assumptions \ref{ass:observation_model}, \ref{ass:selection_on_fourier}, and \ref{ass:donor_set_identification} hold. Given $\bbeta_{\Pi_u}^{\pi}$ and $\bw_u^{n}$ defined in Assumption \ref{ass:donor_set_identification}, we have
\noindent

\noindent (a) Donor units: For $u \in \mathcal{I}$, and $\pi \in \Pi$, 
$
 \E[Y^{(\pi)}_{u} ~ | ~ \mathcal{A}]  =  \sum_{\pi_u \in \Pi_{u}} \beta_{\pi_{u}}^{\pi}
 \E[Y_{u,\pi_{u}} \ | \  \mathcal{A}, \ \mathcal{D}].
$

\noindent
(b) Non-donor units: For $n \in [N] \setminus \mathcal{I}$, and $\pi \notin \Pi_n$,
$
\E[Y^{(\pi)}_{n} ~ | ~ \mathcal{A}] = \sum_{u \in \mathcal{I},\pi_u \in \Pi_u}w_{u}^{n}  \beta_{\pi_{u}}^{\pi}  \E[Y_{u,\pi_u} \ | \  \mathcal{A}, \ \mathcal{D}].
$
\end{theorem}

\noindent Theorem \ref{thm:identification} gives conditions under which the donor set $\mathcal{I}$ and the treatment assignments $\mathcal{D}$ are sufficient to recover the full set of unit specific potential outcomes $\E[Y_n^{(\pi)}|\mathcal{A}]$ in the noise-free limit.
Part (a) establishes that for every donor unit $u \in \mathcal{I}$, the causal estimand can be written as a function of its \emph{own} observed outcomes $\E[\bY_{\Pi_u}]$, given knowledge of $\bbeta^{\pi}_{\Pi_u}$. 
Part (b) states that the target causal estimand $\E[Y_n^{(\pi)}]$ for a non-donor unit and combination $\pi$ can be written as a linear combination of the outcomes of the donor set $\mathcal{I}$, given knowledge of $\bw_n^n$.
Previous work that establishes identification under a latent factor model  requires a growing number of donor units to be observed under all treatments \cite{agarwal2021causalmatrix}.
This is infeasible in our setting because \emph{the vast majority of combinations have no units that receive it}.
As a result, one has to first identify the outcomes of donor units under all combinations (part (a)), before transferring them to non-donor units (part (b)).
In order to do so, Theorem \ref{thm:identification} suggests that the key quantities in estimating $\E[Y_n^{(\pi)}]$ for any unit-combination pair $(n,\pi)$ are $\bbeta_{\Pi_u}^{\pi}$ and $\bw_u^n$. 
In the following section, we propose an algorithm to estimate both $\bbeta_{\Pi_u}^{\pi}$ and $\bw_u^n$, as well as concrete ways of determining the donor set $\mathcal{I}$.

\section{\method~Estimator}\label{sec:estimator_descripton}
\begin{figure}[htbp]
    \centering
    \includegraphics[width = \textwidth]{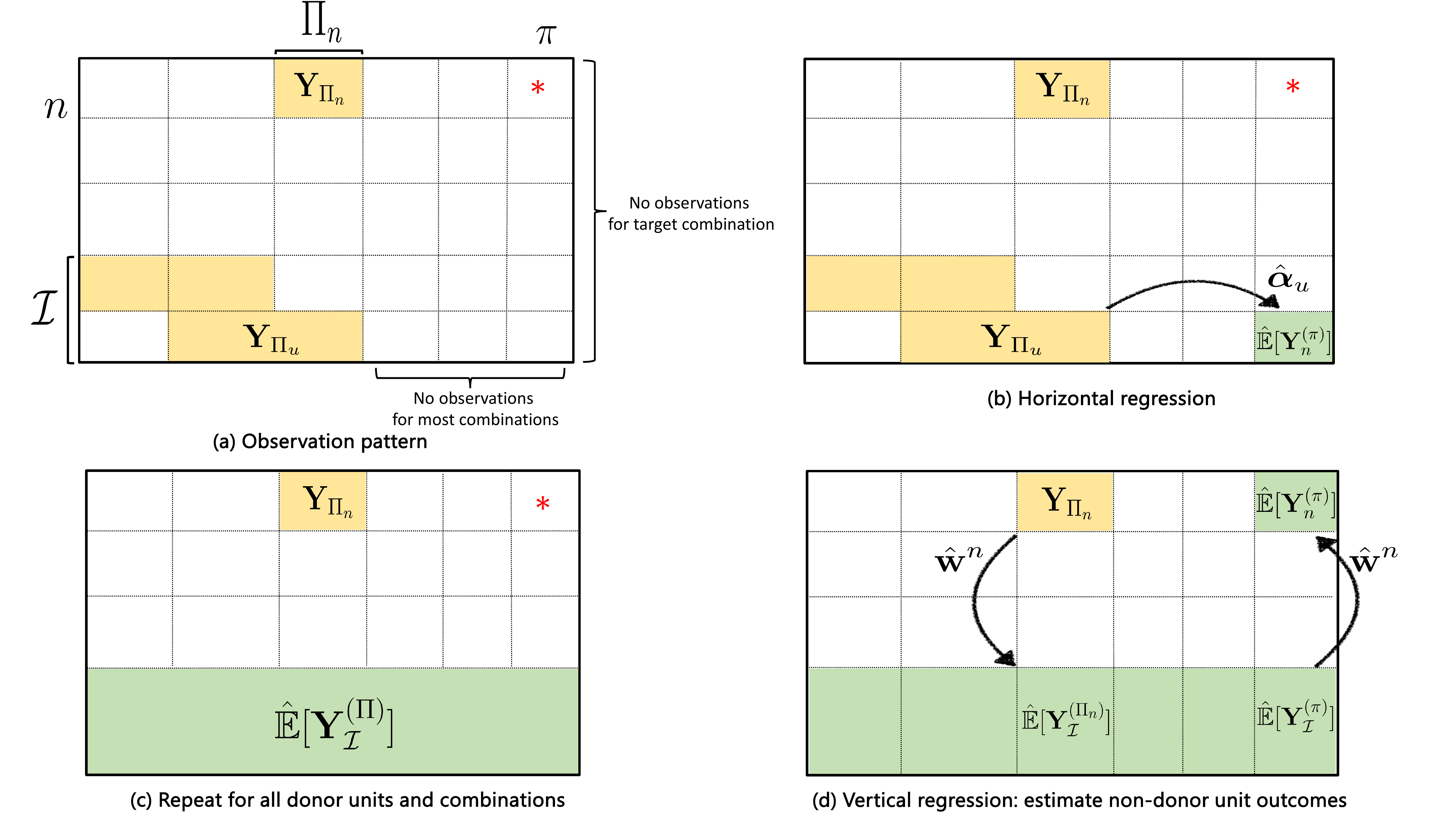}   
     \caption{
     A visual description of \method. Figure \ref{fig:estimator}(a) depicts an example of a particular observation pattern with outcome for unit-combination pair $(n,\pi)$ missing. 
     Figure \ref{fig:estimator}(b) demonstrates horizontal regression for donor unit $u$ to estimate all potential outcomes $\E[Y_u^{(\Pi)}]$.
     Figure \ref{fig:estimator}(c) displays repeating the horizontal regression for the entire donor set $\mathcal{I}$. 
     Figure \ref{fig:estimator}(d) visualizes vertical regression the  estimated outcomes from the donor set $\mathcal{I}$ are extrapolated to estimate outcomes for the unit-combination pair $(n,\pi)$.}
    \label{fig:estimator}
\end{figure}

We now describe the \method~estimator, a simple and flexible two-step procedure for estimating our target causal parameter. 
A pictorial representation of the estimator is presented in Figure \ref{fig:estimator}.

\vspace{2mm}
\noindent \textbf{Step 1: Horizontal Regression.}
For notational simplicity, denote the vector of observed responses $\bY_{n,\Pi_n} = [Y_{n\pi} : \pi \in \Pi_n] \in \mathbb{R}^{|\Pi_n|}$ for any unit $n$ as $\bY_{\Pi_n}$.
Then, for every unit $u$ in the donor set $\mathcal{I}$, $\E[Y_u^{(\pi)}]$ is estimated via the Lasso, i.e., by solving the following convex program with penalty parameter $\lambda_u$: 
\begin{align}\label{eq:Lasso_estimator}
\hat{\balpha}_u =  \argmin_{\balpha} \ \frac{1}{|\Pi_u|}\lVert \bY_{\Pi_u} - \bchi(\Pi_u)\balpha \rVert^2_2 + \lambda_u \lVert \balpha \rVert_1
\end{align}
where recall that $\bchi(\Pi_u) = [\bchi^\pi: \pi \in \Pi_u] \in \mathbb{R}^{|\Pi_u| \times 2^p}$.
Then, for any donor unit-combination pair $(u,\pi)$, let $\hat{\E}[Y_u^{(\pi)}] = \langle  \hat{\balpha}_u, \bchi^{\pi} \rangle$ denote the estimate of the potential outcome $\E[Y_u^{(\pi)}]$.
 %

\vspace{2mm}
\noindent \textbf{Step 2: Vertical Regression.} 
Next, estimate potential outcomes for all units $n \in [N] \setminus \mathcal{I}$.
To do so, some additional notation is required. 
For $\Pi_S \subseteq \Pi$, define the vector of estimated potential outcomes $\hat{\E}[\bY^{(\Pi_S)}_{u}] = [ \hat{\E}[Y_u^{(\pi)}]: \pi \in \Pi^S] \in \R^{|\Pi_S|}$. 
Additionally, let $\hat{\E}[\bY^{(\Pi_S)}_{\mathcal{I}}] = [\hat{\E}[\bY^{(\Pi_S)}_{u}]: u \in \mathcal{I}] \in \mathbb{R}^{|\Pi_S| \times |\mathcal{I}|}$.

\vspace{2mm}
\noindent \emph{Step 2(a): Principal Component Regression.} 
Perform a singular value decomposition (SVD) of $\hat{\E}[\bY^{(\Pi_n)}_{\mathcal{I}}]$ to get $\hat{\E}[\bY^{(\Pi_n)}_{\mathcal{I}}] = \sum^{\min(|\Pi_n|,|\mathcal{I}|)}_{l = 1} \hat{s_l}\hat{\bmu}_{l}\hat{\bnu}^T_{l}$. 
Using a hyper-parameter $\kappa_n \leq \min(|\Pi_n|,|\mathcal{I}|)$
\footnote{ 
Both $\lambda_u$ and $\kappa_n$ can be chosen in a data-driven manner (e.g., via cross-validation).}, 
compute $\hat{\bw}^{n} \in \mathbb{R}^{|\mathcal{I}|}$ as follows:
\begin{equation}
\label{eq:pcr_linear_model_def}
    \hat{\bw}^{n} = \left(\sum^{\kappa_n}_{l = 1} \hat{s_l}^{-1}\hat{\bnu}_{l}\hat{\bmu}^T_{l}\right)\bY_{\Pi_n} 
\end{equation}

\noindent \emph{Step 2(b): Estimation.} Using $\hat{\bw}^{n} = [\hat{w}_u^n : u \in \mathcal{I}]$, we have the following estimate for any intervention $\pi \in \Pi$
\begin{equation}
\label{eq:potential_outcome_estimate_vertical_regression}
     \hat{\E}[Y_n^{(\pi)}] = \sum_{u \in \mathcal{I}} \hat{w}_u^{n} \hat{\E}[Y_u^{(\pi)}]
\end{equation}

\noindent \textbf{Suitability of Lasso and PCR.} Lasso is appropriate for the horizontal regression step because it adapts to the sparsity of $\balpha_u$. However, it can be replaced with other algorithms that adapt to sparsity (see below for a larger discussion).
For vertical regression, PCR is appropriate because  $\mathcal{A}$ is low rank. 
As \cite{agarwal2019robustness,agarwal2020principal} show, PCR implicitly regularizes the regression by adapting to the rank of the covariates, i.e., $\E[\bY^{(\Pi_n)}_{\mathcal{I}}]$. 
As a result, the out-of-sample error of PCR scales with $r$ rather than the ambient covariate dimension, which is given by $\mathcal{I}$. 
\vspace{1mm}

\noindent \textbf{Determining Donor Set $\mathcal{I}$.} 
\method~requires the existence of a subset of units $\mathcal{I} \subset [N]$ such that one can (i) accurately estimate their potential outcomes under all possible combinations, and (ii) transfer these estimated outcomes to a unit $n \in [N] \setminus \mathcal{I}$.
Theoretically, we give sufficient conditions on the observation pattern such that we can perform (i) and (ii) accurately via the Lasso and PCR, respectively. 
In practice, the following practical guidance to determine the donor set $\mathcal{I}$ is recommended.
For every unit $n \in [N]$, learn a separate Lasso model $\hat{\balpha}_n$ and assess its performance through $k$-fold cross-validation (CV). 
Assign units with low CV error (with a pre-determined threshold) into the donor set $\mathcal{I}$, and estimate outcomes $\hat{\E}[Y_u^{(\pi)}]$ for every unit $u \in \mathcal{I}$ and $\pi \in \Pi$. 
For non-donor units, PCR performance can also be assessed via k-fold CV. 
For units with low PCR error, linear span inclusion (Assumption \ref{ass:donor_set_identification}(b)) and the assumptions required for the generalization for PCR likely hold, and hence we estimate their potential outcomes as in \eqref{eq:potential_outcome_estimate_vertical_regression}. 
For units with large PCR error, it is either unlikely that this set of assumptions holds or that $|\Pi_n|$ is not large enough (i.e., additional experiments need to be run for this unit), and hence we do not recommend estimating their counterfactuals. 
In our real-world empirics in Section \ref{sec:real_world_case_study}, we choose donor units via thie approach, and find \method~outperforms other methods. 

\vspace{1mm}
\noindent \textbf{Horizontal Regression Model Selection.} \method~allows for any ML algorithm (e.g., random forests, neural networks, ensemble methods) to be used in the first step. 
We provide an example of this flexibility by showing how the horizontal regression can also be done via CART in Section \ref{subsec:CART_finite_sample}.
Theorem \ref{thm:informal_CART_potential_outcome_convergence_rate} shows CART leads to better finite-sample rates when $\E[Y_n^{(\pi)}]$ is a $k$-Junta. 
This model-agnostic approach allows the analyst to tailor the horizontal learning procedure to the data at hand and include additional structural information for better performance. 
%
\section{Theoretical Analysis}\label{sec:theoretical_analysis}
In this section, finite-sample consistency of \method~is established, starting with a discussion of the additional assumptions required for the results. 

\subsection{Additional Assumptions}\label{subsec:finite_sample_analysis_assumptions}
\begin{assumption}[Bounded Potential Outcomes]\label{ass:boundedness_potential_outcome}
$\E[Y_{n}^{(\pi)}] \in [-1,1]$ for any unit-combination pair $(n,\pi)$.
\end{assumption}
\begin{assumption}[Sub-Gaussian Noise]\label{ass:subgaussian_noise}
Conditioned on $\mathcal{A}$, for any unit-combination pair $(n,\pi)$, $\epsilon_n^{\pi}$ are independent mean-zero sub-Gaussian random variables with $\text{Var}[\epsilon_n^{\pi} ~ | ~  \mathcal{A}] \le \sigma^2$ and $\lVert \epsilon_n^{\pi} ~ | ~ \mathcal{A} \rVert_{\psi_2} \leq C\sigma$ for some constant $C > 0$. 
\end{assumption}
\begin{assumption} [Incoherence of Donor Fourier characteristics]
\label{ass:incoherence}
For every unit $u \in \mathcal{I}$, assume $\bchi(\Pi_u)$ satisfies incoherence:
$
\left\lVert \frac{\bchi(\Pi_u)^T\bchi(\Pi_u)}{|\Pi_u|} - \mathbf{I}_{2^p}\right\rVert_{\infty} \leq \frac{C'}{s}$, 
where $\lVert \cdot \rVert_{\infty}$ denotes the maximum element, and $\mathbf{I}_{2^p} \in \mathbb{R}^{2^p \times 2^p}$ denotes the identity matrix, and $C'$ is a positive constant. 
\end{assumption}
To define the next set of assumptions,necessary notation is introduced. 
For any subset of combinations $\Pi_S \subset \Pi$, let $\E[\bY_{\mathcal{I}}^{(\Pi_S)}] = [\E[\bY_u^{(\Pi_S)}]: u \in \mathcal{I}] \in \mathbb{R}^{|\Pi_S| \times |\mathcal{I}|}$.
\begin{assumption} [Donor Unit Balanced Spectrum]\label{ass:balanced_spectrum}
For a given unit $n \in [N] \setminus \mathcal{I}$, let $r_n$ and $s_{1} \ldots s_{r_n}$ denote the rank and non-zero singular values of $ \E[\bY_{\mathcal{I}}^{(\Pi_n)} \ | \ \mathcal{A}]$, respectively. 
The singular values are well-balanced, i.e., for universal constants $c,c' > 0$, $s_{r_n}/s_{1} \geq c$, and $\lVert \E[\bY^{(\Pi_n)}_{\mathcal{I}} ~ | ~ \mathcal{A} ]\rVert^2_F \geq c'|\Pi_n||\mathcal{I}|$.
\end{assumption}

\begin{assumption} [Subspace Inclusion]\label{ass:rowspace_inclusion} 
For a given unit $n \in [N] \setminus \mathcal{I}$ and intervention $\pi \in \Pi \setminus \Pi_n$, $\E[\bY^{(\pi)}_{\mathcal{I}}]$ lies within the row-span of $\E[\bY_{\mathcal{I}}^{(\Pi_n)}]$
\end{assumption}
\vspace{-2mm}
Assumption \ref{ass:incoherence} is necessary for finite-sample consistency when estimating $\balpha_n$ via the Lasso estimator \eqref{eq:Lasso_estimator}, and is commonly made when studying the Lasso \citep{rigollet2015high}. 
Incoherence can also be seen as an inclusion criteria for a unit $n$ to be included in the donor set $\mathcal{I}$.  
Lemma 2 in \cite{negahban2012learning} shows that the Assumption \ref{ass:incoherence} is satisfied with high probability if $\Pi_u$ is chosen uniformly at random and grows as $\omega(s^2p)$. 
If Assumption \ref{ass:incoherence} does not hold, then the Lasso may not estimate $\balpha_u$ accurately, and alternative horizontal regression algorithms may be required instead. 
Assumption \ref{ass:balanced_spectrum} requires that the non-zero singular values of $\E[\bY_{\mathcal{I}}^{(\Pi_n)} ~ | ~ \mathcal{A}]$ are well-balanced. 
This assumption is standard when studying PCR \cite{agarwal2019robustness,agarwal2020synthetic}, and within the econometrics literature \cite{bai2021matrix,fan2018eigenvector}. 
It can also be empirically validated by plotting the spectrum of $\hat{\E}[\bY_{\mathcal{I}}^{(\Pi_n)}]$; if the singular spectrum of $\hat{\E}[\bY_{\mathcal{I}}^{(\Pi_n)}]$ displays a natural elbow point, then Assumption \ref{ass:balanced_spectrum} is likely to hold. 
Assumption \ref{ass:rowspace_inclusion} is also commonly made when analyzing PCR \cite{agarwal2020principal,agarwal2020synthetic,agarwal2021causal}.
It can be thought of as a ``causal transportability'' condition from the model learned using $\Pi_n$ to the interventions $\pi \in \Pi \setminus \Pi_n$. 
That is, subspace inclusion enables accurate estimation of $\E[Y_n^{(\pi)} ~ | ~ \mathcal{A} ]$ using $\langle \hat{\E}[\bY_{\mathcal{I}}^{(\Pi_n)}], \hat{\bw}^n \rangle$. 
In Section \ref{sec:experimental_design}, we propose a simple experimental design mechanism that ensures that Assumptions \ref{ass:incoherence}, \ref{ass:balanced_spectrum}, \ref{ass:rowspace_inclusion} (and Assumption  \ref{ass:donor_set_identification})---the key conditions for \method~to work---hold with high probability. 
\vspace{-2mm}

\subsection{Finite Sample Consistency}\label{subsec:finite_sample_consistency}
\vspace{-2mm}
The following result establishes finite-sample consistency of \method. 
Without loss of generality, we will focus on estimating the pair of quantities $(\E[Y_u^{(\pi)}],\E[Y_n^{(\pi)}])$ for a given donor unit $u \in \mathcal{I}$, and non-donor unit $n \in [N] \setminus \mathcal{I}$ for combination $\pi \in \Pi$.
To simplify notation, we utilize $O_p$ notation:
for any sequence of random vectors $X_n$, $X_n$ = $O_p(\gamma_n)$ if, for any $\epsilon > 0$, there exists constants $c_\epsilon$ and $n_\epsilon$ such that $\mathbb{P}(\lVert X_n \rVert_2 \geq c_\epsilon\gamma_n) \leq \epsilon$ for every $n \geq n_\epsilon$. 
Additionally, let $O_p$ absorb any dependencies on $\sigma$.
For notational simplicity, define $\tilde{O_p}(\gamma_n)$ which suppresses logarithmic terms. 

\begin{theorem} [Finite Sample Consistency of \method]
\label{thm:potential_outcome_convergence_rate}
Conditioned on $\mathcal{A}$, let Assumptions \ref{ass:observation_model}, \ref{ass:selection_on_fourier}, \ref{ass:donor_set_identification}, \ref{ass:boundedness_potential_outcome}, \ref{ass:subgaussian_noise}, and \ref{ass:incoherence} hold. 
Then, the following statements hold. 
\begin{itemize}
    \item [(a)] 
    For any donor unit-combination pair $(u,\pi)$, let the Lasso regularization parameter satisfy $\lambda_u = \Omega(\sqrt{\frac{p}{|\Pi_u|}})$, then 
        $$
            |\hat{\E}[Y_u^{(\pi)}] - \E[Y_{u}^{(\pi)}]| = O_p\left(s\sqrt{\frac{p}{|\Pi_u|}}\right).
        $$
    \item [(b)] 
    Additionally, let Assumptions \ref{ass:balanced_spectrum}, and \ref{ass:rowspace_inclusion} hold. 
    Then, for any non-donor unit-combination pair $(n,\pi)$ where $n \in [N] \setminus \mathcal{I}$, if $\kappa_n = \text{rank}(\E[\bY_{\mathcal{I}}^{(\Pi_n)}]) \coloneqq r_{n}$, and $ \min_{u \in \mathcal{I}} |\Pi_u| \coloneqq M = \omega(r_ns^2p)$, then
        $$
            \left |\hat{\E}[Y_n^{(\pi)}] - \E[Y_{n}^{(\pi)}]\right|  = \Tilde{O_p}\left(sr^2_n\sqrt{\frac{{p}}{{M}}} +  \frac{r_n}{|\Pi_n|^{1/4}} \right).
        $$
\end{itemize}

\end{theorem}
\vspace{-2.5mm}
Establishing Theorem \ref{thm:potential_outcome_convergence_rate} requires a novel analysis of error-in-variables (EIV) linear regression.
Specifically, the general EIV linear model is as follows: $Y = \bX \beta + \epsilon$, $\mathbf{Z} = \bX + \mathbf{H}$, where $Y$ and $\mathbf{Z}$ are observed.
In our case, $Y = \bY_{\Pi_n}$, $\bX = \E[\bY_{\mathcal{I}}^{(\Pi_n)}]$, $\beta = \bw^n$ , $\bZ = \hat{\E}[\bY_{\mathcal{I}}^{(\Pi_n)}]$, and $\mathbf{H}$ is the error arising in estimating $\E[\bY_{\mathcal{I}}^{(\Pi_n)}]$ via the Lasso.
Typically one assumes that $\mathbf{H}$ is is a matrix of independent sub-gaussian noise. 
Our analysis requires a novel worst-case analysis of $\mathbf{H}$ (due to the 2-step regression of Lasso and then PCR), in which each entry of is  $\mathbf{H}$ simply bounded.



We describe the conditions required on $|\Pi_n|$, $M$, for \method~to consistently estimate $(\E[Y_u^{(\pi)}], \ \E[Y_n^{(\pi)}])$. 
Recall $|\Pi_n|$ are the number of observations for the non-donor unit $n$ of interest, and $M$ is the minimum number of observations for a donor unit.
If $|\Pi_n| = \omega(r^4_n)$ and $M = \omega(r^4_ns^2p)$, then $\max\left(|\hat{\E}[Y_u^{(\pi)}] - \E[Y_{u}^{(\pi)}]|, \\ |\hat{\E}[Y_n^{(\pi)}] - \E[Y_{n}^{(\pi)}]|\right) =\tilde{o}_p(1)$.
Conversely, the corollary below quantifies how quickly the parameters $r_n$, $s$, and $p$ can grow with the number of observations $|\Pi_n|$ and $M$.
\begin{corollary}\label{cor:potential_outcome_convergence_rate_simplified}
With the set-up of Theorem \ref{thm:potential_outcome_convergence_rate}, if the following conditions hold: $r_n = o(|\Pi_n|^{1/4})$, and $s = o\left(\sqrt{\frac{M}{pr^4_n}} \right)$, then 
$
\max\left(|\hat{\E}[Y_u^{(\pi)}] - \E[Y_{u}^{(\pi)}]|, \\ \ |\hat{\E}[Y_n^{(\pi)}] - \E[Y_{n}^{(\pi)}]|\right) =\tilde{o}_p(1) 
$ as $M,|\Pi_n| \to \infty$. 
\end{corollary}

\vspace{-1.5mm}
\noindent Corollary \ref{cor:potential_outcome_convergence_rate_simplified} quantifies how $s, r_n$ can scale with the number of observations to achieve consistency. 
That is, it reflects the maximum ``complexity'' allowed for a given sample size. 

\subsection{Finite-Sample Consistency of CART}
\label{subsec:CART_finite_sample}

\noindent As discussed in Section \ref{sec:estimator_descripton}, \method~allows for any ML algorithm to be used for horizontal regression. 
In this section, we present theoretical results when the horizontal regression is done via CART. 
When the potential outcome function is a $k$-Junta (see equation \eqref{eq:k_junta} for the definition), CART is able to achieve stronger sample complexity guarantees than the Lasso. 
See Appendix \ref{sec:CART_horizontal_regression} for a detailed description of the CART algorithm.
Below, a brief overview of the CART-based horizontal regression procedure is given. 

\noindent \textbf{Step 1: Feature Selection.} 
For each donor unit $u \in \mathcal{I}$, divide the observed interventions $\Pi_u$ equally into two sets $\Pi^{a}_u$ and $\Pi^{b}_u$. 
Fit a CART model $\mathcal{\hat{T}}$ on the dataset $\mathcal{D}^a_u = [\{\bv(\pi),Y_{u\pi}\}: \pi \in \Pi^a_u]$. 
Let $\hat{K}_u = \{\bv(\pi)_j : \bv(\pi)_j$ $\in \mathcal{\hat{T}}\}$ denote every feature that the CART splits on. 

\noindent \textbf{Step 2: Estimation.} For every subset $S \subset \hat{K}_u$, compute $\hat{\alpha}_{u,S} = \frac{1}{|\Pi^b_u|}\sum_{\pi \in \Pi^b_u }Y_{u\pi}\chi^{\pi}_{S}$. The target causal parameter is then estimated as follows: $\hat{\E}[Y_u^{(\pi)}] = \sum_{S \in \hat{K}_u}\hat{\alpha}_{u,S}\chi^{\pi}_{S}$.

\noindent CART is able to take advantage of the $k$-Junta structure of the potential outcomes by first performing feature selection (i.e., learning the relevant feature set $\hat{K}_u$), before estimating the Fourier coefficient for each relevant subset.
In contrast, Lasso estimates the Fourier coefficient of all $2^p$ subsets simultaneously. 

Next, we present an informal theorem for the horizontal regression finite sample error when it is done via CART. 
The formal result (Theorem \ref{thm:CART_convergence_rate}) can be found in Appendix \ref{supp:CART_finite_sample}. 

\begin{theorem} [Informal]
\label{thm:informal_CART_potential_outcome_convergence_rate}
For a given donor unit $u$, suppose $\E[Y_u^{(\pi)}]$ is a $k$-Junta. Then, if the horizontal regression is done via CART, 
        $
            |\hat{\E}[Y_u^{(\pi)}] - \E[Y_{u}^{(\pi)}]| = \Tilde{O}_p\left(\frac{s}{\sqrt{|\Pi_u|}}\right).
       $
\end{theorem}
Compared to the Lasso horizontal regression error (see Theorem \ref{thm:potential_outcome_convergence_rate} (a)), CART removes an additional factor of $p$ by exploiting the $k$-Junta structure. 
It can be verified that CART also removes a factor of $p$ for the vertical regression error (see Theorem \ref{thm:potential_outcome_convergence_rate} (b)). 
As a result,  if $|\Pi_n| = \omega(r^4_n)$, and $M = \omega(r^4_ns^2)$, then $\max\left(|\hat{\E}[Y_u^{(\pi)}] - \E[Y_{u}^{(\pi)}]|, \ |\hat{\E}[Y_n^{(\pi)}] - \E[Y_{n}^{(\pi)}]|\right) =\tilde{o}_p(1)$.
That is, CART reduces the number of required observations for donor units by a factor of $p$.

%
%
%
%
%
%
%


\subsection{Sample Complexity}
\label{subsec:sample_complexity_synth_combo}
\vspace{-2.5mm}
We discuss the sample complexity of \method~to estimate all $N \times 2^p$ causal parameters, and compare it to that of other methods.
To ease our discussion, we ignore dependence on logarithmic factors, and assume that the horizontal regression is done via the Lasso. 

Even if potential outcomes $Y_n^{(\pi)}$ were observed for all unit-combination pairs, consistently estimating $\E[\bY_N^{(\Pi)}]$ is not trivial. 
This is because, we only get to observe a single and noisy version $Y_{n\pi} = \langle \balpha_n, \bchi^{\pi} \rangle + \epsilon_n^{\pi}$. 
Hypothetically, if $K$ independent samples of $Y_{n\pi}$ for a given $(n,\pi)$, denoted by $Y^{1}_{n\pi}, \ldots, Y^{K}_{n\pi}$ are observed, then the maximum likelihood estimator would be the empirical average $\frac{1}{K}\sum^K_{i=1}Y^{i}_{n\pi}$. 
The empirical average would concentrate around $\E[Y_n^{(\pi)}]$ at a rate $O(1/\sqrt{K})$ and hence would require $K = \Omega(\delta^{-2})$ samples to estimate $\E[Y_n^{(\pi)}]$ within error $O(\delta)$. 
Therefore, this naive (unimplementable) solution would require $N \times 2^p \times \delta^{-2}$ observations to estimate $\E[\bY_N^{(\Pi)}]$. 

On the other hand, \method~produces consistent estimates of the potential outcome despite being given {\em at most only a single noisy sample} of each potential outcome. 
As the discussion after Theorem \ref{thm:potential_outcome_convergence_rate} shows, \method~requires $|\mathcal{I}| \times r^4s^2p/\delta^2$ observations for the donor set, and $(N - |\mathcal{I}|) \times r^4/\delta^4$ observations for the non-donor units to achieve an estimation error of $O(\delta)$ for all $N \times 2^p$ causal parameters. 
To satisfy linear span inclusion (Assumption \ref{ass:donor_set_identification} (b)), and the necessary conditions for PCR (i.e., Assumption \ref{ass:rowspace_inclusion}), the donor set needs to have size $|\mathcal{I}| = \omega(r)$. 
Our experimental design mechanism in Section \ref{sec:experimental_design} shows that uniformly choosing a donor set of size $|\mathcal{I}| = O(r)$ satisfies these assumptions.
Hence, the number of observations required to achieve an estimation error of $O(\delta)$ for all pairs $(n,\pi)$  scales as $O\left(\text{poly}(r/\delta) \times \left(N + s^2p \right) \right)$.
%
%
%

\vspace{2mm}
\noindent \textbf{Sample Complexity Comparison to Other Methods.}  

\noindent \emph{Horizontal regression:} An alternative algorithm would be to learn an individual model for each unit $n \in [N]$. 
That is, run a separate horizontal regression via the Lasso for every unit.
This alternative algorithm has sample complexity that scales at least as $O(N \times s^2p/\delta^2)$  rather than $O\left(\text{poly}(r)/\delta^4  \times \left(N + s^2p \right) \right)$ required by \method.  
It suffers because it does not utilize any structure across units (i.e., the low-rank property of $\mathcal{A}$), whereas \method~captures the similarity between units via PCR. 

\vspace{1mm}
\noindent \emph{Matrix completion:} \method~can be thought of as a matrix completion method; estimating $\E[Y_n^{(\pi)}]$ is equivalent to imputing $(n,\pi)$-th entry of the observation matrix $\bY \in \{\mathbb{R} \cup \star\}^{N \times 2^p}$, where recall $\star$ denotes an unobserved unit-combination outcome. 
Under the low-rank property (Assumption \ref{ass:observation_model}(b)) and various models of missingness (i.e., observation patterns), recent works on matrix completion \cite{candes2010matrix,ma2019missing,agarwal2021causalmatrix} (see Section \ref{sec:related_work} for an overview) have established that estimating $\E[Y_n^{(\pi)}]$ to an accuracy $O(\delta)$ requires at least $O\left(\text{poly}(r/\delta) \times \left(N + 2^p \right)\right)$ samples.
This is because matrix completion techniques do not leverage the sparsity of $\balpha_n$.
%
Moreover, matrix completion results typically report error in the Frobenius norm, whereas we give entry-wise guarantees. 
This leads to an extra factor of $s$ in our analysis as it requires proving convergence for $\lVert \hat{\balpha_n} - \balpha_n \rVert_{1}$ rather than $\lVert \hat{\balpha_n} - \balpha_n \rVert_{2}$. 
%

\vspace{2mm}
\noindent \textbf{Natural Lower Bound on Sample Complexity.} We provide an informal discussion on the lower bound sample-complexity to estimate all $N \times 2^p$ potential outcomes. 
As established in Lemma \ref{lem:number_nonzeros}, $\mathcal{A}$ has at most $rs$ non-zero columns. 
Counting the parameters in the singular value decomposition of $\mathcal{A}$, only $r\times (N + rs)$ free parameters are required to be estimated. 
A natural lower bound on the sample complexity scales as $O(Nr + r^2s)$.  
Hence, \method~is only sub-optimal by a factor (ignoring logarithmic factors) of $sp$ and $\text{poly}(r)$.  
As discussed earlier, an additional factor of $s$ can be removed if we focus on deriving Frobenius norm error bounds.
Further, the use of non-linear methods such as CART can remove the factor of $p$ (Theorem \ref{thm:informal_CART_potential_outcome_convergence_rate}) under stronger assumptions on the potential outcomes, e.g., $\E[Y_n^{(\pi)}]$ is a $k$-Junta.
It remains as interesting future work to derive estimation procedures that are able to achieve this lower bound.

\section{Experiment Design}
\label{sec:experimental_design}
In this section, we show how \method~can be used to design experiments that allow for combinatorial inference (i.e., learning all $N \times 2^p$ causal parameters). 

\vspace{2mm}
\noindent \textbf{Key Assumptions Behind Finite Sample Consistency of \method.} \method~requires the existence of a donor set $\mathcal{I}$ such that we are able to perform accurate horizontal regression for all donor units, and then transfer these estimated outcomes to non-donor units via PCR. 
The enabling conditions for accurate horizontal regression  are (i) horizontal span inclusion (Assumption \ref{ass:donor_set_identification} (a)), and (ii) incoherence of the Fourier characteristics (Assumption \ref{ass:incoherence}).   
Similarly, the critical assumptions required for consistency of PCR are (i) linear span inclusion (Assumption  \ref{ass:donor_set_identification} (b)), (ii) well-balanced spectrum (Assumption \ref{ass:balanced_spectrum}), and (iii) subspace inclusion (Assumption \ref{ass:rowspace_inclusion}). 
In terms of experiment design, the donor set and treatment assignments must be carefully chosen such that these key assumptions hold.
To that end, we introduce the following design. 
%
%

\begin{figure}[htbp]
    \centering
    \includegraphics [width = \textwidth]{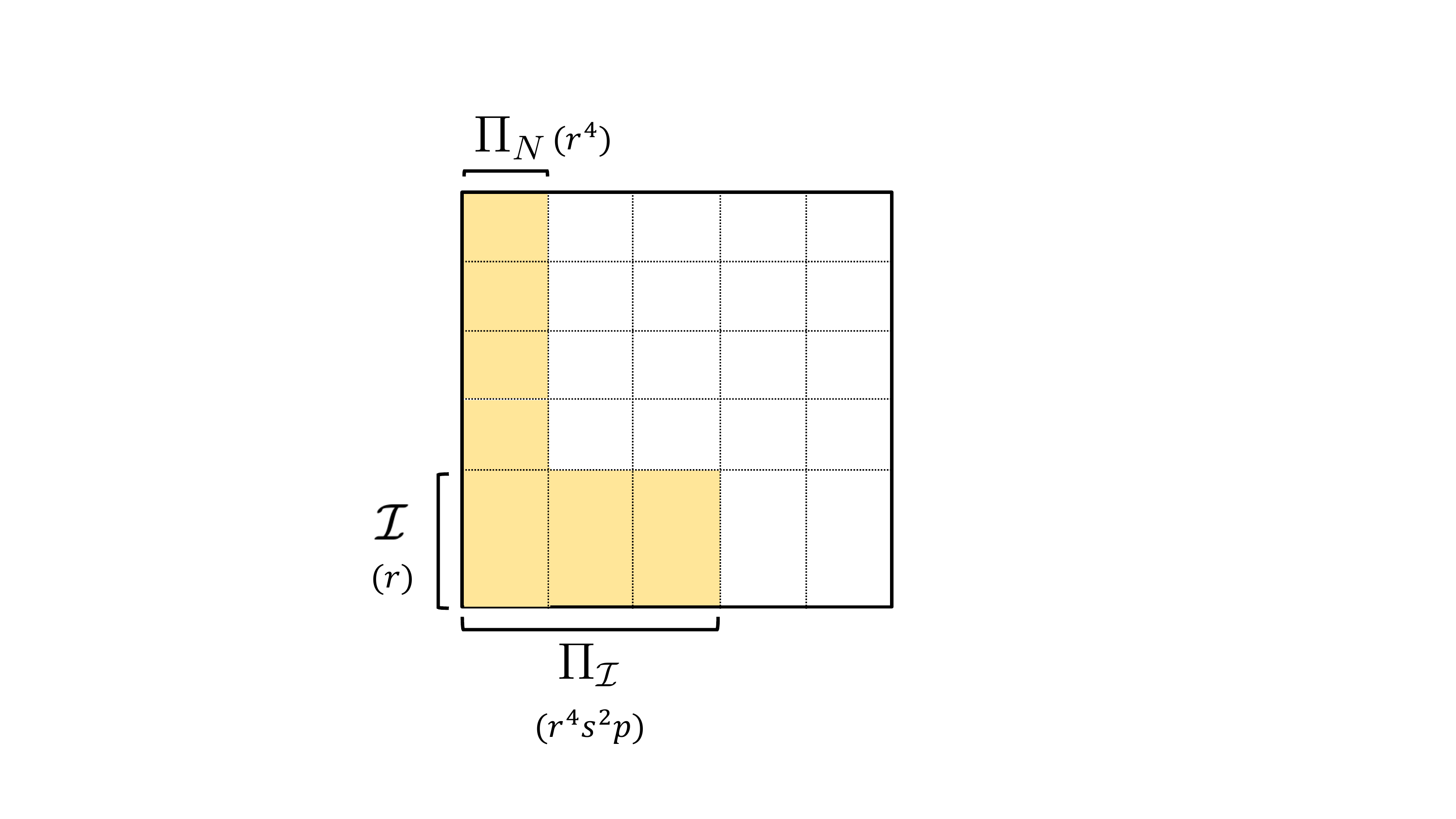}
    \caption{Observation pattern induced by experiment design mechanism. }
    \label{fig:experiment_design_observation_pattern}
\end{figure}

\vspace{2mm}
\noindent \textbf{Experimental Design Mechanism.} Fix a probability threshold $\gamma \in (0,1)$ and estimation error threshold $\delta \in (0,1)$. Our design mechanism (see Figure \ref{fig:experiment_design_observation_pattern} for a a visual description) then proceeds as follows.

\vspace{1mm}
\noindent
 \emph{Step 1: Donor set selection.} Choose the donor set $\mathcal{I} \subset [N]$ by sampling a subset of units independently and uniformly at random with size satisfying $\Omega\left(r\log(rs/\gamma) \right)$. 

\vspace{1mm}
\noindent 
\emph{Step 2: Donor set treatment assignment.} Sample $\Pi_{\mathcal{I}} \subset \Pi$ combinations independently and uniformly at random with size satisfying $|\Pi_{\mathcal{I}}| = \Omega\left(\frac{r^3s^2\log(|\mathcal{I}|2^p/\gamma)}{\delta^2}\right)$. Assign all donor units $u \in \mathcal{I}$ this set of combinations. 

\vspace{1mm}
\noindent
\emph{Step 3: Non-donor unit treatment assignment.} Randomly sample $\Pi_N \subset \Pi$ combinations independently and uniformly at random  of size $|\Pi_N| = 
\Omega\left(r\log(|\mathcal{I}|/\gamma) \vee r^4/\delta^4 \right)$. Assign all non-donor units $n \in [N] \setminus \mathcal{I}$ combinations $\Pi_N$. 

\vspace{1mm}
\noindent
\emph{Potential outcome estimation.} Given the observation pattern described above, estimate outcomes for each unit-combination pair via \method~as described in Section \ref{sec:estimator_descripton}. 
\vspace{0.5mm}

It turns out that this simple design mechanism satisfies the key assumptions presented above with high probability. 
In fact, the proposed mechanism ensures these key conditions hold under very limited assumptions.
The only required assumptions are that: (i) the potential outcome model is satisfied (Assumption \ref{ass:observation_model}), (ii) $\E[Y_n^{(\pi)}]$ is bounded (Assumption \ref{ass:boundedness_potential_outcome}). 
Additionally, only a weakened version of the balanced spectrum condition is needed (Assumption \ref{ass:balanced_spectrum}):
\begin{assumption} [Restricted Balanced Spectrum] 
\label{ass:restricted_balanced_spectrum}
Let $s_{1} \ldots s_{r}$ denote the non-zero singular values of $\E[\bY_N^{(\Pi)}]$. 
Assume that its singular values are well-balanced, i.e., for universal constants  $c,c' > 0$, we have that $s_{r}/s_{1} \geq c$, and $\lVert \E[\bY^{(\Pi)}_{N} ~ | ~ \mathcal{A} ]\rVert^2_F \geq c'N2^p$
\end{assumption}
\noindent As compared to the original balanced spectrum condition, Assumption \ref{ass:restricted_balanced_spectrum} only requires that the singular values are balanced for the entire potential outcome matrix as opposed to a collection of submatrices. 
We then have the following result.
\begin{theorem} 
\label{thm:experiment_design_assumptions_hold}
Let Assumptions \ref{ass:observation_model}, \ref{ass:boundedness_potential_outcome}, and \ref{ass:restricted_balanced_spectrum} hold. 
Then, the proposed experimental design mechanism ensures satisfies Assumption \ref{ass:selection_on_fourier}, and the following conditions simultaneously with probability at least $1 - \gamma$: 
(i) horizontal and linear span inclusion (Assumption \ref{ass:donor_set_identification}), 
(ii) incoherence of donor unit Fourier characteristics (Assumption \ref{ass:incoherence}), well-balanced spectrum (Assumption \ref{ass:balanced_spectrum}) and subspace inclusion (Assumption \ref{ass:rowspace_inclusion}).
\end{theorem}
\begin{corollary} 
\label{cor:experimental_design_error_rate}
Let the set-up of Theorem \ref{thm:experiment_design_assumptions_hold} hold. Then, for every unit-combination pair $(n,\pi)$, we have $|\E[Y_n^{(\pi)}] - \hat{\E}[Y^{(\pi)}_n]| = \Tilde{O}_p (\delta)$.  
\end{corollary}
\noindent  Theorem \ref{thm:experiment_design_assumptions_hold}  implies that the key enabling conditions for \method~are satisfied for every unit-combination pair $(n,\pi)$. 
%
%
Corollary \ref{cor:experimental_design_error_rate} further establishes that with this design, $O(\delta)$ error is achievable for all $N \times 2^p$ causal parameters.  
In contrast, the results in the observational setting do not guarantee accurate estimation of {\em all} $N \times 2^p$ parameters.
Instead, they establish that $\E[Y_n^{(\pi)}]$ can be learned for any specific unit-combination pair $(n,\pi)$ that satisfies the required assumptions on the observation pattern. 
Additionally, given the discussion in Section \ref{subsec:sample_complexity_synth_combo}, one can verify that the number of observations required by this experiment design mechanism scales as $\tilde{O}\left(\text{poly}(r/\delta)\times \left(N + s^2p \right) \right)$. 
In practice, this design requires knowledge of $r,$ and $s$.
To overcome this, one can sequentially sample donor units and their observations until the rank and lasso error stabilizes. 
This provides an estimate of $r,$ and $s$; a formal analysis of this procedure is left as future work.


\section{Asymptotic Normality}
\label{sec:asymptotic_normality}
We now establish asymptotic normality of \method~under additional conditions.
Since \method~is agnostic to the learning algorithm used in the horizontal regression, as demonstrated by the discussion on CART, the Lasso can be replaced by any regression technique that achieves asymptotic normality. 
For example, previous works have studied variants of the Lasso \cite{liu2013asymptotic,zou2006adaptive}, where it has been established that the predictions are asymptotically normal.
Here, we propose replacing the Lasso with the ``Select + Ridge'' (SR) procedure of \cite{liu2013asymptotic}, and specify the conditions on the donor units' observation pattern under which SR achieves asymptotic normality. 
Next, using asymptotic normality of the horizontal regression predictions, asymptotic normality of the vertical regression step of \method~is established.

\subsubsection{Horizontal Regression Asymptotic Normality}
\label{subsec:horizontal_regression_asymptotic_normlity}
The SR estimator consists of the following two steps. 

\noindent \textbf{Step 1: Subset Selection.} For every unit $u$ in the donor set $\mathcal{I}$, fit a Lasso model with regularization parameter $\lambda_u$ using  $\bchi(\Pi_u)$ as described in Step 1 of \method~to obtain $\hat{\balpha}_u$.
Let $\hat{\mathcal{S}}_u = [S \subset [p]: \hat{\alpha}_{u,S} \neq 0]$ denote the non-zero coefficients of $\hat{\balpha}_u$.

\noindent Some additional notation needs to be defined for the next step.
For a unit $u \in \mathcal{I}$, and combination $\pi$, let $\bchi_{\hat{\mathcal{S}}_u}^{\pi} = [\chi^{\pi}_{S} : S \in \hat{\mathcal{S}}_u] \in \{-1,1\}^{|\hat{\mathcal{S}}_u|}$. 
Let $\bchi_{\hat{\mathcal{S}}_u}(\Pi_u) = [\bchi_{\hat{\mathcal{S}}_u}^{\pi}: \pi \in \Pi_{u}] \in \{-1,1\}^{|\Pi_{u}| \times |\hat{\mathcal{S}}_u|}$.

\vspace{1mm}

\noindent \textbf{Step 2: Ridge Regression.} In this step, SR forms predictions using the selected subsets in Step 1, and ridge regression. 
Specifically, for every $u \in \mathcal{I}$, compute the Fourier coefficient as follows,
\begin{align}\label{eq:Ridge_estimator}
\hat{\balpha}^{SR}_{u} =  \left(\left(\bchi_{\hat{\mathcal{S}}_u}(\Pi_u)\right)^T\bchi_{\hat{\mathcal{S}}_u}(\Pi_u) + \frac{1}{|\Pi_u|}\mathbf{I}_{|\hat{\mathcal{S}}_u|} \right)^{-1} \left(\bchi_{\hat{\mathcal{S}}_u}(\Pi_u)\right)^T \bY_{\Pi_u},
\end{align}
where $\mathbf{I}_{|\hat{\mathcal{S}}_u|} \in \mathbb{R}^{|\hat{\mathcal{S}}_u| \times |\hat{\mathcal{S}}_u|}$ is the identity matrix of size $|\hat{\mathcal{S}}_u|$. 
Then, let $\hat{\E}_{SR}[Y_u^{(\pi)}] = \langle  \hat{\balpha}^{SR}_{u}, \bchi^{\pi} \rangle$ denote the estimated potential outcome $\E[Y_u^{(\pi)}]$ for a combination $\pi$.

%

Next, we present our result establishing asymptotic normality of SR predictions.
Some additional notation is required for our result. 
For a given  unit $u \in \mathcal{I}$,  let $\mathcal{S}_u = \{S \subset [p] ~ | ~ \alpha_{u,S} \neq 0 \}$. 
Like above, define $\bchi_{\mathcal{S}_u}^{\pi} = [\chi^{\pi}_{S} : S \in \mathcal{S}_u] \in \{-1,1\}^{|\mathcal{S}_u|}$ and $\bchi_{\mathcal{S}_u}(\Pi_u) = [\bchi_{\mathcal{S}_u}^{\pi}: \pi \in \Pi_{u}] \in \{-1,1\}^{|\Pi_{u}| \times |\mathcal{S}_u|}$.
Denote $\mathbf{K}_{u} = \frac{1}{|\Pi_u|}((\bchi_{\mathcal{S}_u}(\Pi_u))^T\bchi_{\mathcal{S}_u}(\Pi_u))$ as the covariance matrix of the observed Fourier characteristics for a donor unit $u$.
For a square matrix $\bX$, let $\lambda_{\min}(\bX)$ denote its smallest eigenvalue.

\begin{proposition} 
\label{prop:horizontal_asymptotic_normality}

Conditioned on $\mathcal{A}$, let Assumptions \ref{ass:observation_model} and \ref{ass:boundedness_potential_outcome} hold. Additionally, assume the following conditions hold for every donor unit $u$ and combination $\pi$,

\begin{enumerate}
    \item[(a)] $\epsilon_u^{\pi}$ are independent mean-zero independent Gaussian random variables with variance $\sigma^2$,
    \item [(b)] For every $S \subset [p]$, $\sum_{\pi \in \Pi_u} \chi_{S}(\pi) = 0$,
    \item [(c)]  $\lambda_{\min}(\mathbf{K}_u) \geq c_1$ for a positive constant $c_1$,
    \item [(d)] There exists constants $ 0 < c_2 \leq 1 $ and $0 < c_3 < 1 - c_2$ such that the sparsity $s = O(|\Pi_u|^{c_2})$ and $p = O(|\Pi_u|^{c_3})$,
    \item [(e)] $\P(\hat{\mathcal{S}_u} \neq \mathcal{S}_u) = o(e^{-|\Pi_u|^{c_3}})$
   
\end{enumerate}
Then, as $|\Pi_u| \rightarrow \infty$, we have
\begin{equation}
    \sqrt{\frac{|\Pi_u|}{\sigma^2 (\bchi^{\pi})^TK_u^{-1}\bchi^\pi }} \left(\hat{\E}_{SR}[Y_u^{(\pi)}] - \E[Y_u^{(\pi)}] \right)  \xrightarrow{d} \mathcal{N}(0,1)
\end{equation}
\end{proposition}

\noindent Proposition \ref{prop:horizontal_asymptotic_normality} builds upon Theorem 3 of \cite{liu2013asymptotic} to establish asymptotic normality and as a result allows for construction of valid confidence intervals for the donor units. 
Condition (b) can be enforced by simply normalizing each row of $\bchi(\Pi_u)$ to have mean zero. 
Condition (c) is mild, and is required in order to ensure that the model is identifiable even if $\mathcal{S}_u$ is known \cite{wainwright2019high}. 
Condition (d) limits the maximum sparsity and number of interventions allowed in terms of the sample size $|\Pi_u|$.
Equivalently, condition (d) states that $|\Pi_u| = \Omega(s^{1/c_2})$ and $|\Pi_u| = \Omega(p^{1/c_3})$, i.e., at least a polynomial number of samples is required in both $s$ and $p$. 
Condition (e) states the probability of recovering the wrong support set (i.e., $\hat{\mathcal{S}}_u \neq \mathcal{S}_u$) decays exponentially quickly with $|\Pi_u|$.
Consistency of recovering the support is necessary for asymptotic normality, since $\hat{\balpha}^{SR}_u$ needs to be unbiased, which requires $\hat{\mathcal{S}}_u = \mathcal{S}_u$.  
Previous works \cite{wainwright2019high,zhao2006model} have established that $\P(\hat{\mathcal{S}_u} \neq \mathcal{S}_u)$ decays exponentially quickly for the Lasso under certain regularity conditions. 
Next, we use normality of the horizontal regression predictions to establish asymptotic normality for non-donor units. 

\subsubsection{Vertical Regression Asymptotic Normality}
This section establishes asymptotic normality for non-donor units assuming that the horizontal regression predictions are asymptotically normal.
This can be achieved, for example, by using the SR procedure described above. 
However, the result in this section does not assume the use of any particular horizontal regression method since step 1 of \method~is method agnostic. 
Rather, it is just assumed that the donor unit predictions are normal, allowing for the use of any horizontal regression method with asymptotically normal predictions---the variant of Lasso in Section \ref{subsec:horizontal_regression_asymptotic_normlity} being one such concrete example.
Intuitively, since non-donor unit predictions are a linear combination of donor set estimates, normality of the horizontal regression is necessary to ensure asymptotic normality of the vertical regression. 

We define some necessary notation for our result. 
Let $\Tilde{\bw}^n = \bV_{\mathcal{I}}^{(\Pi_n)}(\bV_{\mathcal{I}}^{(\Pi_n)})^T\bw^n \in \mathbb{R}^{|\mathcal{I}|}$, where $\bV_{\mathcal{I}}^{(\Pi_n)}$ are the right singular vectors of $\E[\bY_{\mathcal{I}}^{(\Pi_n)}]$.
That is, $\Tilde{\bw}^n$ denotes the orthogonal projection of $\bw^n$ onto the rowspace of $\E[\bY_{\mathcal{I}}^{(\Pi_n)}]$.
Let $\Tilde{w}_u^n \in \mathbb{R}$ denote a coordinate of $\Tilde{\bw}^n$ corresponding to a given donor unit $u \in \mathcal{I}$.
Additionally, recall that $M \coloneqq \min_{u \in \mathcal{I}} |\Pi_u| $.
%

\begin{theorem} 
\label{thm:vertical_regression_normality}

For a given non-donor unit and combination pair $(n,\pi)$, let the set-up of Theorem \ref{thm:potential_outcome_convergence_rate} hold. 
Define 
\begin{equation*}
    \Tilde{\sigma}_u= \left(\sqrt{\frac{\sigma^2 (\bchi^{\pi})^TK_u^{-1}\bchi^\pi }{|\Pi_u|}} \right)  \Tilde{w}_u^n
\end{equation*}
Additionally, let the following conditions hold,
\begin{enumerate}
    \item [(a)] $|\Pi_n|, \ M \rightarrow \infty $,
    \item [(b)]
    \begin{equation} 
    \label{eq:asymptotic_normality_condition_1}
       \frac{\log^{3}(|\Pi_n||\mathcal{I}|)}{\lVert \Tilde{\bw}_n \rVert_2}  \left(sr^2_n\sqrt{\frac{p}{{M}}} +  \frac{r_n}{|\Pi_n|^{1/4}}\right) = o(1)
    \end{equation}
    
\end{enumerate}
\begin{enumerate}
    \item [(c)] For every donor unit $u \in \mathcal{I}$, as $|\Pi_u| \rightarrow \infty$, assume
    \begin{equation} 
    \label{eq:asymptotic_normality_condition_2}
        \sqrt{\frac{|\Pi_u|}{\sigma^2 (\bchi^{\pi})^TK_u^{-1}\bchi^\pi }} \left(\hat{\E}[Y_u^{(\pi)}] - \E[Y_u^{(\pi)}] \right)  \xrightarrow{d} \mathcal{N}(0,1),
    \end{equation}
\end{enumerate}
Then, we have that
\begin{equation} \label{eq:asymptotic_normality_vertical_regression}
    \frac{\hat{\E}[Y_n^{(\pi)}] - \E[Y_{n}^{(\pi)}]}{\sqrt{\sum_{u \in \mathcal{I}} \hspace{0.5mm} \Tilde{\sigma}^2_u}} \xrightarrow{d} \mathcal{N}(0,1).
\end{equation}
\end{theorem} 

\noindent Theorem \ref{thm:vertical_regression_normality} establishes asymptotic normality for non-donor unit and combination pair $(n,\pi)$, allowing for construction of valid confidence intervals. 
For \eqref{eq:asymptotic_normality_condition_1} to hold,  $\lVert \Tilde{\bw}_n \rVert_2$ needs to be sufficiently large (e.g., $\lVert \Tilde{\bw}_n \rVert_2 \geq c$, for a positive constant $c$), and (ignoring $\log$ factors) if  $M = \omega(r^{4}_n s^2p)$, and $|\Pi_n| = \omega(r_n^4)$. 
Recall from the discussion in Section \ref{sec:theoretical_analysis}, $M = \omega(r^{4}_n s^2p)$, and $|\Pi_n| = \omega(r_n^4)$ are precisely the conditions required for consistency as well. 
As shown in Section \ref{subsec:horizontal_regression_asymptotic_normlity}, \eqref{eq:asymptotic_normality_condition_2} holds when using the SR estimator for horizontal regression. 
%


%



%
\section{Simulations}
\label{sec:sims}
In this section, we corroborate our theoretical findings with numerical simulations in both the observational and experiment design setting.
\method~is compared to the Lasso (i.e., running a separate horizontal regression  for every unit) and two benchmark matrix completion algorithms: \texttt{SoftImpute} \citep{mazumder2010spectral}, and  \texttt{IterativeSVD} \citep{troyanskaya2001missing} . 
We also try running our experiments using the nuclear-norm minimization method introduced in \citep{candes2012exact}, but the method was unable to be run within a reasonable time frame (6 hours).\footnote{Code for \method~can be found at \url{https://github.com/aagarwal1996/synth_combo}.}

\subsection{Observational Setting}
\label{subsec:observational_sims}
For this numerical experiment, we simulate an observation pattern commonly found in applications such as recommendation engines, where most users tend to provide ratings for combinations of goods they either particularly liked or disliked, i.e., there is confounding.
The precise experimental set-up for this setting is as follows.

\noindent \textbf{Experimental Set-up.} We consider $N = 100$ units, and vary the number of interventions $p \in \{10,11, \ldots 15\}$. 
Further, let $r = 3$, and $s = rp^{3/2}$. 
Next, we describe how we generate the potential outcomes and observation. 


\noindent \emph{Generating potential outcomes.} For every $i \in [r],$ $\balpha_i$ is generated by sampling $p^{3/2}$ non-zero coefficients at random.  
Every non-zero coefficient of $\balpha_i$ is then sampled from a standard normal distribution. 
Denote $\mathcal{A}_r = [\balpha_i : i \in [r]] \in \mathbb{R}^{r \times 2^p}$.  
Next, let $\mathcal{A}_{N - r}= \mathbf{B}\mathcal{A}_r \in \mathbb{R}^{(N - r) \times 2^p}$, where $\mathbf{B} \in \mathbb{R}^{(N - r) \times r}$ is sampled i.i.d from a standard Dirichlet distribution. 
Let $\mathcal{A} = [\mathcal{A}_r, \mathcal{A}_{N-r}]$; by construction,  $\text{rank}(\mathcal{A}) = r$, and $\lVert \balpha_n \rVert_0 \leq  rp^{3/2} = s$ for all $n \in [N]$.
Normally distributed noise $\epsilon \sim N(0,\sigma^2)$ is added to the potential outcomes $\E[Y_n^{(\pi)}]$, where $\sigma^2$ is chosen such that the  signal-to-noise ratio is 1.0.

\noindent \emph{Observation pattern.} Denote $p_{n,\pi}$ as the probability that $\E[Y_n^{(\pi)}]$ is observed.
Let $p_{n,\pi} = |\E[Y_n^{(\pi)}]|/\sum_{\pi \in \Pi} |\E[Y_n^{(\pi)}]|$; this definition induces the missingness pattern where outcomes with larger absolute values are more likely to be seen.
$|\mathcal{I}| = 2r$ donor units are chosen uniformly at random. 
For each donor unit $u \in \mathcal{I}$, randomly sample $|\Pi_u| = 2p^{5/2}$ combinations without replacement, where each combination $\pi$ is chosen with probability $p_{u,\pi}$.
That is, each donor unit $u$ is assigned $2p^{5/2}$ treatments, where the outcome $Y_u^{(\pi)}$ is observed with probability $p_{u,\pi}$.
For each non-donor unit $n$, generate $2r^4$ observation according to the same procedure. That is, $|\Pi_n| = 2r^{4}$ combinations are randomly sampled without replacement, where each combination $\pi$ is chosen with probability $p_{n,\pi}$.

\noindent \emph{Hyper-parameter choices}. The hyper-parameters of the sub-procedures of~\method, i.e., Lasso and PCR, are tuned via $5-$fold CV. 
\texttt{SoftImpute} and \texttt{IterativeSVD} require that the rank $r$ of the underlying matrix to be recovered is provided as a hyper-parameter.
We provide the true rank $r = 3$ to both algorithms.

\noindent \textbf{Results.} We measure mean squared error (MSE) averaged over 5 repetitions between the estimated potential outcome matrix and the true potential outcome matrix for each method. 
The results are displayed in Figure \ref{fig:sim_results} (a).
\method~outperforms other approaches as $p$ grows.  
Further, the gap in performance between \method~and the Lasso enforces the utility of using PCR for non-donor units that do not have sufficient measurements.

\subsection{Experimental Design Simulations}
\label{subsec:experimental_design_sims}

\noindent \textbf{Experimental Set-up.} Outcomes are generated as in the observational setting. 
The observation pattern is generated by the experimental design mechanism described in Section \ref{sec:experimental_design}.
%

\begin{figure}[htbp]
    \centering
    \begin{minipage}{0.5\textwidth}
        \centering
        \includegraphics[width=1\textwidth]{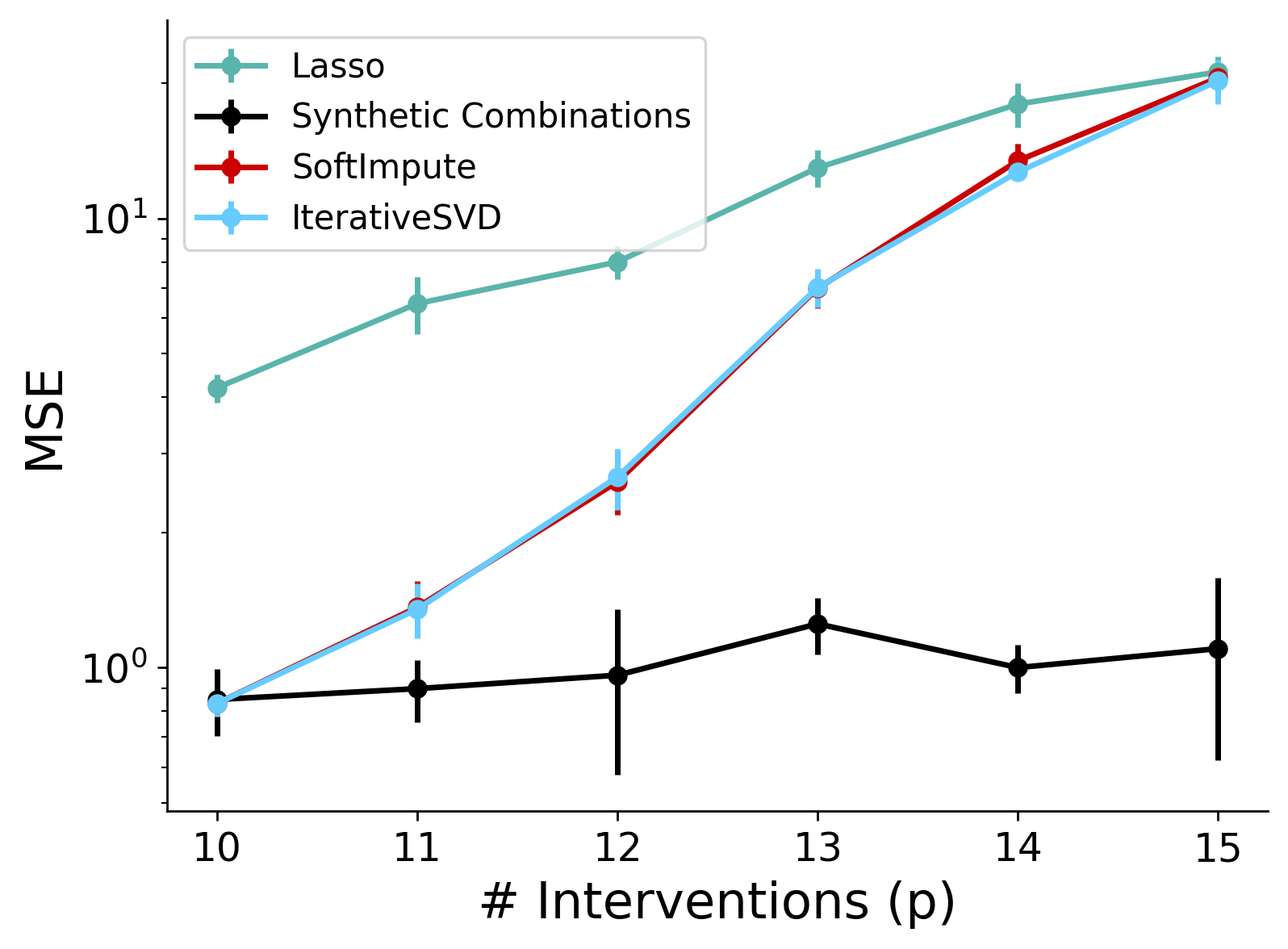}  
        \caption*{(a) Observational setting simulations.}
    \end{minipage}\hfill
    \begin{minipage}{0.5\textwidth}
        \centering
        \includegraphics[width=1\textwidth]{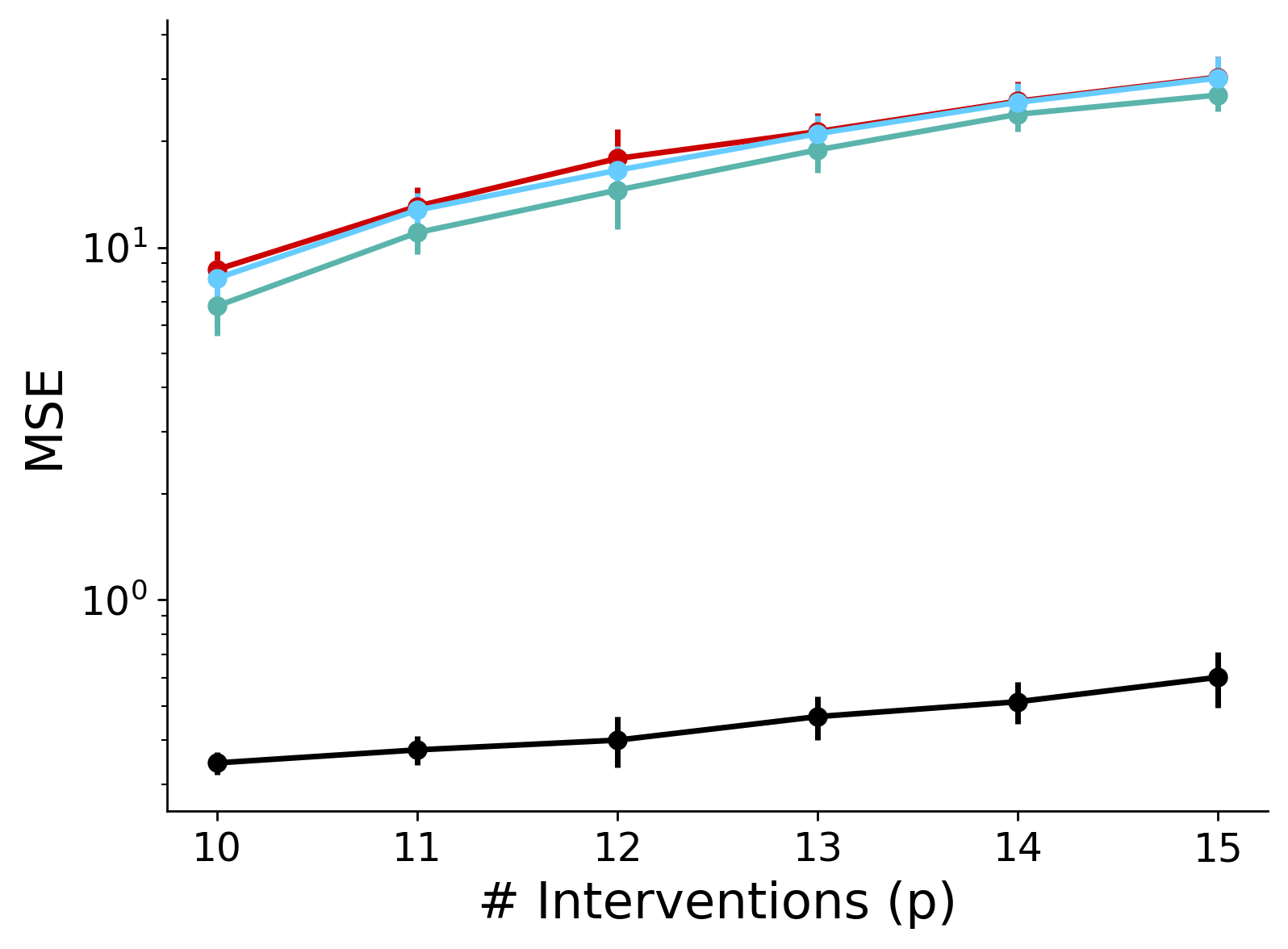} 
        \caption*{(b) Experimental design simulations.}
    \end{minipage}
    \caption{We simulate data in an observational setting (a) and under our experimental design mechanism (b). \method~outperforms both  Lasso (green) and matrix completion methods (red and blue) as measured by MSE in both settings.}
    \label{fig:sim_results}
\end{figure}

\noindent \textbf{Results.} We plot the MSE (averaged over 5 repetitions) for the different methods as the number of interventions is increased $p \in \{10,11,\ldots 15\}$ in Figure \ref{fig:sim_results} (b). 
\method~significantly outperforms other methods (and itself in the observational setting), which corroborates our theoretical findings that this experimental design utilizes the strengths of the estimator effectively. 

\section{Real World Case Study}
\label{sec:real_world_case_study}
This section details a real-world data experiment on recommendation systems for sets of movie ratings, comparing \method~to the algorithms listed above.
Further, we empirically validate that our key modeling assumptions (i.e., low-rank condition of $\mathcal{A}$ and sparsity of donor unit Fourier coefficients) hold in this dataset. 
For all methods, hyper-parameters are chosen via $5-$fold CV.
Additionally, the donor set $\mathcal{I}$ is chosen via the procedure described in Section \ref{sec:estimator_descripton}.

\noindent \textbf{Data and Experimental Set-up.} We use data collected in \cite{sharma2019sets} which consists of user ratings of sets of movies. 
Specifically, users were asked to provide a rating of $1$-$5$ on a set of $5$ movies chosen at random. 
This resulted in a total of ratings from $854$ users over $29,516$ sets containing $12,549$ movies.
$80\%$ of each user's ratings are chosen as the training set, and the other $20\%$ as the test set.

\noindent \textbf{Results.} We measure the RMSE averaged over 5 repetitions for all methods, and display the results in the Table below. 
with  \method~outperforming all other approaches.
The performance gap between \method~and other methods demonstrates the benefit of only performing the horizontal regression on the units with sufficient observations (i.e., the donor units), and using PCR for the non-donor units that have an insufficient number of measurements.

\begin{table} [H]
\centering
\footnotesize
\begin{tabular}{lrrrr}
\toprule
Method   & \textbf{\method} & \texttt{SoftImpute} & \texttt{IterativeSVD} & Lasso \\
\midrule
RMSE & \textbf{0.30} $\pm$ 0.03 &  0.38 $\pm$ 0.02  &  0.38 $\pm$ 0.02 & 0.43 $\pm$ 0.05 \\
\bottomrule
\end{tabular}
\label{tab:ratings_results}
\caption{\method~outperforms other approaches for the real-world experiment on sets of ratings for movies.}
\end{table}

\begin{figure}[htbp]
    \centering
    \includegraphics [width = 0.6\textwidth]{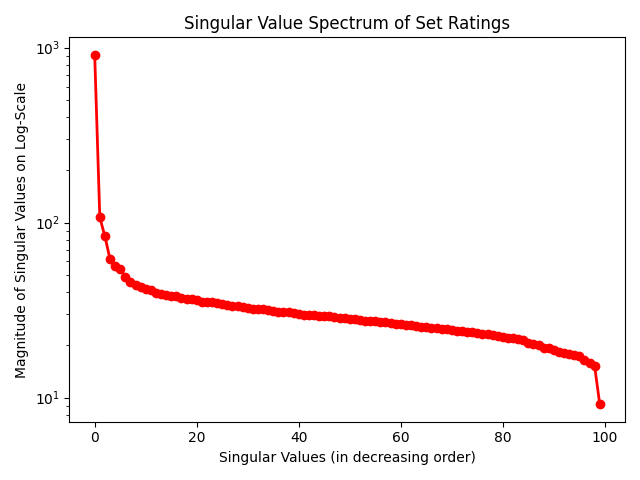}
    \caption{Singular value spectrum of user ratings on sets of movies. Inspecting the singular spectrum shows that the matrix of observed ratings is low-rank.}
    \label{fig:singular_spectrum}
\end{figure}

\noindent \textbf{Key Assumptions of \method~hold.} We also verify that our two key modeling assumptions (i.e., low-rankness and sparsity of $\mathcal{A}$) hold in this dataset. 
For the low-rank condition,the singular value spectrum of movies rated by all users is plotted on a log-scale in Figure \ref{fig:singular_spectrum}. 
The plot shows the outcomes (and hence the Fourier coefficients) are low-rank. 
To investigate sparsity, we examine $\hat{\balpha}_u$ for the donor set. 
The MSE averaged across all donor units on the test set was 0.22, indicating that the estimated Fourier coefficient is an accurate representation of the true underlying Fourier coefficient. 
Further, the estimated donor unit Fourier coefficients are indeed sparse, and on average have 8.7\% non-zero coefficients.

%

%
%
%
%

%
\section{Conclusion}\label{sec:discussion}
This paper introduces a causal inference framework for combinatorial interventions, a setting that is ubiquitous in practice. 
We propose a model that imposes both unit-specific combinatorial structure and latent similarity across units.  
Under this model, \method, an estimation procedure, is introduced. 
\method~leverages the sparsity and low-rankness of the Fourier coefficients to efficiently estimate all $N \times 2^p$ causal parameters while implicitly allowing for unobserved confounding.
Theoretically, finite-sample consistency and asymptotic normality of \method~is established. 
A novel experiment design mechanism is proposed, which ensures that the key assumptions required for the estimator to accurately recover all $N \times 2^p$ mean potential outcomes of interest hold. 
The empirical effectiveness of \method~is demonstrated through numerical simulations and a real-world case study on movie ratings.
We discuss how \method~can be adapted to estimate counterfactuals under different permutations of items, i.e., rankings. 
This work suggests future directions for research such as providing an analysis of \method~that is agnostic to the horizontal regression algorithm used and deriving estimators that can achieve the sample complexity lower bound discussed.
More broadly, we hope this work serves as a bridge between causal inference and the Fourier analysis of Boolean functions.

%
%
%
%
%
%
%
\if1\blind{
    \section{Acknowledgements}
\label{sec:acknowledgements}
We thank Alberto Abadie, Peng Ding, Giles Hooker, Devavrat Shah, Vasilis Syrgkanis, and Bin Yu for
useful discussions and feedback.
We also thank Austin Serif for his help in implementing \method. 
} \fi

\bibliography{rf.bib}

\newpage

\newpage
\bibliographystyleappendix{agsm}

\appendix
\allowdisplaybreaks

\section{Proof of Theorem \ref{thm:identification}}
\label{subsec:identification_thm_proof}
Below, the symbol $\stackrel{AX}{=}$ and $\stackrel{DX}{=}$ imply that the equality follows from Assumption $X$ and Definition $X$, respectively. 
We begin with the proof of Theorem \ref{thm:identification} (a). 
For a donor unit $u \in \mathcal{I}$ and $\pi \in \Pi \setminus \Pi_u$, we have
\begin{align*}
    \E[Y^{(\pi)}_{u}~|~\mathcal{A}] & \stackrel{A\ref{ass:observation_model}}{=} \E[\langle \balpha_u, \bchi^{\pi} \rangle + \epsilon_u^{\pi} ~|~\mathcal{A}] \\ 
    & \stackrel{A\ref{ass:observation_model}(c)}{=} \langle \balpha_u, \bchi^\pi \rangle ~| ~\mathcal{A} \\
    & \stackrel{}{=} \langle \balpha_u, \bchi^\pi \rangle ~| ~\mathcal{A}, \ \mathcal{D} \\
    &  \stackrel{A\ref{ass:observation_model}(b)}{=} \langle \balpha_{u}, \tilde{\bchi}_{u}^{\pi} \rangle ~|~ \mathcal{A}, \ \mathcal{D}\\
    & \stackrel{A\ref{ass:donor_set_identification}(a)}{=} \langle \balpha_{u}, \sum_{\pi_u \in  \Pi_u} \beta_{\pi_{u}}^{\pi}\tilde{\bchi}_{u}^{\pi_u} \rangle  ~ | ~  \mathcal{A}, \ \mathcal{D} \\ 
    & = \sum_{\pi_u \in \Pi_u} \beta^{\pi}_{\pi_u} \langle \balpha_{u}, \tilde{\bchi}_{u}^{\pi_u}  \rangle  ~|~ \mathcal{A}, \ \mathcal{D} \\
    & \stackrel{A\ref{ass:observation_model}(c), A\ref{ass:selection_on_fourier}}{=}  \sum_{\pi_u \in \Pi_u} \beta^{\pi}_{\pi_u} \E[\langle \balpha_{u}, \tilde{\bchi}_{u}^{\pi_u}  \rangle + \epsilon_u^{\pi_u} ~ | ~ \mathcal{A}, \ \mathcal{D} ] \\
    & \stackrel{A\ref{ass:observation_model}}{=} \sum_{\pi_u \in \Pi_u} \beta^{\pi}_{\pi_u} \E[Y_{u,\pi_u} ~ | ~ \mathcal{A}, \ \mathcal{D}]
\end{align*}
\noindent Next, we establish the  proof of Theorem \ref{thm:identification} (b). For a non-donor unit $n$ and  $\pi \in \Pi \setminus \Pi_n$, we have
\begin{align*}
    \E[Y^{(\pi)}_{n}~|~\mathcal{A}] & \stackrel{A\ref{ass:observation_model}}{=} \E[\langle \balpha_n, \bchi^{\pi} \rangle + \epsilon_n^{\pi} ~|~\mathcal{A}] \\ 
    & \stackrel{A\ref{ass:observation_model}(c)}{=} \langle \balpha_n, \bchi^\pi \rangle ~| ~\mathcal{A} \\
    & \stackrel{}{=} \langle \balpha_n, \bchi^\pi \rangle ~| ~\mathcal{A}, \  \mathcal{D} \\
    & \stackrel{A\ref{ass:donor_set_identification}(b)}{=} \langle \sum_{u \in \mathcal{I}}w_u^n\balpha_u, \bchi^{\pi} \rangle  ~ | ~  \mathcal{A}, \ \mathcal{D} \\ 
    & = \sum_{u \in \mathcal{I}}w_u^n \langle \balpha_{u}, \bchi^{\pi} \rangle  ~|~ \mathcal{A}, \ \mathcal{D} \\
    & \stackrel{A\ref{ass:observation_model}(c), A\ref{ass:selection_on_fourier}}{=}  \sum_{u \in \mathcal{I}} w_u^n \E[\langle \balpha_{u}, \bchi^{\pi} \rangle + \epsilon_u^{\pi} ~ | ~ \mathcal{A}, \ \mathcal{D} ] \\
    & \stackrel{A\ref{ass:observation_model}}{=} \sum_{u \in \mathcal{I}} w_u^n \E[Y^{(\pi)}_{u} ~ | ~ \mathcal{A}, \ \mathcal{D}] \\
    & = \sum_{u \in \mathcal{I}} \sum_{\pi_u \in \Pi_u}  w_u^n \beta^{\pi}_{\pi_u} \E[Y_{u,\pi_u} ~ | ~ \mathcal{A},\  \mathcal{D}]
\end{align*}
where the last equality follows from Theorem \ref{thm:identification} (a).

\section{Proof of Theorem \ref{thm:potential_outcome_convergence_rate}}

\subsection{Proof of Theorem  \ref{thm:potential_outcome_convergence_rate} (a)}
We have that 
\label{subsec:horizontal_regression_theorem_proof}
\begin{align} 
     \hat{\E}[Y_u^{(\pi)}] - \E[Y_{u}^{(\pi)}] & = \langle  \hat{\balpha}^u, \bchi^{\pi} \rangle - \langle  \balpha_u, \bchi^{\pi} \rangle \nonumber\\
     & = \langle  \hat{\balpha}_u - \balpha_u, \bchi^{\pi} \rangle \nonumber \\
     & \leq \lVert \hat{\balpha}_u - \balpha_u \rVert_1 \lVert \bchi^{\pi} \rVert_{\infty} \nonumber \\ 
     & = \lVert \hat{\balpha}_u - \balpha_u \rVert_1 \label{eq:lasso_proof_holder_inequality}
\end{align}

\noindent To finish the proof, we quote the following Theorem which we adapt to our notation. 

\begin{theorem}[Theorem 2.18 in \cite{rigollet2015high}] 
\label{thm:rigollet_lasso_high_prob_bound}
Fix the number of samples $n \geq 2$. Assume that the linear model $Y = \bX\theta^* + \epsilon$, where $\bX \in \mathbb{R}^{n \times d}$ and $\epsilon$ is a sub-gaussian random variable with noise variance $\sigma^2$. Moreover, assume that $\lVert \theta^* \rVert_0 \leq k$, and that 
$\bX$ satisfies the incoherence condition (Assumption \ref{ass:incoherence}) with parameter $k$. Then, the lasso estimator $\hat{\theta}^L$ with regularization parameter defined by 
\begin{equation*}
    2\tau = 8\sigma\sqrt{\log(2d)/n} +  8\sigma\sqrt{\log(1/\delta)/n}
\end{equation*}
satisfies 
\begin{equation}
    \lVert \theta^* - \hat{\theta}^L \rVert^2_2 \leq k\sigma^2\frac{\log(2d/\delta)}{n}
\end{equation}
with probability at least $1 - \delta$. 
\end{theorem}
Further, as in established in the proof of Theorem 2.18 in \cite{rigollet2015high}, $\lVert \theta^* - \hat{\theta}^L \rVert_1 \leq \sqrt{k} \lVert \theta^* - \hat{\theta}^L \rVert_2$. Note that the set-up of Theorem \ref{thm:rigollet_lasso_high_prob_bound} holds in our setting with the following notational changes: $Y = \bY_{\Pi_u}$, $\bX = \bchi(\Pi_u)$, $\theta^* = \balpha_u$, $\hat{\theta}^L = \hat{\balpha}_u$,  $k = s$ as well as our assumptions on the regularization parameter $\lambda_u$ and that $\bepsilon_u^{\pi}$ is sub-gaussian (Assumption \ref{ass:subgaussian_noise}). Applying Theorem \ref{thm:rigollet_lasso_high_prob_bound} gives us
\begin{equation*}
    \lVert \hat{\balpha}_u - \balpha_u \rVert_1 = O_p \left(s \sqrt{\frac{p}{|\Pi_u|}}\right)
\end{equation*}
Substituting this bound into \eqref{eq:lasso_proof_holder_inequality} yields the claimed result.



\subsection{Proof of Theorem  \ref{thm:potential_outcome_convergence_rate} (b)}
For any matrix $\mathbf{A}$ with orthonormal columns, let $\mathcal{P}_{A} = \mathbf{A}\mathbf{A}^T$ denote the projection matrix on the subspace spanned by the columns of $\mathbf{A}$. Define $\Tilde{\bw}^n = \mathcal{P}_{V_{\mathcal{I}}^{(\Pi_n)}}\bw^n$, where $\bV_{\mathcal{I}}^{(\Pi_n)}$ are the right singular vectors of $\E[\bY_{\mathcal{I}}^{(\Pi_n)}]$. Let $\Delta^n_w = \hat{\bw}^n - \Tilde{\bw}^n \in \R^{|\mathcal{I}|}$, and $\Delta_{\mathcal{I}}^{\pi} = \hat{\E}[\bY_{\mathcal{I}}^{(\pi)}] - \E[\bY_{\mathcal{I}}^{(\pi)}] \in \R^{|\mathcal{I}|}$. Denote $\Delta_E = \max_{u \in \mathcal{I}, \pi \in \Pi} \left |\hat{\E}[Y_u^{(\pi)}] - \E[Y_{u}^{(\pi)}] \right |$. Throughout the proof, we use $C$ to refer to constants that might change from line to line. In order to proceed, we first state the following result,  
\begin{lemma} 
\label{lem:w_tilde_transfer_outcomes}
Let the set-up of Theorem \ref{thm:potential_outcome_convergence_rate} hold. Then, we have 
\begin{equation*}
    \E[Y_{n}^{(\pi)}] =  \langle \E[\bY_{\mathcal{I}}^{(\pi)}], \Tilde{\bw}^n \rangle
\end{equation*}
\end{lemma}

\noindent Using Lemma \ref{lem:w_tilde_transfer_outcomes}, and the notation established above, we have
\begin{align}
     \left |\hat{\E}[Y_n^{(\pi)}] - \E[Y_{n}^{(\pi)}]\right| & = \left| \langle \hat{\E}[\bY_{\mathcal{I}}^{(\pi)}], \hat{\bw}^n \rangle - \langle \E[\bY_{\mathcal{I}}^{(\pi)}], \Tilde{\bw}^n \rangle \right| \nonumber \\
     & \leq \left| \langle \Delta_{\mathcal{I}}^{\pi}, \Tilde{\bw}^n \rangle \right| +  \left| \langle \E[\bY_{\mathcal{I}}^{(\pi)}] ,\Delta^n_w   \rangle \right| + \left| \langle \Delta^n_w, \Delta_{\mathcal{I}}^{\pi} \rangle \right| \label{eq:three_term_bound_pre_projection}
\end{align}
From Assumption \ref{ass:rowspace_inclusion}, it follows that $\E[\bY_{\mathcal{I}}^{(\pi)}] = \mathcal{P}_{V_{\mathcal{I}}^{(\Pi_n)}} \E[\bY_{\mathcal{I}}^{(\pi)}]  $ , where $\bV_{\mathcal{I}}^{(\Pi_n)}$ are the right singular vectors of $\E[\bY_{\mathcal{I}}^{(\Pi_n)}]$. Using this in \eqref{eq:three_term_bound_pre_projection} gives us
\begin{equation}
 \label{eq:three_term_bound_post_projection}
   \left |\hat{\E}[Y_n^{(\pi)}] - \E[Y_{n}^{(\pi)}]\right| \leq  \left| \langle \Delta_{\mathcal{I}}^{\pi}, \Tilde{\bw}^n \rangle \right| +  \left| \langle \E[\bY_{\mathcal{I}}^{(\pi)}] ,\mathcal{P}_{V_{\mathcal{I}}^{(\Pi_n)}}\Delta^n_w   \rangle \right| + \left| \langle \Delta^n_w, \Delta_{\mathcal{I}}^{\pi} \rangle \right|
\end{equation}
Below, we bound the three terms on the right-hand-side of \eqref{eq:three_term_bound_post_projection} separately. Before we bound each term, we state a Lemma that will be useful to help us establish the results. 

\begin{lemma} 
\label{lem:l1_bound_w_tilde}
Let Assumptions \ref{ass:observation_model}, \ref{ass:donor_set_identification}, \ref{ass:boundedness_potential_outcome}, and \ref{ass:balanced_spectrum} hold. Then, $\lVert \tilde{\bw}^n \rVert_2 \leq C\sqrt{\frac{r_n}{|\mathcal{I}|}}$
\end{lemma}

\noindent \emph{Bounding Term 1.} By Holder's inequality and Lemma \ref{lem:l1_bound_w_tilde}, we have that 
\begin{equation}
\label{eq:Term1_Op_bound}
    \left| \langle \Delta_{\mathcal{I}}^{\pi}, \Tilde{\bw}^n \rangle \right| \leq \lVert \Tilde{\bw}^n \rVert_1 \lVert \Delta_{\mathcal{I}}^{\pi} \rVert_\infty \leq \lVert \Tilde{\bw}^n \rVert_1 \Delta_E \leq  \sqrt{|\mathcal{I}|}\lVert \Tilde{\bw}^n \rVert_2 \Delta_E \leq C\sqrt{r_n}\Delta_E
\end{equation}
\noindent This concludes the analysis for the first term. \\ 

\noindent \emph{Bounding Term 2}. By Cauchy-Schwarz and Assumption \ref{ass:boundedness_potential_outcome} we have,  
\begin{align}
    \left| \langle \E[\bY_{\mathcal{I}}^{(\pi)}] ,\mathcal{P}_{V_{\mathcal{I}}^{(\Pi_n)}}\Delta^n_w \rangle \right | & \leq \lVert \E[\bY_{\mathcal{I}}^{(\pi)}] \rVert_2 \lVert \mathcal{P}_{V_{\mathcal{I}}^{(\Pi_n)}}\Delta^n_w  \rVert_2 \nonumber \\
    & \leq \sqrt{|\mathcal{I}|} \lVert \mathcal{P}_{V_{\mathcal{I}}^{(\Pi_n)}}\Delta^n_w  \rVert_2 \label{eq:term2_cauchy_schwarz}
\end{align}
\noindent We now state a lemma that will help us conclude our bound of Term 2. The proof is given in Appendix \ref{subsubsec:proof_lemma_projection_delta_w_bound}. 
\begin{lemma} 
\label{lem:projection_delta_w_bound}
Let the set-up of Theorem \ref{thm:potential_outcome_convergence_rate} hold. Then, 
\begin{align}
& \lVert \mathcal{P}_{V_{\mathcal{I}}^{(\Pi_n)}}\Delta^n_w \rVert_2 \nonumber \\
& = O_p\left(\log^{3}\left(|\Pi_n||\mathcal{I}|\right)  r_n^{2}\left[\frac{\Delta_E  }{\sqrt{|\mathcal{I}|}\min\{\sqrt{|\Pi_n|},\sqrt{|\mathcal{I}|\}}}  + \frac{\Delta^2_E}{\sqrt{|\mathcal{I}|}} \right] + \frac{r_n^{3/2}\Delta_E}{\sqrt{|\mathcal{I}|}} + \frac{r_n\left( 1 + \sqrt{\Delta_E}r^{1/4}_n\right)}{\sqrt{|\mathcal{I}|}|\Pi_n|^{1/4}}  \right) \label{eq:projection_delta_w_bound}
\end{align}
\end{lemma}
\noindent Incorporating the result of the lemma above into \eqref{eq:term2_cauchy_schwarz} gives us
\begin{align}
     & \left| \langle \E[\bY_{\mathcal{I}}^{(\pi)}] \nonumber ,\mathcal{P}_{V_{\mathcal{I}}^{(\Pi_n)}}\Delta^n_w \rangle  \right| \\
     & = O_p \left(\log^{3}\left(|\Pi_n||\mathcal{I}|\right) r_n^{2}\left[\frac{\Delta_E  }{\min\{\sqrt{|\Pi_n|},\sqrt{|\mathcal{I}|\}}}  + \Delta^2_E \right]  + r_n^{3/2}\Delta_E + \frac{r_n\left( 1 + \sqrt{\Delta_E}r^{1/4}_n\right)}{|\Pi_n|^{1/4}}  \right)  \label{eq:Term2_Op_bound}
\end{align}

\noindent \emph{Bounding Term 3.} By Holder's inequality,  we have that
\begin{align}
    \left| \langle \Delta^n_w, \Delta_{\mathcal{I}}^{\pi} \rangle \right| & \leq \lVert \Delta^n_w \rVert_2 \lVert \Delta_{\mathcal{I}}^{\pi} \rVert_2 \nonumber \\
    & \leq \sqrt{|\mathcal{I}|} \lVert \Delta^n_w \rVert_2 \lVert \Delta_{\mathcal{I}}^{\pi} \rVert_\infty \nonumber \\
    & \leq  \sqrt{|\mathcal{I}|} \Delta_E \lVert \Delta^n_w \rVert_2 \label{eq:term3_holder}  
\end{align}
\noindent We now state a proposition that will help us conclude our proof of Term 3. The proof is given in Appendix \ref{subsec:pcr_proofs}.

\begin{proposition}
\label{prop:pcr_convergence_rate} Let the set-up of Theorem \ref{thm:potential_outcome_convergence_rate} hold. Then, conditioned on $\mathcal{A}$, we have 
\begin{equation}
    \hat{\bw}^n - \Tilde{\bw}^n=  {O}_p \left(\log^{3}(|\Pi_n||\mathcal{I}|)r_n \lVert \Tilde{\bw}^{n} \rVert_2 \left[\frac{1 }{\min\{\sqrt{|\Pi_n|},\sqrt{|\mathcal{I}|\}}}  + \Delta_E \right] \right) 
\end{equation}
\end{proposition}

\noindent As a result of Proposition \ref{prop:pcr_convergence_rate} and Lemma \ref{lem:l1_bound_w_tilde}, we have the following bound for Term 3,  
\begin{equation}
\label{eq:Term3_Op_bound}
     \left| \langle \Delta^n_w, \Delta_{\mathcal{I}}^{\pi} \rangle \right| = O_p \left(\log^{3}(|\Pi_n||\mathcal{I}|)r^{3/2}_n \left[\frac{\Delta_E  }{\min\{\sqrt{|\Pi_n|},\sqrt{|\mathcal{I}|\}}}  + \Delta^2_E \right] \right)
\end{equation}

\noindent \emph{Collecting Terms.} Combining equations \eqref{eq:Term1_Op_bound}, \eqref{eq:Term2_Op_bound}, \eqref{eq:Term3_Op_bound} gives us
\begin{equation}
\label{eq:collecting_terms}
\begin{split}
&\left |\hat{\E}[Y_n^{(\pi)}] - \E[Y_{n}^{(\pi)}]\right|   
\\ &= O_p\left(\log^{3}(|\Pi_n||\mathcal{I}|) r_n^{2}\left[\frac{\Delta_E  }{\min\{\sqrt{|\Pi_n|},\sqrt{|\mathcal{I}|\}}}  + \Delta^2_E \right] + r_n^{3/2}\Delta_E + \frac{r_n\left( 1 + \sqrt{\Delta_E}r^{1/4}_n\right)}{|\Pi_n|^{1/4}}  \right)  
\end{split}
\end{equation} 

\noindent By Theorem \ref{thm:potential_outcome_convergence_rate} (a), we have that 
\begin{equation}
\label{eq:delta_E_subsitution}
    \Delta_E = \max_{u \in \mathcal{I}} \ O_p \left(s\sqrt{\frac{p}{|\Pi_u|}}\right) = O_p \left(s\sqrt{\frac{p}{M}}\right)
\end{equation}
\noindent where we remind the reader that $M = \min_{u \in \mathcal{I}} |\Pi_u|$. Substituting \eqref{eq:delta_E_subsitution} into \eqref{eq:collecting_terms}, and simplifying, then we get 
\begin{equation}
\label{eq:collecting_terms_simplified_1}
\begin{split}
\left |\hat{\E}[Y_n^{(\pi)}] - \E[Y_{n}^{(\pi)}]\right|  = O_p\left(\log^{3}(|\Pi_n||\mathcal{I}|)r^2_n\left(\sqrt{\frac{s^2p}{M}} \vee \frac{s^2p}{M} \right) +  \frac{r_n}{|\Pi_n|^{1/4}}\left(1 \vee \frac{r_n^{1/4}\sqrt{s}p^{1/4}}{M^{1/4}}\right) \right)  
\end{split}
\end{equation}

\noindent Substituting our assumption that $M \coloneqq \omega(r^2_ns^2p)$, we get 
\begin{equation}
\label{eq:collecting_terms_simplified_log}
\begin{split}
\left |\hat{\E}[Y_n^{(\pi)}] - \E[Y_{n}^{(\pi)}]\right|  = O_p\left(\log^{3}(|\Pi_n||\mathcal{I}|) r^2_n\sqrt{\frac{{s^2p}}{{M}}} +  \frac{r_n}{|\Pi_n|^{1/4}} \right)
\end{split}
\end{equation}
\noindent Finally, absorbing the logarithmic factors, we get the claimed result, 
\begin{equation}
\label{eq:collecting_terms_simplified}
\begin{split}
\left |\hat{\E}[Y_n^{(\pi)}] - \E[Y_{n}^{(\pi)}]\right|  = \Tilde{O}_p\left(r^2_n\sqrt{\frac{{s^2p}}{{M}}} +  \frac{r_n}{|\Pi_n|^{1/4}} \right)
\end{split}
\end{equation}



\section{Proofs of Helper Lemmas for Theorem  \ref{thm:potential_outcome_convergence_rate}.}
\label{sec:helper_lemma_proofs}
In this section we provide proofs of Lemmas \ref{lem:w_tilde_transfer_outcomes}, \ref{lem:l1_bound_w_tilde}, \ref{lem:projection_delta_w_bound}, and Proposition \ref{prop:pcr_convergence_rate} which were required for the Proof of Theorem \ref{thm:potential_outcome_convergence_rate}.  

\subsection{Proof of Lemma \ref{lem:w_tilde_transfer_outcomes}}
\label{subsec:proof_of_w_tilde_transfer_outcomes}

By Assumption \ref{ass:observation_model} and \ref{ass:donor_set_identification} (b), we have
\begin{align*}
    \E[Y_n^{(\pi)}] & = \E[\langle \balpha_n, \bchi^{\pi} \rangle + \epsilon_n^{\pi}] \\
    & =  \E[\langle \balpha_n, \bchi^{\pi} \rangle ] \\
    & =  \sum_{u \in \mathcal{I}}w^n_u\E[\langle \balpha_u, \bchi^{\pi} \rangle] \\
    & = \sum_{u \in \mathcal{I}}w^n_u \E[Y_u^{(\pi)}] \\
    & = \E[\bY_{\mathcal{I}}^{(\pi)}]^T \bw^{n}
\end{align*}
From Assumption \ref{ass:rowspace_inclusion}, we have that  $\E[\bY_{\mathcal{I}}^{(\pi)}] = \mathcal{P}_{V_{\mathcal{I}}^{(\Pi_n)}} \E[\bY_{\mathcal{I}}^{(\pi)}]$. Substituting this into the equation above completes the proof.

\subsection{Proof of Lemma \ref{lem:l1_bound_w_tilde}}
\label{subsec:proof_of_l1_bound_w_tilde}
For simplicity, denote $\E[\bY_n^{(\Pi_n)}] = \E[\bY^{(\Pi_n)}]$.  By definition, $\tilde{\bw}^n$ is the solution to the following optimization program 
\begin{align}
\min_{ \bw \in \mathbb{R}^{|\mathcal{I}|}} \quad & ||{\bw}||_2 \nonumber \\
\textrm{s.t.} \quad & \E[\bY^{(\Pi_n)}] = \E[\bY_{\mathcal{I}}^{(\Pi_n)}]\bw
\label{eq:linear_system_donor_set}
\end{align}

\noindent Let $\bU_{\mathcal{I}}^{(\Pi_n)}, \mathbf{\Sigma}_{\mathcal{I}}^{(\Pi_n)}, \bV_{\mathcal{I}}^{(\Pi_n)}$ denote the SVD of $\E[\bY_{\mathcal{I}}^{(\Pi_n)}]$. Further, let $\bU_{\mathcal{I}}^{(\Pi_n),r_n}, \mathbf{\Sigma}_{\mathcal{I}}^{(\Pi_n),r_n}, \bV_{\mathcal{I}}^{(\Pi_n),r_n}$ denote the rank $r_n$ truncation of the SVD. Then, define $\bw_{r_n} =  \bV_{\mathcal{I}}^{(\Pi_n),r_n} (\mathbf{\Sigma}_{\mathcal{I}}^{(\Pi_n),r_n})^{\dagger} (\bU_{\mathcal{I}}^{(\Pi_n),r_n})^T \E[\bY^{(\Pi_n)}]$, where $\dagger$ is pseudo-inverse. We first show that $\bw_{r_n}$ is a solution to \eqref{eq:linear_system_donor_set}.

\begin{align*}
    \E[\bY_{\mathcal{I}}^{(\Pi_n)}] \bw_{r_n} &= \left( \bU_{\mathcal{I}}^{(\Pi_n)} \mathbf{\Sigma}_{\mathcal{I}}^{(\Pi_n)} (\bV_{\mathcal{I}}^{(\Pi_n)})^T \right) \bV_{\mathcal{I}}^{(\Pi_n),r_n} (\mathbf{\Sigma}_{\mathcal{I}}^{(\Pi_n),r_n})^{\dagger} (\bU_{\mathcal{I}}^{(\Pi_n),r_n})^T \E[\bY^{(\Pi_n)}] \\
    & = \left(\sum^{r_n}_{i=1} s_i u_i v^{T}_i\right) \left(\sum^{r_n}_{j=1} \frac{1}{s_j}v_j u^T_j \E[\bY^{(\Pi_n)}]\right) \\
    & = \sum^{r_n}_{i,j=1} \frac{s_i}{s_j} u_i v^T_i v_j u_j^T \E[\bY^{(\Pi_n)}] \\
    & = \sum^{r_n}_{i=1} u_i u^T_i \E[\bY^{(\Pi_n)}] \\
    & = \E[\bY^{(\Pi_n)}] 
\end{align*}

Next, we bound $\lVert \bw_{r_n} \rVert_2$ using Assumptions \ref{ass:boundedness_potential_outcome} and \ref{ass:balanced_spectrum} as follows 
\begin{align*}
    \lVert \bw_{r_n} \rVert_2 & \leq \lVert (\mathbf{\Sigma}_{\mathcal{I}}^{(\Pi_n),r_n})^{\dagger} \rVert_2  \lVert \E[\bY^{(\Pi_n)}] \rVert_2 \\
    & \leq \frac{\sqrt{|\Pi_n|}}{s_{r_n}(\E[\bY_{\mathcal{I}}^{(\Pi_n)}])} \\
    & \leq \sqrt{\frac{cr_n}{|\mathcal{I}|}}
\end{align*}
Therefore, we have 
\begin{align*}
    \lVert \Tilde{\bw}^n \rVert_1 \leq \sqrt{|\mathcal{I}|} \lVert \Tilde{\bw}_{r_n} \rVert_2 \leq \sqrt{cr_n}
\end{align*}

\subsection{Proof of Lemma \ref{lem:projection_delta_w_bound}}
\label{subsubsec:proof_lemma_projection_delta_w_bound}

\noindent First, we introduce some necessary notation required for the proof. Let $\hat{\E}\left[\bY_{\mathcal{I}}^{(\Pi_n)} \right] = \hat{\bU}_{\mathcal{I}}^{(\Pi_n)}\hat{\bS}_{\mathcal{I}}^{(\Pi_n)}\hat{\bV}_{\mathcal{I}}^{(\Pi_n)}$ denote the rank $r_n$ SVD of $\hat{\E}\left[\bY_{\mathcal{I}}^{(\Pi_n)} \right]$. Then, to establish Lemma \ref{lem:projection_delta_w_bound}, consider the following decomposition:
\begin{equation*}
    \mathcal{P}_{V_{\mathcal{I}}^{(\Pi_n)}}\Delta^n_w = \left(\mathcal{P}_{V_{\mathcal{I}}^{(\Pi_n)}} - \mathcal{P}_{\hat{V}_{\mathcal{I}}^{(\Pi_n)}}  \right)\Delta^n_w  + \mathcal{P}_{\hat{V}_{\mathcal{I}}^{(\Pi_n)}}\Delta^n_w 
\end{equation*}
\noindent We bound each of these terms separately again. \\

\noindent \emph{Bounding Term 1}. We have
\begin{equation}
\label{eq:projection_operator_error_cauchy}
     \lVert \left(\mathcal{P}_{V_{\mathcal{I}}^{(\Pi_n)}} - \mathcal{P}_{\hat{V}_{\mathcal{I}}^{(\Pi_n)}}  \right)\Delta^n_w \rVert_2  \leq \left\lVert \mathcal{P}_{V_{\mathcal{I}}^{(\Pi_n)}} - \mathcal{P}_{\hat{V}_{\mathcal{I}}^{(\Pi_n)}}  \right\rVert_{\text{op}} \lVert \Delta^n_w \rVert_2
\end{equation}

\begin{theorem} [Wedin's Theorem \cite{wedin1972perturbation}]
\label{thm:wedin_theorem}
Given $\mathbf{A},\mathbf{B} \in \R^{m \times n}$, let $(\bU,\bV),(\hat{\bU},\hat{\bV})$ denote their respective left and right singular vectors. Further, let $(\bU_k,\bV_k) \in $(respectively, $(\hat{\bU}_k,\hat{\bV}_k)$) correspond to the truncation of $(\bU,\bV)$ (respectively, $(\hat{\bU},\hat{\bV})$), respectively, that retains the columns correspondng to the top $k$ singular values of $\mathbf{A}$ (respectively, $\mathbf{B}$). Let $s_i$ represent the $i$-th singular values of $A$. Then, $\max(\lVert \mathcal{P}_{U_k} - \mathcal{P}_{\hat{U}_k}\rVert_{\text{op}}, \lVert \mathcal{P}_{V_k} - \mathcal{P}_{\hat{V}_k}\rVert_{\text{op}}) \leq \frac{2\lVert \mathbf{A} - \mathbf{B} \rVert_{\text{op}}}{s_{k}-s_{k+1}}$   
\end{theorem}

\noindent Applying Theorem \ref{thm:wedin_theorem} gives us 
\begin{align}
\max\left(\left\lVert \mathcal{P}_{U_{\mathcal{I}}^{(\Pi_n)}} - \mathcal{P}_{\hat{U}_{\mathcal{I}}^{(\Pi_n)}}  \right\rVert_{\text{op}}, \left\lVert \mathcal{P}_{V_{\mathcal{I}}^{(\Pi_n)}} - \mathcal{P}_{\hat{V}_{\mathcal{I}}^{(\Pi_n)}}  \right\rVert_{\text{op}} \right) & \leq  \frac{2\lVert \E[\bY_{\mathcal{I}}^{(\Pi_n)}] - \hat{\E}[\bY_{\mathcal{I}}^{(\Pi_n)}] \rVert_{\text{op}}}{s_{r_n} - s_{r_n + 1}}  \nonumber \\
& \leq \frac{2\sqrt{|\mathcal{I}||\Pi_n|}\lVert \E[\bY_{\mathcal{I}}^{(\Pi_n)}] - \hat{\E}[\bY_{\mathcal{I}}^{(\Pi_n)}] \rVert_{\text{max}}}{s_{r_n} - s_{r_n + 1}} \nonumber  \\
& = \frac{2\sqrt{|\mathcal{I}||\Pi_n|}\Delta_E}{s_{r_n} - s_{r_n + 1}} \nonumber \\ 
& = \frac{2\sqrt{|\mathcal{I}||\Pi_n|}\Delta_E}{s_{r_n}} \label{eq:projection_operator_error_operator_bound}
\end{align}
\noindent where the last equality follows from the fact that  $\text{rank}(\E[\bY_{\mathcal{I}}^{\Pi_n}]) = r_n$, hence $s_{r_n + 1} = 0$. Now, plugging Assumption \ref{ass:balanced_spectrum}, \eqref{eq:projection_operator_error_operator_bound}  into \eqref{eq:projection_operator_error_cauchy} gives us
\begin{equation}
\label{eq:projection_operator_error_bound_simplified}
    \max\left(\left\lVert \mathcal{P}_{V_{\mathcal{I}}^{(\Pi_n)}} - \mathcal{P}_{\hat{V}_{\mathcal{I}}^{(\Pi_n)}}  \right\rVert_{\text{op}}, \left\lVert \mathcal{P}_{V_{\mathcal{I}}^{(\Pi_n)}} - \mathcal{P}_{\hat{V}_{\mathcal{I}}^{(\Pi_n)}}  \right\rVert_{\text{op}} \right) \leq C\sqrt{r_n}\Delta_E 
\end{equation}
Substituting the result of Proposition \ref{prop:pcr_convergence_rate} and \eqref{eq:projection_operator_error_bound_simplified} into \eqref{eq:projection_operator_error_cauchy} gives us
\begin{equation}
\label{eq:projection_operator_term1_bound}
      \left \lVert \left(\mathcal{P}_{V_{\mathcal{I}}^{(\Pi_n)}} - \mathcal{P}_{\hat{V}_{\mathcal{I}}^{(\Pi_n)}}  \right)\Delta^n_w \right \rVert_2 = O_p \left(\log^3\left(|\Pi_n||\mathcal{I}|\right) r_n^{3/2} \lVert \Tilde{\bw}^{n} \rVert_2 \left[\frac{ \Delta_E  }{\min\{\sqrt{|\Pi_n|},\sqrt{|\mathcal{I}|\}}}  + \Delta^2_E \right] \right)
\end{equation}

\noindent Substituting Lemma \ref{lem:l1_bound_w_tilde} into \eqref{eq:projection_operator_term1_bound} gives us
\begin{equation}
\label{eq:projection_operator_term1_bound_simplified}
      \left \lVert \left(\mathcal{P}_{V_{\mathcal{I}}^{(\Pi_n)}} - \mathcal{P}_{\hat{V}_{\mathcal{I}}^{(\Pi_n)}}  \right)\Delta^n_w \right \rVert_2 = {O_p} \left( \log^3\left(|\Pi_n||\mathcal{I}|\right) r_n^{2}\left[\frac{\Delta_E  }{\sqrt{|\mathcal{I}|}\min\{\sqrt{|\Pi_n|},\sqrt{|\mathcal{I}|\}}}  + \frac{\Delta^2_E}{\sqrt{|\mathcal{I}|}} \right] \right)
\end{equation}

\noindent  \emph{Bounding Term 2.}  We introduce some necessary notation required to bound the second term. 
Let $\hat{\E}[\bY_{\mathcal{I}}^{(\Pi_n),{r_n}}] = \sum^{r_n}_{l=1}\hat{s}_l\hat{\mu}_l\hat{v}'_l$ denote the $r_n$ decomposition of $\hat{\E}[\bY_{\mathcal{I}}^{(\Pi_n)}]$. Let, $\hat{\E}[\bY_{\mathcal{I}}^{(\Pi_n),{r_n}}] = \hat{\bU}_{\mathcal{I}}^{(\Pi_n)}\hat{\bS}_{\mathcal{I}}^{(\Pi_n)}(\hat{\bV}_{\mathcal{I}}^{(\Pi_n)})^T$. 
Further, define $\bepsilon^{\Pi_n} = [\epsilon_n^{\pi}: \pi \in \Pi_n] \in \R^{|\Pi_n|}$. \\
\noindent Then to begin, note that since $\hat{\bV}^{(\Pi_n)}_{\mathcal{I}}$ is a isometry, we have that
\begin{equation*}
    \lVert \mathcal{P}_{\hat{V}_{\mathcal{I}}^{(\Pi_n)}}\Delta^n_w  \rVert^2_2  = \lVert (\hat{\bV}^{(\Pi_n)}_{\mathcal{I}})^T\Delta^n_w \rVert^2_2
\end{equation*}

\noindent To upper bound $\lVert (\hat{\bV}^{(\Pi_n)}_{\mathcal{I}})^T \Delta^n_w \rVert^2_2$ as follows, consider 
\begin{equation*}
   \lVert \hat{\E}[\bY^{(\Pi_n)}_{\mathcal{I}}] \Delta^n_w\rVert^2_2 = \left( (\hat{\bV}^{(\Pi_n)}_{\mathcal{I}})^T\Delta^n_w \right)^T \left(\hat{\bS}_{\mathcal{I}}^{(\Pi_n)}\right)^2 \left( (\hat{\bV}^{(\Pi_n)}_{\mathcal{I}})^T \Delta^n_w \right) \geq \hat{s}^2_{r_n} \lVert (\hat{\bV}^{(\Pi_n)}_{\mathcal{I}})^T \Delta^n_w  \rVert^2_2
\end{equation*}

\noindent Using the two equations above gives us
\begin{equation}
\label{eq:projection_term2_intermediate_bound1}
    \lVert \mathcal{P}_{\hat{V}_{\mathcal{I}}^{(\Pi_n)}}\Delta^n_w  \rVert^2_2   \leq \frac{ \lVert \hat{\E}[\bY^{(\Pi_n),r_n}_{\mathcal{I}}] \Delta^n_w\rVert^2_2}{\hat{s}^2_{r_n}}
\end{equation}

\noindent To bound the numerator in \eqref{eq:projection_term2_intermediate_bound1}, note that  by definition $\E[\bY^{(\Pi_n)}] = \E[\bY_{\mathcal{I}}^{(\Pi_n)}] \Tilde{\bw}^n$. Using this observation, we have
\begin{align}
    \lVert \hat{\E}[\bY^{(\Pi_n),r_n}_{\mathcal{I}}] \Delta^n_w\rVert^2_2 & \leq 2 \lVert \hat{\E}[\bY^{(\Pi_n),r_n}_{\mathcal{I}}] \hat{\bw}^n - \E[\bY^{(\Pi_n)}] \rVert^2_2 + 2\lVert \hat{\E}[\bY^{(\Pi_n),r_n}_{\mathcal{I}}] \tilde{\bw}^n - \E[\bY^{(\Pi_n)}] \rVert^2_2 \nonumber \\
    & = 2\lVert \hat{\E}[\bY^{(\Pi_n),r_n}_{\mathcal{I}}] \hat{\bw}^n - \E[\bY^{(\Pi_n)}] \rVert^2_2 + 2\lVert (\hat{\E}[\bY^{(\Pi_n),r_n}_{\mathcal{I}}] - \E[\bY_{\mathcal{I}}^{(\Pi_n)}])\tilde{\bw}^n  \rVert^2_2\label{eq:term2_numerator_bound}
\end{align} 
\noindent To proceed, we then use the following inequality: for any $\mathbf{A} \in \mathbb{R}^{a \times b}, v \in \mathbb{R}^b$, we have
\begin{equation}
\label{eq:l2_infty_inequality}
    \lVert \mathbf{A}v \rVert_2 = \lVert \sum^b_{j =1} \mathbf{A}_{.j} v_j \rVert_2 \leq \left( \max_{j \leq b} \lVert \mathbf{A}_{.j} \rVert_2 \right) \left( \sum^b_{j=1} v_j \right) = \lVert \mathbf{A} \rVert_{2,\infty} \lVert v \rVert_1
\end{equation} 
\noindent Substituting \eqref{eq:term2_numerator_bound} into \eqref{eq:projection_term2_intermediate_bound1} and then applying inequality \eqref{eq:l2_infty_inequality} gives us  
\begin{equation}
\label{eq:projection_term2_intermediate_bound2}
     \lVert \mathcal{P}_{\hat{V}_{\mathcal{I}}^{(\Pi_n)}}\Delta^n_w  \rVert^2_2  \leq \frac{2}{\hat{s}^2_{r_n}}\left(\lVert \hat{\E}[\bY^{(\Pi_n),r_n}_{\mathcal{I}}] \hat{\bw}^n - \E[\bY^{(\Pi_n)}] \rVert^2_2  + \lVert \hat{\E}[\bY^{(\Pi_n),r_n}_{\mathcal{I}}] - \E[\bY_{\mathcal{I}}^{(\Pi_n)}] \rVert^2_{2,\infty} \lVert \tilde{\bw}^n \rVert^2_1\right)
\end{equation}
Next, we bound $\lVert \hat{\E}[\bY^{(\Pi_n),r_n}_{\mathcal{I}}] \hat{\bw}^n - \E[\bY^{(\Pi_n)}] \rVert^2_2$. To this end, note that Assumption \ref{ass:observation_model} implies that
\begin{align}
     &\lVert \hat{\E}[\bY^{(\Pi_n),r_n}_{\mathcal{I}}] \hat{\bw}^n - \bY^{(\Pi_n)} \rVert^2_2 \nonumber \\
     & = \lVert \hat{\E}[\bY^{(\Pi_n),r_n}_{\mathcal{I}}] \hat{\bw}^n - \E[\bY^{(\Pi_n)}] - \bepsilon^{\Pi_n} \rVert^2_2 \nonumber \\
    & =  \lVert \hat{\E}[\bY^{(\Pi_n),r_n}_{\mathcal{I}}] \hat{\bw}^n - \E[\bY^{(\Pi_n)}] \rVert^2_2 + \lVert \bepsilon^{\Pi_n} \rVert^2_2 - 2\langle  \hat{\E}[\bY^{(\Pi_n),r_n}_{\mathcal{I}}] \hat{\bw}^n - \E[\bY^{(\Pi_n)}], \bepsilon^{\Pi_n} \rangle \label{eq:pcr_training_error_bound_noiseless}
\end{align}
\noindent Next, we proceed by calling upon Property 4.1 of \cite{agarwal2020principal} which states that $\hat{\bw}^n$ as given by \eqref{eq:pcr_linear_model_def} is the unique solution to the following convex program: 
\begin{align}
\label{eq:hat_w_min_l2_property}
& \min_{\bw \in \mathbb{R}^{|\mathcal{I}|}} \quad  ||{\bw}||_2 \nonumber \\
& \textrm{s.t. } \bw \in \argmin_{\bw \in \mathbb{R}^{|\mathcal{I}|}} \lVert \bY^{(\Pi_n)} - \hat{\E}[\bY^{(\Pi_n),r_n}_{\mathcal{I}}]\bw \rVert^2_2
\end{align}
Using this property, we have that
\begin{align}
    & \lVert \hat{\E}[\bY^{(\Pi_n),r_n}_{\mathcal{I}}] \hat{\bw}^n - \bY^{(\Pi_n)} \rVert^2_2 \nonumber 
    \\ & \leq   \lVert \hat{\E}[\bY^{(\Pi_n),r_n}_{\mathcal{I}}] \tilde{\bw}^n - \bY^{(\Pi_n)} \rVert^2_2 \nonumber \\
    & = \lVert \hat{\E}[\bY^{(\Pi_n),r_n}_{\mathcal{I}}] \tilde{\bw}^n - \E[\bY^{(\Pi_n)}] - \bepsilon^{\Pi_n}  \rVert^2_2 \nonumber\\
    & = \lVert \hat{\E}[\bY^{(\Pi_n),r_n}_{\mathcal{I}}] \tilde{\bw}^n - \E[\bY_{\mathcal{I}}^{(\Pi_n)}]\tilde{\bw}^n  - \bepsilon^{\Pi_n}  \rVert^2_2 \nonumber \\
    & = \lVert (\hat{\E}[\bY^{(\Pi_n),r_n}_{\mathcal{I}}] - \E[\bY_{\mathcal{I}}^{(\Pi_n)}])\tilde{\bw}^n \rVert^2_2 + \lVert \bepsilon^{\Pi_n} \rVert^2_2 - 2 \langle (\hat{\E}[\bY^{(\Pi_n),r_n}_{\mathcal{I}}] - \E[\bY_{\mathcal{I}}^{(\Pi_n)}])\tilde{\bw}^n, \bepsilon^{\Pi_n} \rangle \label{eq:pcr_training_error_bound}
\end{align}

\noindent Substituting \eqref{eq:pcr_training_error_bound_noiseless} and \eqref{eq:pcr_training_error_bound} into \eqref{eq:projection_term2_intermediate_bound2} and using \eqref{eq:l2_infty_inequality}, we get
\begin{align*}
& \lVert \hat{\E}[\bY^{(\Pi_n),r_n}_{\mathcal{I}}] \hat{\bw}^n - \E[\bY^{(\Pi_n)}] \rVert^2_2 \\
& \leq \lVert (\hat{\E}[\bY_{\mathcal{I}}^{(\Pi_n),r_n}] - \E[\bY_{\mathcal{I}}^{(\Pi_n)}])\tilde{\bw}^n \rVert^2_2 + 2 \langle (\hat{\E}[\bY^{(\Pi_n),r_n}_{\mathcal{I}}])\Delta_w^n, \bepsilon^{\Pi_n} \rangle \\
& \leq \lVert (\hat{\E}[\bY_{\mathcal{I}}^{(\Pi_n),r_n}] - \E[\bY_{\mathcal{I}}^{(\Pi_n)}])\rVert^2_{2,\infty} \lVert \tilde{\bw}^n  \rVert^2_1 + 2 \langle (\hat{\E}[\bY^{(\Pi_n),r_n}_{\mathcal{I}}])\Delta_w^n, \bepsilon^{\Pi_n} \rangle 
\end{align*}
    
\noindent Then substituting this equation into \eqref{eq:projection_term2_intermediate_bound2} gives us 
\begin{equation}
\label{eq:projection_operator_term2_intermediate_bound}
   \lVert \mathcal{P}_{\hat{V}_{\mathcal{I}}^{(\Pi_n)}}\Delta^n_w \rVert^2_2 \leq \frac{4}{\hat{s}^2_{r_n}}\left(\lVert \hat{\E}[\bY_{\mathcal{I}}^{(\Pi_n),{r_n}}] - \E[\bY_{\mathcal{I}}^{(\Pi_n)}] \rVert^2_{2,\infty} \lVert \Tilde{\bw}^n \rVert^2_1 +  \langle  \hat{\E}[\bY_{\mathcal{I}}^{(\Pi_n),{r_n}}] \Delta^n_w, \bepsilon^{\Pi_n} \rangle \right)
\end{equation}
We state three lemmas that help us bound the equation above with their proofs given in Appendices \ref{subsubsec:proof_singular_value_lower_bound}, \ref{subsubsec:proof_of_lemma_rank_r_approximation_l2_infty_bound} and \ref{subsubsec:proof_of_lemma_rank_r_approximaton_noise_inner_product_bound} respectively. 
\begin{lemma}
\label{lem:singular_value_lower_bound}
Let the set-up of Theorem \ref{thm:potential_outcome_convergence_rate} hold. Then, 
\begin{equation*}
    \hat{s}_{r_n} - s_{r_n} = O_p(1)
\end{equation*}
\end{lemma}

\begin{lemma}
\label{lem:rank_r_approximation_l2_infty_bound}
Let the set-up of Theorem \ref{thm:potential_outcome_convergence_rate} hold. Then for a universal constant $C > 0$, we have 
\begin{equation}
   \lVert \hat{\E}[\bY_{\mathcal{I}}^{(\Pi_n),{r_n}}] - \E[\bY_{\mathcal{I}}^{(\Pi_n)}] \rVert_{2,\infty} \leq C \sqrt{r_n|\Pi_n|} \Delta_E
\end{equation}
\end{lemma}

\begin{lemma} Let the set-up of Theorem \ref{thm:potential_outcome_convergence_rate} hold. Then, 
\label{lem:rank_r_approximaton_noise_inner_product_bound}
 \begin{equation}
  \langle  \hat{\E}[\bY_{\mathcal{I}}^{(\Pi_n),{r_n}}] \Delta^n_w, \bepsilon^{\Pi_n} \rangle = O_p \left(\sqrt{|\Pi_n|} + r_n +  \lVert \hat{\E}[\bY_{\mathcal{I}}^{(\Pi_n),{r_n}}] - \E[\bY_{\mathcal{I}}^{(\Pi_n)}] \rVert_{2,\infty} \lVert \Tilde{\bw}^n \rVert_1 \right)
\end{equation}
\end{lemma}
\noindent Using the results of three lemmas above, applying Assumption \ref{ass:balanced_spectrum} in \eqref{eq:projection_operator_term2_intermediate_bound}, and then simplifying gives us 
\begin{equation}
\label{eq:projection_operator_term2_bound}
    \lVert \mathcal{P}_{\hat{V}_{\mathcal{I}}^{(\Pi_n)}}\Delta^n_w \rVert_2  = O_p \left(\frac{r_n\Delta_E\lVert \Tilde{\bw}^n \rVert_1}{\sqrt{|\mathcal{I}|}} + \frac{r_n\left( 1 + \sqrt{\Delta_E\lVert \Tilde{\bw}^n \rVert_1}\right)}{\sqrt{|\mathcal{I}|}|\Pi_n|^{1/4}}\right)
\end{equation}
\noindent This concludes the proof for term 2. Using the result of Lemma \ref{lem:l1_bound_w_tilde}, we have that $\lVert \Tilde{\bw}^n \rVert_1 \leq c\sqrt{r_n}$. Substituting this into \eqref{eq:projection_operator_term2_bound} gives us
\begin{equation}
\label{eq:projection_operator_term2_bound_simplified}
    \lVert \mathcal{P}_{\hat{V}_{\mathcal{I}}^{(\Pi_n)}}\Delta^n_w \rVert_2  = O_p \left(\frac{r_n^{3/2}\Delta_E}{\sqrt{|\mathcal{I}|}} + \frac{r_n\left( 1 + \sqrt{\Delta_E}r^{1/4}_n\right)}{\sqrt{|\mathcal{I}|}|\Pi_n|^{1/4}}\right)
\end{equation}

\noindent \emph{Collecting Terms. }Combining the results of \eqref{eq:projection_operator_term1_bound_simplified} and \eqref{eq:projection_operator_term2_bound_simplified}, gives us
\begin{align*}
& \lVert \mathcal{P}_{V_{\mathcal{I}}^{(\Pi_n)}}\Delta^n_w \rVert_2 \nonumber \\
& = O_p \left( \log^{3}(|\Pi_n||\mathcal{I}|)r_n^{2}\left[\frac{\Delta_E  }{\sqrt{|\mathcal{I}|}\min\{\sqrt{|\Pi_n|},\sqrt{|\mathcal{I}|\}}}  + \frac{\Delta^2_E}{\sqrt{|\mathcal{I}|}} \right]  + \frac{r_n^{3/2}\Delta_E}{\sqrt{|\mathcal{I}|}} + \frac{r_n\left( 1 + \sqrt{\Delta_E}r^{1/4}_n\right)}{\sqrt{|\mathcal{I}|}|\Pi_n|^{1/4}}  \right) 
\end{align*}

\subsubsection{Proof of Lemma \ref{lem:singular_value_lower_bound}}
\label{subsubsec:proof_singular_value_lower_bound}
We first state Weyl's inequality which will be useful for us to establish the results. 
\begin{theorem}[Weyl's Inequality]
\label{thm:weyl_inequality}
Given two matrices $\mathbf{A},\mathbf{B} \in \R^{m \times n}$, let $s_i$ and $\hat{s}_i$ denote the $i$-th singular values of $\mathbf{A}$ and $\mathbf{B}$ respectively. Then, for all, $i \leq \min\{n,m\}$, we have $\left| s_i - \hat{s}_i \right| \leq \left\lVert\mathbf{A} - \mathbf{B} \right\rVert_{\text{op}}$
\end{theorem}
\noindent Using Weyl's inequality gives us 
\begin{align*}
    | \hat{s}_{r_n} - s_{r_n} | & \leq \lVert \E[\bY_{\mathcal{I}}^{(\Pi_n)}] - \hat{E}[\bY_{\mathcal{I}}^{(\Pi_n)}] \rVert_{\text{op}} \\
    & \leq \sqrt{|\mathcal{I}||\Pi_n|}\Delta_E
\end{align*}
\noindent Using the inequality above and Assumption \ref{ass:balanced_spectrum}, we have 
\begin{align*}
\hat{s}_{r_n} & \geq s_{r_n} - \sqrt{|\mathcal{I}|\Pi_n}\Delta_E \\
        & = s_{r_n}\left(1 - \frac{\sqrt{|\mathcal{I}|\Pi_n}\Delta_E}{s_{r_n}}\right) \\
        & \geq s_{r_n}\left(1 - \sqrt{r_n}\Delta_E \right) \\
\end{align*}
Then, substituting \eqref{eq:delta_E_subsitution} into the equation above gives us that
\begin{equation*}
    \frac{\hat{s}_{r_n}}{s_{r_n}} \geq \left(1 - C\sqrt{\frac{r_ns^2p}{M}}\right)
\end{equation*}
holds with high probability for some universal constant $C \geq 0$. Finally, using the assumption that $M = \omega(r_n^2s^2p)$ yields the claimed result.

\subsubsection{Proof of Lemma \ref{lem:rank_r_approximation_l2_infty_bound}}
\label{subsubsec:proof_of_lemma_rank_r_approximation_l2_infty_bound}
For notational simplicity, let $\hat{\bU}_{r_n},\hat{\bS}_{r_n},\hat{\bV}_{r_n}$ denote $\hat{\bU}_{\mathcal{I}}^{(\Pi_n)},\hat{\bS}_{\mathcal{I}}^{(\Pi_n)},\hat{\bV}_{\mathcal{I}}^{(\Pi_n)}$ respectively.  For a matrix $\mathbf{A}$, let $A_{.,j}$ denote its $j$-th column. Additionally, denote  $\Delta_{j} = \hat{\E}[\bY_{\mathcal{I}}^{(\Pi_n),{r_n}}]_{.,j} - \E[\bY_{\mathcal{I}}^{(\Pi_n)}]_{.,j}$. Then,
\begin{align*}
    & \hat{\E}[\bY_{\mathcal{I}}^{(\Pi_n),{r_n}}]_{.,j} - \E[\bY_{\mathcal{I}}^{(\Pi_n)}]_{.,j} \\
    & = \left(\hat{\E}[\bY_{\mathcal{I}}^{(\Pi_n),{r_n}}]_{.,j} - \hat{\bU}_{r_n}\hat{\bU}^T_{r_n} \E[\bY_{\mathcal{I}}^{(\Pi_n)}]_{.,j} \right) + \left(\hat{\bU}_{r_n}\hat{\bU}^T_{r_n} \E[\bY_{\mathcal{I}}^{(\Pi_n)}]_{.,j}  - \E[\bY_{\mathcal{I}}^{(\Pi_n)}]_{.,j} \right)
\end{align*}
\noindent We have that $\left(\hat{\E}[\bY_{\mathcal{I}}^{(\Pi_n),{r_n}}]_{.,j} - \hat{\bU}_{r_n}\hat{\bU}^T_{r_n} \E[\bY_{\mathcal{I}}^{(\Pi_n)}]_{.,j} \right)$ belongs to the subspace spanned by the column vectors of $\hat{\bU}_{r_n}$, while $\left(\hat{\bU}_{r_n}\hat{\bU}^T_{r_n} \E[\bY_{\mathcal{I}}^{(\Pi_n)}]_{.,j}  - \E[\bY_{\mathcal{I}}^{(\Pi_n)}]_{.,j} \right)$ belongs to its orthogonal complement. Therefore, 
\begin{align*}
     & \lVert \hat{\E}[\bY_{\mathcal{I}}^{(\Pi_n),{r_n}}]_{.,j} - \E[\bY_{\mathcal{I}}^{(\Pi_n)}]_{.,j} \rVert^2_2 \\
      & = \left\lVert\hat{\E}[\bY_{\mathcal{I}}^{(\Pi_n),{r_n}}]_{.,j} - \hat{\bU}_{r_n}\hat{\bU}^T_{r_n} \E[\bY_{\mathcal{I}}^{(\Pi_n)}]_{.,j} \right\rVert^2_2 + \left\lVert\hat{\bU}_{r_n}\hat{\bU}^T_{r_n} \E[\bY_{\mathcal{I}}^{(\Pi_n)}]_{.,j}  - \E[\bY_{\mathcal{I}}^{(\Pi_n)}]_{.,j} \right\rVert^2_2
\end{align*}

\noindent \emph{Bounding $\left\lVert\hat{\E}[\bY_{\mathcal{I}}^{(\Pi_n),{r_n}}]_{.,j} - \hat{\bU}_{r_n}\hat{\bU}^T_{r_n} \E[\bY_{\mathcal{I}}^{(\Pi_n)}]_{.,j} \right\rVert^2_2$}. 
Observe that, we have 
\begin{align*}
    \hat{\bU}_{r_n}\hat{\bU}^T_{r_n}\hat{\E}[\bY_{\mathcal{I}}^{(\Pi_n)}]_{.,j} & =   \hat{\bU}_{r_n}\hat{\bU}^T_{r_n} \hat{\E}[\bY_{\mathcal{I}}^{(\Pi_n)}]\mathbf{e_{j}}  = \hat{\bU}_{r_n}\hat{\bU}^T_{r_n}\hat{\bU}_{\mathcal{I}}^{(\Pi_n)}\hat{\bS}_{\mathcal{I}}^{(\Pi_n)}\hat{\bV}_{\mathcal{I}}^{(\Pi_n)} \mathbf{e_{j}} \\
    & =  \hat{\bU}_{r_n}\hat{\bS}_{r_n}\hat{\bV}^T_{r_n}\mathbf{e}_j =    \hat{\E}[\bY_{\mathcal{I}}^{(\Pi_n),{r_n}}]_{.,j}
\end{align*}
\noindent Therefore, we have
\begin{align*}
    \left\lVert\hat{\E}[\bY_{\mathcal{I}}^{(\Pi_n),{r_n}}]_{.,j} - \hat{\bU}_{r_n}\hat{\bU}^T_{r_n} \E[\bY_{\mathcal{I}}^{(\Pi_n)}]_{.,j} \right\rVert^2_2  & = \left\lVert\hat{\bU}_{r_n}\hat{\bU}^T_{r_n}\hat{\E}[\bY_{\mathcal{I}}^{(\Pi_n)}]_{.,j} - \hat{\bU}_{r_n}\hat{\bU}^T_{r_n} \E[\bY_{\mathcal{I}}^{(\Pi_n)}]_{.,j} \right\rVert^2_2 \\
    & \leq \left\lVert\hat{\bU}_{r_n}\hat{\bU}^T_{r_n} \right\rVert^2_2 \left \lVert \hat{\E}[\bY_{\mathcal{I}}^{(\Pi_n)}]_{.,j} - \E[\bY_{\mathcal{I}}^{(\Pi_n)}]_{.,j}\right\rVert^2_2 \\ 
    & \leq |\Pi_n|\Delta^2_E
\end{align*} 

\noindent \emph{Bounding $\left\lVert\hat{\bU}_{r_n}\hat{\bU}^T_{r_n} \E[\bY_{\mathcal{I}}^{(\Pi_n)}]_{.,j}  - \E[\bY_{\mathcal{I}}^{(\Pi_n)}]_{.,j} \right\rVert^2_2$.} Note that $ \E[\bY_{\mathcal{I}}^{(\Pi_n)}]_{.,j} = \bU_{\mathcal{I}}^{(\Pi_n)}\left(\bU_{\mathcal{I}}^{(\Pi_n)}\right)^T\E[\bY_{\mathcal{I}}^{(\Pi_n)}]_{.,j}$. Using Assumption \ref{ass:boundedness_potential_outcome} and \eqref{eq:projection_operator_error_bound_simplified}, we have 
\begin{align*}
    \left\lVert\hat{\bU}_{r_n}\hat{\bU}^T_{r_n} \E[\bY_{\mathcal{I}}^{(\Pi_n)}]_{.,j}  - \E[\bY_{\mathcal{I}}^{(\Pi_n)}]_{.,j} \right\rVert^2_2 & \leq \left\lVert \hat{\bU}_{r_n}\hat{\bU}^T_{r_n} -  \bU_{\mathcal{I}}^{(\Pi_n)}\left(\bU_{\mathcal{I}}^{(\Pi_n)}\right)^T \right\rVert^2_2 \lVert \E[\bY_{\mathcal{I}}^{(\Pi_n)}]_{.,j} \rVert^2_2 \\
    & \leq \left\lVert \hat{\bU}_{r_n}\hat{\bU}^T_{r_n} -  \bU_{\mathcal{I}}^{(\Pi_n)}\left(\bU_{\mathcal{I}}^{(\Pi_n)}\right)^T \right\rVert^2_2 \left|\Pi_n \right| \\
    & \leq C r_n |\Pi_n| \Delta^2_E
\end{align*}
\noindent for a universal constant $C > 0$. 
\noindent Combining the two bounds gives us, 
\begin{equation*}
    \max_{j} \ \Delta_j \leq C\sqrt{r_n|\Pi_n|}\Delta_E
\end{equation*}
\noindent for a universal constant $C > 0.$

\subsubsection{Proof of Lemma \ref{lem:rank_r_approximaton_noise_inner_product_bound}}
\label{subsubsec:proof_of_lemma_rank_r_approximaton_noise_inner_product_bound}

The proof of Lemma \ref{lem:rank_r_approximaton_noise_inner_product_bound} follows from adapting the notation of Lemma H.5 in \cite{agarwal2020synthetic} to that used in this paper. Specifically, we have the notational changes established in Table \ref{tab:notational_changes}, as well as the following changes: $r_{pre} = r_n, \bY_{pre,\mathcal{I}^{(d)}}^{r_{pre}} = \hat{\E}[\bY_{\mathcal{I}}^{(\Pi_n),{r_n}}], \E[\bY_{pre,\mathcal{I}^{(d)}}], \epsilon_{pre,n} = \bepsilon^{\Pi_n}, T_0 = \Pi_n$  where the left hand side of each equality is the notation used in \cite{agarwal2020synthetic} , and the right hand side is the notation used in this work. Using the notation established, we can use the result of Lemma H.5 in \cite{agarwal2020synthetic} to give us 
\begin{equation*}
      \langle  \hat{\E}[\bY_{\mathcal{I}}^{(\Pi_n),{r_n}}] \Delta^n_w, \bepsilon^{\Pi_n} \rangle = O_p \left(\sqrt{|\Pi_n|} + r_n +  \lVert \hat{\E}[\bY_{\mathcal{I}}^{(\Pi_n),{r_n}}] - \E[\bY_{\mathcal{I}}^{(\Pi_n)}] \rVert_{2,\infty} \lVert \Tilde{\bw}^n \rVert_1 \right)
\end{equation*}
which completes the proof.

\subsection{Proof of Proposition \ref{prop:pcr_convergence_rate}}
\label{subsec:pcr_proofs}
We prove Proposition \ref{prop:pcr_convergence_rate}, which also serves as the finite-sample analysis of the vertical regression procedure (step 2 of \method) done via principal component regression. The proof follows from adapting the notation of Proposition E.3 in \cite{agarwal2021causal} to that used in this paper. Specifically, we present Table \ref{tab:notational_changes} below that matches their notation to ours. 
This gives the result,
\begin{equation*}
    \lVert \hat{\bw}^n - \Tilde{\bw}^n \rVert_2 = O_p \left(\log^{3}(|\Pi_n||\mathcal{I}|)r_n\lVert \Tilde{\bw}^{n} \rVert_2\left[\frac{ 1 }{\min\{\sqrt{|\Pi_n|},\sqrt{|\mathcal{I}|\}}}  + \Delta_E \right] \right)
\end{equation*}
\begin{table}[H]
\centering
\begin{tabular}{||c c||} 
 \hline
 Notation of \cite{agarwal2021causal} & Our Notation \\ [1ex] 
 \hline
 $\bY$ & $\bY_{\Pi_n}$  \\ [1ex]
 \hline
 $\bX$ & $\E[\bY_{\mathcal{I}}^{(\Pi_n)}]$ \\  [1ex]
 \hline
 $\bZ$ &  $\hat{\E}[\bY_{\mathcal{I}}^{(\Pi_n)}]$  \\ [1ex]
 \hline 
 $n$ & $|\Pi_n|$ \\
 \hline 
 $p$ & $|\mathcal{I}|$ \\
 \hline 
$\bbeta^*$ & $\Tilde{\bw}$ \\
\hline
$\hat{\bbeta}$ & $\hat{\bw}$ \\
\hline
$\Delta_E$ & $\Delta_E$ \\
\hline 
$r$ & $r_n$ \\
\hline 
$\phi^{lr}$ & $0$ \\ [1ex]
 \hline
 $\Bar{A}$ & $1$ \\ 
 \hline
 $\Bar{K}$ & 0  \\ [1ex] 
 \hline
 $K_{a}, \kappa, \Bar{\sigma}$ & $C\sigma$ \\
 \hline 
 $\rho_{min}$ & 1 \\
 \hline 
 \end{tabular}
 \vspace{1mm}
\caption{A summary of the main notational differences between our setting and that of \cite{agarwal2021causal}.}
\label{tab:notational_changes}
\end{table}



\section{CART Horizontal Regression}
\label{sec:CART_horizontal_regression}

\subsection{CART Background}
\label{sec:CART_background}

Assume we have access to a training set $\data = \{(\bx_1,y_1) \ldots (\bx_n,y_n)\}$ where $\bx_i$ is sampled from a distribution $\mathcal{D}_x$ with support $\{-1,1\}^p$, and the responses are generated as $y = f(\bx_i) + \epsilon$.  In the context of combinations, $\bx_i$ can correspond to the binary vector $\bv(\pi)$. 

Let $S \subset [p]$ denote a subset of coordinate indices, and $\bz \in \{-1,1\}^{|S|}$. Then, we define a \emph{node} $\cell \subset \{-1,1\}^p$ as a subcube of the form $\cell(S,\bz) = \braces{\bx \in \{-1,1\}^p ~\colon~ x_j = z_j~\text{for}~j \in S}$. Let $N(\cell) \coloneqq | \braces{i :  \bx_{i} \in \cell } |$ denote the size of this set. Define the sub-cell $\cell_{j,1} = \{\bx \in \cell, x_j = 1\}$ with respect to a feature $j$.  Analogously, define the sub-cell $\cell_{j,-1} = \{\bx \in \cell, x_j = -1\}$. Furthermore, denote the mean response over the node by
$$
\hat\E_{\cell}[y] \coloneqq \hat\E[y|\bx \in \cell] = \frac{1}{N(\cell)}\sum_{\bx_i \in \cell} y_{i}.
$$



A \emph{partition} $\partition = \braces{\cell_1,\ldots,\cell_m}$ is a collection of nodes with disjoint interiors, whose union is the Boolean hypercube $\{-1,1\}^p$. For any $\bx \in \{-1,1\}^p$, we define $\partition(\bx)$ to be the cell of $\partition$ that contains $\bx$.  Given the training set $\data$, every partition yields an estimator $\hat f(-; \partition, \data)$ for $f$ via \emph{leaf-only averaging}: for every input $\bx$, the estimator outputs the mean response over the node containing $\bx$. In other words, we define
$$
\hat f(\bx; \partition, \data) \coloneqq \sum_{\cell \in \partition} \hat{\E}_\cell[y]~ \indicator\braces{\bx \in \cell}.
$$




\noindent The CART algorithm constructs a partition $\mathcal{\hat{T}} = \braces{\cell_1,\ldots,\cell_m}$, by choosing a feature $j$ to split on at each intermediate node in a greedy fashion according to the impurity decrease split criterion defined as follows
\begin{equation*}
    \hat\Delta(\cell,j) \coloneqq \frac{1}{N(\cell)} \left( \sum_{\bx_i \in \cell}\left(y_{i} - \hat\E_\cell[y] \right)^2 - \sum_{\bx_i \in \cell_{j,1}}\left(y_{i} - \hat\E_{\cell_{j,1}}[y] \right)^2 - \sum_{\bx_i \in \cell_{j,-1}}\left(y_{i} - \hat\E_{\cell_{j,-1}}[y] \right)^2 \right).
\end{equation*}

The tree is stopped from growing upon reaching a stopping condition. For example, common stopping conditions include minimum number of samples in a node, maximum depth, or when the number of nodes exceeds a pre-sepcified threshold.




\subsection{CART Finite-Sample Analysis}
\label{supp:CART_finite_sample}
\noindent We now define the additional assumptions required for finite-sample analysis. Throughout our analysis, we follow the notation used in \cite{syrgkanis2020estimation}. 
%
%
%

\begin{definition}
For a given partition $\partition$, and a cell $\node \in \partition$, define 
\begin{equation}
    \Bar{V}(\cell,\partition) = \E_{\bx \sim \mathcal{D}_x}\left[\left( \E_{\bz \sim \mathcal{D}_x }[f(\bz) | z \in \partition(\bx)] \right)^2   \ | \  \bx \in \cell \right]
\end{equation}
\end{definition}

\noindent In order to introduce the next assumption, we first introduce some additional notation. For a partition $\partition$, and cell $\node \in \partition$, define the split operator $\mathcal{S}\left(\partition,\cell,j\right)$ that outputs a partition with the cell $\cell$ split on feature $j$. That is, $\mathcal{S}\left(\partition,\cell,j\right)$ = $(\partition \setminus \{\cell\}) \cup \{\cell_{j,-1}, \cell_{j,1}\}$. The split operator for a partition $\partition$ and cell $\node$ can also be defined for a set of features $T \subset [p]$ by repeatedly splitting the cell $\node$ with respect to all the features in $T$. More formally, this can be defined inductively as follows: for all $j \in T$, let $\mathcal{S}\left(\partition,\cell,T\right) = \mathcal{S}\left(\mathcal{S}\left(\mathcal{S}(\partition, \cell,j), \cell_{j,-1}, T \setminus \{j\}\right), \cell_{j,1}, T \setminus \{j\}\right)$. Moreover, define the short-hand $\Bar{V}(\cell,T) =  \Bar{V}(\cell,\mathcal{S}\left(\partition,\cell,T\right))$

\begin{definition}[Strong Partition Sparsity, Assumption 4.2 in \cite{syrgkanis2020estimation}]
\label{def:strong_partition_sparsity}
A function $f: \{-1,1\}^p \to \mathbb{R}$ is $(\beta,k)$-strong partition sparse if $f$ is a $k-$junta (see \eqref{eq:k_junta}) with relevant features $K$ and the function $\Bar{V}$ satisfies: $\Bar{V}(\cell, T \cup \{j\}) - \Bar{V}(\cell, T) + \beta \leq \Bar{V}(\cell, T \cup \{i\}) - \Bar{V}(\cell, T)$, for all possible cells $\cell \in \partition$, any $T \subset [p]$, all $i \in K$, $j \in [p] \setminus K$. 
\end{definition}

\noindent  A function satisfying Definition \ref{def:strong_partition_sparsity} and using Equation (4.4) in \cite{syrgkanis2020estimation} imply that the reduction in expected mean square error (i.e., the impurity decrease) when splitting a cell on a relevant variable is larger than when splitting on an irrelevant variable by at least $\beta$. This generative assumption is required for greedy algorithms such as CART to identify the relevant feature subset $K$.

\begin{assumption}[Bounded Noise]
\label{ass:bounded_noise}
Conditioned on the Fourier coefficients $\mathcal{A}$, for any unit-intervention pair $(n,\pi)$, $\epsilon_n^{\pi}$ are independent zero-mean bounded random variables with $\epsilon_n^{\pi} \in [-1/2,1/2]$.
\end{assumption}

\begin{theorem}
\label{thm:CART_convergence_rate} Suppose $\Pi_u \subset \Pi$ is chosen uniformly and independently at random. Additionally, let 
Assumptions \ref{ass:observation_model},  \ref{ass:boundedness_potential_outcome}, \ref{ass:bounded_noise}  hold and suppose that $\E[Y_u^{(\pi))}]$ satisfies definition \ref{def:strong_partition_sparsity} with parameters $(\beta,k)$. If $\hat{T}$ is grown with number of nodes\footnote{In practice, the number of nodes can be tuned by cross-validation.} $m = 2^k$, $s = 2^k$, and $|\Pi_u|  = \Omega(2^k\log(p)/\beta^2)$, then we have 
\begin{equation}
\label{eq:CART_convergene_rate}
  \hat{\E}[Y_u^{(\pi)}] - \E[Y_u^{(\pi)}] = O_p\left(\sqrt{\frac{s^2\log(s)}{|\Pi_u|}}  \right) 
\end{equation}
for any donor unit $u \in \mathcal{I}$ and any combination $\pi \in \Pi$.
\end{theorem}

\begin{remark}[CART Donor Set Inclusion] Under the set-up of Theorem \ref{thm:CART_convergence_rate}, any unit $n \in [N]$ that has  $|\Pi_n| =  \Omega(2^k\log(p)/\beta^2)$ randomly sampled observations can be considered a donor unit. 
\label{rem:CART_donor_set_inclusion}
\end{remark}

\noindent CART consistently estimates the target causal parameter for a donor unit $u$ provided that the structural assumptions on the potential outcomes are satisfied and the number of observations $|\Pi_u|$ grows as $\omega\left(s^2\max\{\log(p),
\log(s)\} \right)$. 
On the other hand, the Lasso estimator requires that $|\Pi_u|$ grows as $\omega(s^2p)$. 
The difference in sample complexity can be ascribed to the fact that CART performs feature selection (i.e., learns the subset of features that $\E[Y_u^{(\pi)}]$ depends on), thereby only having to learn the Fourier coefficient over a $2^k$ dimensional space, instead of $2^p$ for the Lasso. 
It remains as interesting future work to analyze the use of other decision tree methods such as those proposed in \cite{tan2023fast,agarwal2022hierarchical,agarwal2023mdi,mazumder2022quant}.

\section{Proofs for CART Horizontal Regression}

\subsection{Proof of Theorem \ref{thm:CART_convergence_rate}}
\label{subsec:CART_horizontal_regression_proof}

 Let $\mathcal{E}_{1}$ denote the event that $\hat{K} = K$. Further, define the event $\mathcal{E}_{S}$ for any subset $S \subset K$ as  
\begin{equation*}
    \mathcal{E}_{S} =  \left\{|\hat{\alpha}_{u,S} - \alpha_{u,S} |\leq \sqrt{\frac{8\log(2^{k+3}/\delta)}{|\Pi_u^b|}}  \right\}
\end{equation*}
\noindent Additionally, let $\mathcal{E} = \mathcal{E}_1 \cap \left(\cap_{S \subset K}\mathcal{E}_S\right) $. We first show that $\mathcal{E}$ occurs with high probability. To establish this claim, we state the following two results proved in Appendix \ref{subsubsec:cart_helper_lemmas}.

\begin{proposition} [Theorem D.9 in \cite{syrgkanis2020estimation}]
\label{prop:CART_feature_selection_high_prob_bound}
Let the set-up of Theorem \ref{thm:CART_convergence_rate} hold. Then, $\hat{K} = K$ with probability $1 - \delta/2$. 
\end{proposition}

\begin{lemma}
\label{lem:CART_fourier_coefficient_convergence}
Let the set-up of Theorem \ref{thm:CART_convergence_rate} hold. Then for any subset $S \subset K$, we have 
\begin{equation}
\label{eq:CART_fourier_coefficient_convergence}
\P\left(|\hat{\alpha}_{u,S} - \alpha_{u,S} |\geq \sqrt{\frac{8\log(2^{k+2}/\delta)}{|\Pi_u^b|}}\right) \leq \frac{\delta}{2^k}
\end{equation}
where $\hat{\alpha}_{u,S} =  \frac{1}{|\Pi^b_u|}\sum_{\pi \in \Pi^b_u }Y_{u\pi}\chi^{\pi}_{S}$
\end{lemma}

\noindent Using Proposition \ref{prop:CART_feature_selection_high_prob_bound} and Lemma \ref{lem:CART_fourier_coefficient_convergence}, and applying the union bound alongside Demorgan's law, we get
\begin{equation*}
    \P\left(\mathcal{E}^c_3 \right) = \P\left(\mathcal{E}^c_1 \cup \left(\cup_{S \subset K}\mathcal{E}^c_S\right) \right) \leq \P\left(\mathcal{E}^c_1\right) + \P\left(\cup_{S \subset K}\mathcal{E}^c_{S}\right) \leq  \frac{\delta}{2} + \sum_{S \subset K} \frac{\delta}{2^{k+1}} \leq \delta
\end{equation*}

\noindent Then conditioned on $\mathcal{E}$, we get 
\begin{align*}
    | \hat{\E}[Y_u^{(\pi)}] - \E[Y_u^{(\pi)}] | & =  \left|\sum_{S \subset \hat{K}}\hat{\alpha}_{u,S}\chi^{\pi}_{S} -  \sum_{S \subset K}\alpha_{u,S}\chi^{\pi}_{S} \right| \\
    & = \left|\sum_{S \subset K}\left(\hat{\alpha}_{u,S} - \alpha_{u,S}\right)\chi^{\pi}_{S} \right| \\
    & \leq \sum_{S \subset K}  \left| \hat{\alpha}_{u,S} - \alpha_{u,S}\right | \\ 
    & \leq \sqrt{\frac{2^{2k+3}\log(2^{k+3}/\delta)}{|\Pi_u^b|}}
\end{align*}
\noindent Therefore, we have 
\begin{equation*}
    | \hat{\E}[Y_u^{(\pi)}] - \E[Y_u^{(\pi)}] | = O_p \left(\sqrt{\frac{2^{2k}k}{|\Pi_u|}} \right)
\end{equation*}

\noindent Substituting $s = 2^k$ concludes the proof.

\subsubsection{Proof of Lemma \ref{lem:CART_fourier_coefficient_convergence}}
\label{subsubsec:cart_helper_lemmas}
Fix any subset $S \subset K$. Then for any $t > 0$, we have 
\begin{align}
&\P\left(|\hat{\alpha}_{u,S} - \alpha_{u,S} |\geq t \right) 
\\& = \P\left(\left| \frac{1}{|\Pi^b_u|}\sum_{\pi \in \Pi_u^{b}}\left(\E[Y_u^{(\pi)}]\chi_S^{\pi} + \epsilon_u^{\pi}\chi_S^{\pi}\right) - \alpha_{u,S} \right| \geq t  \right) \nonumber \\
& \leq \P\left( \left| \frac{1}{|\Pi^b_u|}\sum_{\pi \in \Pi_u^{b}}\E[Y_u^{(\pi)}]\chi_S^{\pi} - \alpha_{u,S} \right| + \left|\frac{1}{|\Pi^b_u|}\sum_{\pi \in \Pi_u^{b}}\epsilon_u^{\pi}\chi_S^{\pi} \right| \geq t \right) \nonumber \\
& \leq \P\left( \left| \frac{1}{|\Pi^b_u|}\sum_{\pi \in \Pi_u^{b}}\E[Y_u^{(\pi)}]\chi_S^{\pi} - \alpha_{u,S} \right| \geq \frac{t}{3} \right) + \P\left(\left|\frac{1}{|\Pi^b_u|}\sum_{\pi \in \Pi_u^{b}}\epsilon_u^{\pi}\chi_S^{\pi} \right| \geq \frac{2t}{3} \right) \nonumber
\\ & \leq \P\left( \left| \frac{1}{|\Pi^b_u|}\sum_{\pi \in \Pi_u^{b}}\E[Y_u^{(\pi)}]\chi_S^{\pi} - \alpha_{u,S} \right| \geq \frac{t}{3} \right) + \P\left(\left|\frac{1}{|\Pi^b_u|}\sum_{\pi \in \Pi_u^{b}}\epsilon_u^{\pi}\chi_S^{\pi} \pm \frac{1}{2^p}\sum_{\pi \in \Pi}\epsilon_u^{\pi}\chi_S^{\pi}\right| \geq \frac{2t}{3} \right) \nonumber
\\ & \leq \P\left( \left| \frac{1}{|\Pi^b_u|}\sum_{\pi \in \Pi_u^{b}}\E[Y_u^{(\pi)}]\chi_S^{\pi} - \alpha_{u,S} \right| \geq \frac{t}{3} \right) + \P\left(\left|\frac{1}{|\Pi^b_u|}\sum_{\pi \in \Pi_u^{b}}\epsilon_u^{\pi}\chi_S^{\pi} - \frac{1}{2^p}\sum_{\pi \in \Pi}\epsilon_u^{\pi}\chi_S^{\pi}\right| \geq \frac{t}{3} \right)  \nonumber
\\& \quad + \P\left(\left| \frac{1}{2^p}\sum_{\pi \in \Pi}\epsilon_u^{\pi}\chi_S^{\pi}\right| \geq \frac{t}{3} \right)
\label{eq:fourier_coefficient_intermediate_bound}
\end{align}
We bound each of the three terms above separately.  \\

\noindent \emph{Bounding Term 1.} We begin by showing that $\E\left[ \frac{1}{|\Pi^b_u|}\sum_{\pi \in \Pi_u^{b}}\E[Y_u^{(\pi)}]\chi_S^{\pi} \right] = \alpha_{u,S}$, where the expectation is taken over the randomness is sampling $\pi$. Since each $\pi \in \Pi^b_{u}$ is chosen uniformly and independently at random from $\Pi$, we have that 
\begin{align*}
    \E\left[ \frac{1}{|\Pi^b_u|}\sum_{\pi \in \Pi_u^{b}}\E[Y_u^{(\pi)}]\chi_S^{\pi} \right] = \frac{1}{2^p} \sum_{\pi \in \Pi} \E[Y_u^{(\pi)}]\chi_S^{\pi} = \langle \E[Y_u^{(\pi)}], \chi_S \rangle_B = \alpha_{u,S}
\end{align*}
Next, since each $\chi_{S}^{\pi} \in \{-1,1\}$, and $\E[Y_{u}^{(\pi)}] \in [-1,1]$, we have that  $\frac{1}{|\Pi^b_u|}\sum_{\pi \in \Pi_u^{b}}\E[Y_u^{(\pi)}]\chi_S^{\pi}$ is a sum of independent bounded random variables. Hence, applying Hoeffding's inequality gives us
\begin{equation*}
    \P\left( \left| \frac{1}{|\Pi^b_u|}\sum_{\pi \in \Pi_u^{b}}\E[Y_u^{(\pi)}]\chi_S^{\pi} - \alpha_{u,S} \right| \geq \frac{t}{3} \right) \leq 2\exp{\left(-\frac{|\Pi_u|t^2}{9} \right)}
\end{equation*} \\

\noindent \emph{Bounding Term 2.} By Assumption \ref{ass:bounded_noise}, and the mutual independence of $
\chi_S^{\pi}$ and $\epsilon_u^{\pi}$ for each $\pi$, we have that $\frac{1}{|\Pi^b_u|}\sum_{\pi \in \Pi_u^{b}}\epsilon_u^{\pi}\chi_S^{\pi}$ is a sum of independent bounded random variables. Applying Hoeffding's inequality once again gives us
\begin{equation*}
    \P\left(\left|\frac{1}{|\Pi^b_u|}\sum_{\pi \in \Pi_u^{b}}\epsilon_u^{\pi}\chi_S^{\pi} - \frac{1}{2^p}\sum_{\pi \in \Pi}\epsilon_u^{\pi}\chi_S^{\pi}\right| \geq \frac{t}{3} \right) \leq 2\exp{\left(-\frac{|\Pi_u|t^2}{9}\right)}
\end{equation*}
\vspace{2mm}

\noindent \emph{Bounding Term 3.} By Assumption \ref{ass:bounded_noise}, $\epsilon_u^{\pi}$ are independent mean-zero bounded random variables for each $\pi$. Hence, applying Hoeffding's inequality gives us
\begin{equation*}
    \P\left(\left| \frac{1}{2^p}\sum_{\pi \in \Pi}\epsilon_u^{\pi}\chi_S^{\pi}\right| \geq \frac{t}{3} \right) \leq 2\exp{\left(-\frac{2^{p+1}t^2}{9} \right)} \leq 2\exp{\left(-\frac{2|\Pi_u|t^2}{9} \right)}
\end{equation*}
where the last inequality follows from the fact that $|\Pi_u| \leq 2^p$

\noindent \emph{Collecting Terms.} Choosing $t = \sqrt{\frac{9\log(2^{k+3}/\delta)}{|\Pi_u|}}$, and plugging the bounds of Term 1,2 and 3 into \eqref{eq:fourier_coefficient_intermediate_bound} gives us
\begin{align*}
    \P\left(|\hat{\alpha}_{u,S} - \alpha_{u,S} |\geq \sqrt{\frac{8\log(2^{k+2}/\delta)}{|\Pi_u^b|}}\right) & \leq
    2\exp{\left(-\frac{|\Pi_u|t^2}{9}\right)} + 2\exp{\left(-\frac{2|\Pi_u|t^2}{9}\right)} + 2\exp{\left(-\frac{|\Pi_u|t^2}{9}\right)} \\
    & \leq 6\exp{\left(-\frac{|\Pi_u|t^2}{9} \right)} \\
    & \leq \frac{\delta}{2^k}
\end{align*}
Observing that the result above holds uniformly across uniformly for all subsets $S \subset [K]$ completes the proof.

\section{Proof of Theorem \ref{thm:experiment_design_assumptions_hold}}
\label{sec:experimental_design_proofs}


Let $\mathcal{E}_{1}, \ \mathcal{E}_{2},\  \mathcal{E}_{3},\  \mathcal{E}_{4},\  \mathcal{E}_{5}$ denote the event that horizontal span inclusion (Assumptions \ref{ass:donor_set_identification} (a)), incoherence of donor units (Assumption \ref{ass:incoherence}), linear span inclusion (\ref{ass:donor_set_identification}) (b), balanced spectrum (Assumption \ref{ass:balanced_spectrum}) and subspace inclusion (Assumption \ref{ass:rowspace_inclusion}) hold, respectively. Let $\mathcal{E}_{H} = \mathcal{E}_1 \cap \mathcal{E}_2$ and $\mathcal{E}_{V} = \mathcal{E}_3 \cap \mathcal{E}_4 \cap \mathcal{E}_5$ denote the events that the conditions required for horizontal and vertical regression are satisfied respectively. We proceed by showing that both events $\mathcal{E}_{H}$ and $\mathcal{E}_{V}$ hold with high probability. Throughout the proof of the Theorem, and all lemmas, we use $c,c',c'',c''',C,C'$ to refer to universal constants that might change from line to line. 
\vspace{2mm}

\noindent \emph{Step 1: $\mathcal{E}_H$ high probability bound.} To show $\mathcal{E}_H$ holds with high probability, we state the following two lemmas proved in Appendix \ref{subsec:experiment_design_lemmas_proof_horizontal_regression}. 

\begin{lemma} [Horizontal Span Inclusion]
\label{lem:experiment_design_horizontal_span_inclusion} 
Let the set-up of Theorem \ref{thm:experiment_design_assumptions_hold} hold. Then, horizontal span inclusion (Assumption \ref{ass:donor_set_identification}(a)) holds for every unit $u \in \mathcal{I}$ with probability at least $1 - \gamma/4$.
\end{lemma}

\begin{lemma} [Incoherence of Donor Units]
\label{lem:experiment_design_incoherence} Let the set-up of Theorem \ref{thm:experiment_design_assumptions_hold} hold. Then, $\bchi(\Pi_{\mathcal{I}})$ satisfies the incoherence condition: 
\begin{equation*}
    \lVert \frac{\bchi(\Pi_{\mathcal{I}})^T\bchi(\Pi_{\mathcal{I}})}{|\Pi_{\mathcal{I}}|} - \mathbf{I}_{2^p} \rVert_{\infty} \leq \frac{C}{s}
\end{equation*}
for a universal constant $C > 0$ with probability at least $1 - \gamma/4$. 
\end{lemma}

\noindent Applying Lemmas \ref{lem:experiment_design_horizontal_span_inclusion} and \ref{lem:experiment_design_incoherence} alongside Demorgan's law and the union bound gives us
\begin{equation}
\label{eq:experiment_design_horizontal_events_high_probability}
    \mathbb{P} \left(\mathcal{E}^c_H \right) =  \mathbb{P} \left(\mathcal{E}^c_1 \cup \mathcal{E}^c_2 \right) \leq \mathbb{P} \left(\mathcal{E}^c_1\right) +\mathbb{P} \left(\mathcal{E}^c_2\right) \leq \gamma/2
\end{equation}

\vspace{2mm}

\noindent \emph{Step 2: $\mathcal{E}_V$ high probability bound.} Next, we show that $\mathcal{E}_V$ holds with high probability. To that end, we state the following two lemmas proved in Appendix \ref{subsec:experiment_design_lemmas_proof_vertical_regression}. We define some necessary notation for the following lemma, let $\mathcal{A}_{\mathcal{I}} = [\balpha_u : u \in \mathcal{I}] \in \mathbb{R}^{|\mathcal{I}| \times 2^p}$. 

\begin{lemma} [Linear Span Inclusion]
\label{lem:experiment_design_linear_span_inclusion}
Let the set-up of Theorem \ref{thm:experiment_design_assumptions_hold} hold. Then, we have that
\begin{equation}
\label{eq:experimental_design_donor_set_balanced_spectrum}
\frac{C|\mathcal{I}|}{r} \geq s_1 \left(\mathcal{A}_{\mathcal{I}}^T \mathcal{A}_{\mathcal{I}} \right)  \geq s_r \left(\mathcal{A}_{\mathcal{I}}^T \mathcal{A}_{\mathcal{I}} \right) \geq \frac{C'|\mathcal{I}|}{r}
\end{equation}
for universal constants $C,C' > 0$ with probability at least $ 1 - \gamma/8$.
\end{lemma}
\noindent 

\noindent Since $\mathcal{A}_{\mathcal{I}}$ has rank $r$, we have that $\text{rank}(\mathcal{A}_{\mathcal{I}}) = \text{rank}(\mathcal{A})$. Further, since  $\mathcal{A}_{\mathcal{I}}$ is a sub-matrix of $\mathcal{A}$, linear span inclusion (i.e., $\mathcal{E}_3$) holds with probability at least $ 1 - \gamma/8$. Next, we show that balanced spectrum and subspace inclusion hold. 

\begin{lemma} [Balanced Spectrum]
\label{lem:experiment_design_balanced_spectrum}
Let the set-up of Theorem \ref{thm:experiment_design_assumptions_hold} and \eqref{eq:experimental_design_donor_set_balanced_spectrum} hold. Then, for universal constants $c,c' > 0$, we have 
$s_r(\E[\bY_{\mathcal{I}}^{(\Pi_N)}])/s_1(\E[\bY_{\mathcal{I}}^{(\Pi_N)}]) \geq c$ and $\lVert \E[\bY_{\mathcal{I}}^{(\Pi_N)}]\rVert^2_F \geq c'|\Pi_N||\mathcal{I}|$ with probability at least $1 - \gamma/8$. That is, balanced spectrum (Assumption \ref{ass:balanced_spectrum}) holds with high probability. 
\end{lemma}

\begin{lemma} [Subspace Inclusion] 
\label{lem:experiment_design_subspace_inclusion}
Let the set-up of Theorem \ref{thm:experiment_design_assumptions_hold} and \eqref{eq:experimental_design_donor_set_balanced_spectrum} hold. Then, $\E[\bY_{\mathcal{I}}^{(\pi)}]$ lies in the row-space of $\E[\bY_{\mathcal{I}}^{(\Pi_N)}]$ holds with probability at least $1 - \gamma/8$. That is, subspace inclusion (Assumption \ref{ass:rowspace_inclusion}) holds with high probability.  
\end{lemma}

\noindent Let $\mathcal{E}_6$ denote the event that \eqref{eq:experimental_design_donor_set_balanced_spectrum} holds. As a result of Lemma \ref{lem:experiment_design_subspace_inclusion}, we have that $\P(\mathcal{E}_4 ~ | ~ \mathcal{E}_6) \geq 1 - \gamma/8$ and  $\P(\mathcal{E}_5 ~ | ~ \mathcal{E}_6) \geq 1 - \gamma/8$. Additionally, note that $\mathcal{E}_3 \subset \mathcal{E}_6$. Hence, we have
\begin{align}
    \P(\mathcal{E}_V) & = \P(\mathcal{E}_3 \cap \mathcal{E}_4 \cap \mathcal{E}_5)  \geq \P(\mathcal{E}_4 \cap \mathcal{E}_5 \cap \mathcal{E}_6) \nonumber \\
    & = \P(\mathcal{E}_4 \cap \mathcal{E}_5  ~ | ~ \mathcal{E}_6) \P( \mathcal{E}_6) \nonumber \\
    & = \left(\P(\mathcal{E}_4   ~ | ~ \mathcal{E}_6) + \P(\mathcal{E}_5   ~ | ~ \mathcal{E}_6) - \P(\mathcal{E}_4 \cup \mathcal{E}_5  ~ | ~ \mathcal{E}_6)\right) \P( \mathcal{E}_6)  \nonumber \\
    & \geq \left(\P(\mathcal{E}_4   ~ | ~ \mathcal{E}_6) + \P(\mathcal{E}_5   ~ | ~ \mathcal{E}_6) \right) \P( \mathcal{E}_6)  - 1 \nonumber \\
    & =  \P(\mathcal{E}_4 ~ | ~ \mathcal{E}_6) \P( \mathcal{E}_6)  +  \P(\mathcal{E}_5 ~ | ~ \mathcal{E}_6) \P( \mathcal{E}_6)  - 1 \nonumber \\
    & \geq 2(1 - \gamma/8)^2 - 1\geq 1 - \gamma/2 \label{eq:experiment_design_vertical_regression_hold}
\end{align}


\noindent \emph{Step 3: Collecting Terms.} Let $\mathcal{E} = \mathcal{E}_H \cap \mathcal{E}_V$. To complete the proof, it suffices to show that $\P(\mathcal{E}) \geq 1 - \gamma$. Applying \eqref{eq:experiment_design_horizontal_events_high_probability} and \eqref{eq:experiment_design_vertical_regression_hold} gives us 
\begin{equation*}
    \P(\mathcal{E}^c) =  \P(\mathcal{E}_H^c \cup \mathcal{E}_V^c) \leq  \P(\mathcal{E}_H^c) + \P(\mathcal{E}_V^c) \leq 1 - \gamma
\end{equation*}
This completes the proof.

\section{Proofs of Helper Lemmas for  Theorem \ref{thm:experiment_design_assumptions_hold}}
\label{subsec:experimental_design_helper_lemmas}

Throughout these proofs, we use $c,c',c'',c''',C,C'$ to refer to positive universal constants that can change from line to line.

\subsection{Proof of Lemmas for Horizontal Regression}
\label{subsec:experiment_design_lemmas_proof_horizontal_regression}

In this section, we provide proofs that horizontal span inclusion and incoherence of the donor unit Fourier characteristics hold with high probability under our experimental design mechanism. That is, we prove Lemmas \ref{lem:experiment_design_horizontal_span_inclusion} and \ref{lem:experiment_design_incoherence} used in Theorem \ref{thm:experiment_design_assumptions_hold}.

\subsubsection{Proof of Lemma \ref{lem:experiment_design_horizontal_span_inclusion}}
We begin by defining some necessary notation. For a given  unit $u \in \mathcal{I}$, and let $\mathcal{S}_u = \{S \subset [p] ~ | ~ \alpha_{u,S} \neq 0 \}$. 
That is, $\mathcal{S}_u$ denotes the subset of coordinates where $\balpha_u \in \mathbb{R}^{2^p}$ is non-zero. 
Note that by Assumption \ref{ass:observation_model},  $|\mathcal{S}_u| \leq s$. 
For any combination $\pi$, let $\bchi_{\mathcal{S}_u}^{\pi} = [\chi^{\pi}_{S} : S \in \mathcal{S}_u] \in \{-1,1\}^{|\mathcal{S}_u|}$ denote the projection of $\bchi^\pi$ to the non-zero coordinates of $\balpha_u$. 
For example, if $\balpha_u = (1,1, \ldots ,0)$ and $\bchi^{\pi} = (1,1,\ldots ,1)$, then $\bchi_{\mathcal{S}_u}^{\pi} = (1,1)$. 
Finally, let $\bchi_{\mathcal{S}_u}(\Pi_{\mathcal{I}}) = [\bchi_{\mathcal{S}_u}^{\pi}: \pi \in \Pi_{\mathcal{I}}]$ $\in \{-1,1\}^{|\Pi_{\mathcal{I}}| \times |\mathcal{S}_u|}$.

\noindent To proceed, we  first show horizontal span inclusion holds for a given unit $u \in \mathcal{I}$, and then show that it extends to the entire donor set $\mathcal{I}$ via the union bound. To that end, fix a donor unit $u \in \mathcal{I}$ and then we proceed by quoting the following result required for our proof, which is based on the matrix Bernstein's inequality \cite[Theorem 1.4]{tropp2012user}.

\begin{theorem}[{\cite[Theorem 5.41]{vershynin_2012}}]
Let $A$ be an $N \times s$ matrix whose rows $A_i$ are independent isotropic random vectors in $\mathbb{R}^s$. Let $m$ be such that $||{A_i}||_2 \le \sqrt{m}$ almost surely for all $i \in [N]$. Then, for every $t \ge 0$, one has
\[\sqrt{N} - t\sqrt{m} \le s_{\text{min}}(A) \le s_{\text{max}}(A) \le \sqrt{N} + t\sqrt{m}\]
with probability at least $1 - 2s\cdot\exp(-ct^2)$.
\end{theorem}

\noindent To apply the theorem, we first show that $\bchi_{\mathcal{S}_u}^{\pi}$ 
is isotropic for any combination $\pi \in \Pi_{\mathcal{I}}$. That is, we show  $\E[\bchi_{\mathcal{S}_u}^{\pi}(\bchi_{\mathcal{S}_u}^{\pi})^T] = \mathbf{I}_{\mathcal{S}_u}$ where the expectation is taken over the randomness in choosing $\pi$ uniformly and independently at random from $\Pi$. The isotropy of $\bchi_{\mathcal{S}_u}^{\pi}$  follows since for any two subsets $S,S'\subset[p]$, we have 
\begin{align*}
\E[\chi^{\pi}_{S}\chi^{\pi}_{S'}] 
&= \frac{1}{2^p}\sum_{\pi \in \Pi} \chi^{\pi}_{S}\chi^{\pi}_{S'} \\
&= \left\langle \chi_{S}, \chi_{S'} \right\rangle_B \\
& = \mathbbm{1}\{S=S'\}.
\end{align*}
Next, we verify that $\|\bchi_{\mathcal{S}_u}^{\pi}\|_2 \le \sqrt{s}$ almost surely. This follows since $\bchi_{\mathcal{S}_u}^{\pi} \in \{1,-1\}^{|\mathcal{S}_u|}$, so
\[\left\|\bchi_{\mathcal{S}_u}^{\pi}\right\|_2 \le \sqrt{|\mathcal{S}_u|}\left\|\bchi_{\mathcal{S}_u}^{\pi}\right\|_\infty \le \sqrt{|\mathcal{S}_u|}.\]
Thus, applying the theorem with $N= |\Pi_{\mathcal{I}}|$ gives
\[\sqrt{|\Pi_{\mathcal{I}}|} - t\sqrt{|\mathcal{S}_u|} \le s_{\text{min}}(\bchi_{\mathcal{S}_u}(\Pi_{\mathcal{I}}))\]
with probability at least $1 - 2|\mathcal{S}_u|\cdot\exp(-ct^2)$.
Choosing $t = \sqrt{|\Pi_{\mathcal{I}}|/4|\mathcal{S}_u|}$ gives 
\begin{equation}
\label{eq:lower_bound_min_singular_value_fourier_characteristic}
    \sqrt{|\Pi_{\mathcal{I}}|/4} \le s_{\text{min}}(\bchi_{\mathcal{S}_u}(\Pi_{\mathcal{I}}))
\end{equation}
with probability at least $1 - 2|\mathcal{S}_u|\cdot\exp(-c|\Pi_{\mathcal{I}}|/4|\mathcal{S}_u|)$. Using our assumption that $|\Pi_{\mathcal{I}}| \geq \frac{Cr^2s^2\log(|\mathcal{I}|2^p/\gamma)}{\delta^2} \geq C|\mathcal{S}_u|\log(|\mathcal{S}_u||\mathcal{I}|/\gamma)$
establishes that $\bchi_{\mathcal{S}_u}(\Pi_{\mathcal{I}})$ has full-rank with probability at least $1 - \gamma/|\mathcal{I}|$. Since, $\bchi_{\mathcal{S}_u}(\Pi_{\mathcal{I}})$ has full-rank, we have that horizontal span inclusion holds for unit $u$. Taking a union bound over all donor units $u \in \mathcal{I}$ completes the proof. 

\subsubsection{Proof of Lemma \ref{lem:experiment_design_incoherence}}
\label{ref:subsubsec:experiment_design_incoherence_proof}
We begin the proof by defining some notation. For a subset $S \subset [p]$, let $\bchi^{\Pi_I}_{S} = [\chi^{\pi}_{S} : \pi \in \Pi_{\mathcal{I}} ] \in \{-1,1 \}^{|\Pi_{\mathcal{I}}|}$. For any two distinct subsets $S,S' \in [p]$, define $z_i = \chi^{\pi_i}_{S}\chi^{\pi_i}_{S'}/|\Pi_{\mathcal{I}}| \in \{-\frac{1}{|\Pi_{\mathcal{I}}|},\frac{1}{|\Pi_{\mathcal{I}}|} \}$ for $1 \leq i \leq |\Pi_\mathcal{I}|$. Let $S \ \Delta \ S'$ denote the symmetric difference of  two subsets $S,S'$.  Then, 
\begin{align*}
    \E[z_i] = \E\left[\frac{1}{|\Pi_\mathcal{I}|} \chi^{\pi_i}_{S}\chi^{\pi_i}_{S'}\right]  & =  \E\left[\frac{1}{|\Pi_\mathcal{I}|} \left(\prod_{j \in S} v(\pi_i)_j \right) \left(\prod_{j' \in S'} v(\pi_i)_{j'} \right)\right] \\
    & = \E\left[\frac{1}{|\Pi_\mathcal{I}|} \left(\prod_{j \in S \ \Delta \ S'} v(\pi_i)_j \right)\right]
\end{align*}
where the expectation is taken over choosing $\pi_i$ uniformly and independently at random. Since each $\pi \in \Pi_{\mathcal{I}}$ is chosen in this fashion, we have that the random variables $v(\pi)_j$ are independent and identically distributed being over over $\{-1,1\}$. Plugging this into the equation above, we have that  
\begin{align*}
    \E[z_i] & = \E\left[\frac{1}{|\Pi_\mathcal{I}|} \left(\prod_{j \in S \ \Delta \ S'} v(\pi_i)_j \right)\right] \\
    & = \frac{1}{|\Pi_\mathcal{I}|} \prod_{j \in S  \Delta  S'} \E[v(\pi_i)_j] = 0
\end{align*}

\noindent Next, observe that since each $\pi \in \Pi_{\mathcal{I}}$ is chosen independently and uniformly and random, we have that $z_1 \ldots z_{|\Pi_\mathcal{I}|}$ are independent and identically distributed bounded random variables. Hence, we can apply Hoeffding's inequality for any $ t > 0$ to give us, 
\begin{equation}
\label{eq:high_probabilty_bound_inner_product_fourier_characteristic}
    \P\left(\sum^{|\Pi_{\mathcal{I}}|}_{i = 1} z_i \geq t \right) \leq 2\exp(-\frac{|\Pi_I\mathcal{}|t^2}{2})
\end{equation}

\noindent Then, for any distinct subsets $S,S' \subset [p]$, we have
\begin{align*}
    \frac{1}{|\Pi_\mathcal{I}|} \langle \bchi^{\Pi_I}_{S}, \bchi^{\Pi_I}_{S'} \rangle = \sum_{\pi_i \in \Pi_{\mathcal{I}}} \chi^{\pi_i}_{S}, \chi^{\pi_i}_{S'} = \sum^{|\Pi_\mathcal{I}|}_{i = 1} z_i
\end{align*}
Therefore, we have that 
\begin{align*}
    \P\left(|\max_{S \neq S'} \frac{1}{|\Pi_\mathcal{I}|}  \langle \bchi^{\Pi_I}_{S}, \bchi^{\Pi_I}_{S'} \rangle | \geq t \right) & \leq \sum_{S \neq S'} \P\left( \frac{1}{|\Pi_\mathcal{I}|}  | \langle \bchi^{\Pi_I}_{S}, \bchi^{\Pi_I}_{S'} \rangle | \geq t \right) \\
     & \leq 2^{2p+1}\exp(-\frac{|\Pi_I\mathcal{}|t^2}{2})
\end{align*}
where we use \eqref{eq:high_probabilty_bound_inner_product_fourier_characteristic} for the last inequality. Choosing $t = C'/s$, and using our assumption that $|\Pi_{\mathcal{I}}| \geq \frac{Cr^2s^2\log(|\mathcal{I}|2^p/\gamma)}{\delta^2} \geq Cs^2\log(2^p/\gamma)$ gives us 
\begin{equation}
\label{eq:high_probabilty_bound_max_inner_product_fourier_characteristic}
     \P\left( \max_{S \neq S'}  |\frac{1}{|\Pi_\mathcal{I}|}  \langle \bchi^{\Pi_I}_{S}, \bchi^{\Pi_I}_{S'} \rangle | \geq \frac{C'}{s} \right) \leq \gamma 
\end{equation}
for an appropriately chosen constant $C > 0$. To proceed, observe that $\langle \bchi^{\Pi_I}_{S}, \bchi^{\Pi_I}_{S} \rangle/|\Pi_{\mathcal{I}}| = 1$. Hence, we have that 
\begin{equation*}
    \lVert \frac{\bchi(\Pi_\mathcal{I})\bchi(\Pi_\mathcal{I})^T}{|\Pi_{\mathcal{I}}|} - \mathbf{I}_{2^p} \rVert_{\infty} = \max_{S \neq S'}  |\frac{1}{|\Pi_\mathcal{I}|}  \langle \bchi^{\Pi_I}_{S}, \bchi^{\Pi_I}_{S'}  \rangle |
\end{equation*}
Substituting this into \eqref{eq:high_probabilty_bound_max_inner_product_fourier_characteristic} completes the proof.


\subsection{Proofs of Lemmas for Vertical Regression}
\label{subsec:experiment_design_lemmas_proof_vertical_regression}

In this section, we provide proofs that vertical span inclusion and subspace inclusion holds with high probability under our experimental design mechanism. That is, we prove Lemmas \ref{lem:experiment_design_linear_span_inclusion}, \ref{lem:experiment_design_balanced_spectrum}, and \ref{lem:experiment_design_subspace_inclusion} used in Theorem \ref{thm:experiment_design_assumptions_hold}.  

\subsubsection{Proof of Lemma \ref{lem:experiment_design_linear_span_inclusion}}
 Let $\mathcal{A}_{\mathcal{I}} = [\balpha_u: u \in \mathcal{I}] \in \mathbb{R}^{|\mathcal{I}| \times 2^p }$ denote the matrix of Fourier coefficients of the uniformly sampled donor set. By Assumption \ref{ass:observation_model} (a), it suffices to show that $\mathcal{A}_{\mathcal{I}}$ has rank $r$ with high probability. 
 To proceed, note that $\bchi(\Pi)/2^{p/2}$ is an orthogonal matrix since $(\bchi(\Pi)/2^{p/2})^T \bchi(\Pi)/2^{p/2} = \mathbf{I}_{2^p}$. Further, observe that $\E[\bY_{N}^{(\Pi)}] = \bchi(\Pi)\mathcal{A}^T$. Then, since multiplication by an orthogonal matrix preserves singular values, we have
\begin{equation}
\label{eq:fourier_coefficient_outcome_matrix_singular_value_equality}
   2^p s_i (\mathcal{A}^T\mathcal{A})  = 2^p s_i \left(\left(\frac{\bchi(\Pi)}{2^{p/2} }\right)\mathcal{A}^T\mathcal{A} \left(\frac{\bchi(\Pi)}{2^{p/2}}\right)^T\right)  = s_i(\E[\bY_{N}^{(\Pi)}]\E[\bY_{N}^{(\Pi)}]^T) 
\end{equation}
Then, by Assumption \ref{ass:restricted_balanced_spectrum} and the equality above, we have 
\begin{equation}
\label{eq:fourier_matrix_singular_value_bound}
    \frac{CN}{r} \ge s_1(\mathcal{A}^T\mathcal{A}) \ge s_r(\mathcal{A}^T\mathcal{A})  \ge \frac{cN}{r}.
\end{equation}

\noindent Now, let $\mathcal{\Tilde{A}} = [\Tilde{\balpha}_n : n \in [N]] \in \mathbb{R}^{|N| \times p'}$ be formed by removing all the zero columns of $\mathcal{A}$. Analogously, define $\mathcal{\Tilde A}_{\mathcal{I}} =  [\Tilde{\balpha}_u : u \in \mathcal{I}] \in \mathbb{R}^{|\mathcal{I}| \times p'}$ by removing all the zero columns of $\mathcal{A}_{\mathcal{I}}$. Then clearly, for every $i \in [r]$, we have that $s_{i}(\mathcal{\tilde{A}}^T\mathcal{\tilde{A}}) = s_i (\mathcal{A}^T\mathcal{A}) $.  Moreover, note that
\begin{equation*}
    \E[\tilde\balpha_u\tilde\balpha_u^\top] = \frac{1}{N}\sum_{n=1}^N\tilde\balpha_n\tilde\balpha_n^\top = \mathcal{\tilde{A}}^T\mathcal{\tilde{A}}/N.
\end{equation*}

\noindent where the expectation is taken over the randomness in uniformly and independently sampling donor units $u$ from $[N]$. That is, the expectation is taken over the randomness in choosing $\balpha_u$ uniformly and random from $\mathcal{A}$. To bound the number of non-zero columns $p'$, we state the following Lemma proved in Appendix \ref{subsubsec:proof_lemma_nonzeros}. 

\begin{lemma}
\label{lem:number_nonzeros}
The number of nonzero columns of $\mathcal{A}$, which we have denoted $p'$, satisfies $p' \le rs$.
\end{lemma}

\noindent Further, since multiplication by an orthogonal matrix preserves the $2$-norm of a vector, for every unit $n \in [N]$, we have
\begin{equation}
\label{eq:2_norm_fourier_coefficient}
    ||\balpha_n||_2  =  2^{-p/2}||\bchi(\Pi)\balpha_n||_2 =  2^{-p/2}||{\E[\bY_n^{(\Pi)}]}||_2 \leq 1 
\end{equation}
where we used the assumption $\E[Y_n^{(\pi)}] \in [-1,1]$ for every unit-combination pair. To proceed, we state the following result. 

\vspace{2mm}

\begin{theorem} [Theorem 5.44 in \cite{vershynin_2012}]
\label{thm:matrix_bernstein}
Let $X$ be an $q \times p'$ matrix whose rows $X_i$ are independent random vectors in $\mathbb{R}^n$ with the common second moment matrix $\Sigma = \mathbb{E}[X_iX_i^{\top}]$. Let $m$ be a number such that $\|X_i\|_2 \le \sqrt{m}$ almost surely for all $i$. Then for every $t\ge 0$, the following inequality holds with probability at least $1-p' \cdot \exp(-ct^2)$:
\[\left\|\frac{1}{q}X^{\top}X - \Sigma \right\| \le t\sqrt{\frac{m||\Sigma||}{q}} \vee \frac{t^2 m}{q}\]
Here $c > 0$ is an absolute constant.
\end{theorem}

\noindent Then, we can apply Theorem \ref{thm:matrix_bernstein} with $X = \mathcal{\tilde A_{\mathcal{I}}}$, $q = |\mathcal{I}|$, $m = 1$ (obtained from \eqref{eq:2_norm_fourier_coefficient}), $\Sigma = \E[\tilde\balpha_u\tilde\balpha_u^\top]$, $||\Sigma||= C/r$ (obtained from \eqref{eq:fourier_matrix_singular_value_bound}), and $t = \sqrt{\log(rs/\gamma)/c'}$ to give us that 
\begin{equation}
\label{eq:matrix_bernstein_fourier_matrix_bound}
    \lVert \frac{1}{|\mathcal{I}|}\mathcal{\tilde A_{\mathcal{I}}^T}\mathcal{\tilde A_{\mathcal{I}}} - \Sigma \rVert \leq \sqrt{\frac{\log(8rs/\gamma)}{c'r|\mathcal{I}|}} \vee \frac{\log(8rs/\gamma)}{c''|\mathcal{I}|}
\end{equation}
holds with probability at least $1-\gamma/8$. Hence, we have that 
\begin{align*}
    s_1(\mathcal{\tilde A_{\mathcal{I}}^T}\mathcal{\tilde A_{\mathcal{I}}}/|\mathcal{I}|) & \leq \frac{C}{r} +  \sqrt{\frac{\log(8rs/\gamma)}{c'r|\mathcal{I}|}} \vee \frac{\log(8rs/\gamma)}{c''|\mathcal{I}|} \\
\end{align*}
Choosing $|\mathcal{I}| \geq Cr\log(8rs/\gamma)$ for an appropriate constant $C > 0$ gives us that 
\begin{equation}
\label{eq:experimental_design_donor_set_largest_singular_value_upper_bound}
     s_1(\mathcal{\tilde A_{\mathcal{I}}^T}\mathcal{\tilde A_{\mathcal{I}}}/|\mathcal{I}|) \leq \frac{C}{r}
\end{equation}
with probability at least $1 - \gamma/8$. To bound the smallest singular value, we use Weyl's inequality (Theorem \ref{thm:weyl_inequality}), \eqref{eq:fourier_matrix_singular_value_bound}, and \eqref{eq:matrix_bernstein_fourier_matrix_bound}, to give us 
\begin{align*}
    s_r(\mathcal{\tilde A_{\mathcal{I}}^T}\mathcal{\tilde A_{\mathcal{I}}}/|\mathcal{I}|) & \geq  s_r(\mathcal{\tilde A^T}\mathcal{\tilde A}/N) - \lVert \frac{1}{|\mathcal{I}|}\mathcal{\tilde A_{\mathcal{I}}^T}\mathcal{\tilde A_{\mathcal{I}}} - \Sigma \rVert \\
    & \geq \frac{c}{r} - \sqrt{\frac{\log(8rs/\gamma)}{c'r|\mathcal{I}|}} \vee \frac{\log(8rs/\gamma)}{c''|\mathcal{I}|}. 
\end{align*}
Substituting $|\mathcal{I}| \geq Cr\log(8rs/\gamma)$ for an appropriate constant $C > 0$ gives us that 
\begin{equation}
\label{eq:experimental_design_donor_set_singular_value}
    s_r(\mathcal{\tilde A_{\mathcal{I}}^T}\mathcal{\tilde A_{\mathcal{I}}}/|\mathcal{I}|) \geq \frac{c'''}{r}
\end{equation}
for some universal constant $c''' > 0$ with probability at least $1 - \gamma/8$. Next, observe that $ s_i(\mathcal{\tilde A_{\mathcal{I}}^T}\mathcal{\tilde A_{\mathcal{I}}}/|\mathcal{I}|) =  s_i(\mathcal{ A_{\mathcal{I}}^T}\mathcal{ A_{\mathcal{I}}}/|\mathcal{I}|)$ for any $i \in [r]$. Combining \eqref{eq:experimental_design_donor_set_largest_singular_value_upper_bound} and \eqref{eq:experimental_design_donor_set_singular_value} completes the proof.

\subsection{Proof of Lemma \ref{lem:experiment_design_balanced_spectrum}}
\label{subsec:proof_experiment_design_balanced_spectrum}

We proceed by quoting the following result proved in \ref{subsubsec:experiment_design_pcr_balanced_spectrum_proof}.  
\begin{lemma}
\label{lem:experiment_design_pcr_balanced_spectrum}
Let the set-up of Theorem \ref{thm:experiment_design_assumptions_hold}, and \eqref{eq:experimental_design_donor_set_balanced_spectrum} hold. Then, 
\begin{equation}
\label{eq:experiment_design_pcr_balanced_spectrum}
\frac{C|\mathcal{I}|}{r} \geq s_{1}(\E[\bY_{\mathcal{I}}^{(\Pi_N)}]^T\E[\bY_{\mathcal{I}}^{(\Pi_N)}]/|\Pi_N|) \geq  s_{r}(\E[\bY_{\mathcal{I}}^{(\Pi_N)}]^T\E[\bY_{\mathcal{I}}^{(\Pi_N)}]/|\Pi_N|) \geq \frac{C'|\mathcal{I}|}{r} 
\end{equation}
holds with probability at least $1 - \gamma/8$. 
\end{lemma}

Conditional on \eqref{eq:experiment_design_pcr_balanced_spectrum}, we have that $s_{r}(\E[\bY_{\mathcal{I}}^{(\Pi_N)}]^T\E[\bY_{\mathcal{I}}^{(\Pi_N)}])/s_{1}(\E[\bY_{\mathcal{I}}^{(\Pi_N)}]^T\E[\bY_{\mathcal{I}}^{(\Pi_N)}])\geq c$ for some universal constant $c > 0$. Further, conditional on \eqref{eq:experiment_design_pcr_balanced_spectrum}, we have that $\lVert \E[\bY_{\mathcal{I}}^{(\Pi_N)} ~ | ~ \mathcal{A}_{\mathcal{I}}] \rVert^2_F \geq r s_r(\E[\bY_{\mathcal{I}}^{(\Pi_N)}]^T\E[\bY_{\mathcal{I}}^{(\Pi_N)}]) \geq c' |\Pi_N| |\mathcal{I} |$. Hence, balanced spectrum (Assumption \ref{ass:balanced_spectrum}) holds with probability at least $1 - \gamma/8$, which completes the proof. 


\subsection{Proof of Lemma \ref{lem:experiment_design_subspace_inclusion}}
It suffices to show that $\E[\bY_{\mathcal{I}}^{(\Pi_N)}] = [\E[\bY_{\mathcal{I}}^{(\pi)}]: \pi \in \Pi_N] \in \mathbb{R}^{|\Pi_N| \times |\mathcal{I}|}$ has rank $r$ with high probability. This immediately follows from \eqref{eq:experiment_design_pcr_balanced_spectrum}, hence completing the proof.

\subsubsection{Proof of Lemma \ref{lem:number_nonzeros}}
\label{subsubsec:proof_lemma_nonzeros}
Let $\mathcal{A}_t$ denote the sub-matrix of $\mathcal{A}$ formed by taking the first $t$ rows. Further, let $p'_t$ denote the number of nonzero columns of $\mathcal{A}_t$. We proceed by induction, that is we show for all $t$, 
\begin{equation*}
    p'_t - s \cdot \mathrm{rank}(\mathcal{A}_t) \le 0
\end{equation*}

In the case of $t=1$, either the first row is the zero vector, in which case $p'_1 = \mathrm{rank}(\mathcal{A}_1) = 0$, or else $\mathrm{rank}(\mathcal{A}_1) = 1$ and $p'_1 \leq s$ since the first row is at most $s$-sparse. Then, for general $t$, note that 
\[\mathrm{rank}(\mathcal{A}_{t-1}) \le  \mathrm{rank}(\mathcal{A}_t) \le \mathrm{rank}(\mathcal{A}_{t-1}) + 1.\] If $\mathrm{rank}(\mathcal{A}_{t}) =  \mathrm{rank}(\mathcal{A}_{t-1})$ then we must also have $p'_{t-1} = p'_t$ and the inductive hypothesis holds. 
Otherwise, $\mathrm{rank}(\mathcal{A}_{t}) =  \mathrm{rank}(\mathcal{A}_{t-1}) + 1$. Note that the $t^\text{th}$ row of $\mathcal{A}$ has only $s$ nonzero entries, so $p'_t \leq p'_{t-1} + s$. In this case, we have that 
\begin{align*}
    p'_{t} -  s \cdot \mathrm{rank}(\mathcal{A}_t) & =  p'_{t} -  s \cdot ( \mathrm{rank}(\mathcal{A}_{t-1}) + 1) \\
    & \leq (p'_{t-1} + s) - s \cdot ( \mathrm{rank}(\mathcal{A}_{t-1}) + 1) \leq 0
\end{align*}
where the last inequality holds due to the inductive hypothesis. Since $\mathrm{rank}(\mathcal{A}) = r$, we have that $p' \leq rs$.  
This completes the proof. 

\subsubsection{Proof of Lemma \ref{lem:experiment_design_pcr_balanced_spectrum}}
\label{subsubsec:experiment_design_pcr_balanced_spectrum_proof}

 We begin by establishing that $\E[\bY_{\mathcal{I}}^{(\Pi)}]$ has a balanced spectrum. To proceed, note the since $\bchi(\Pi)/2^{p/2}$ is an orthogonal matrix,  $ \E[\bY_{\mathcal{I}}^{(\Pi)}] = \bchi(\Pi)\mathcal{A}_{\mathcal{I}}^T$, and that multiplication by an orthogonal matrix preserves singular values, we have that
\begin{equation*}
     2^p s_i (\mathcal{A}_{\mathcal{I}}^T\mathcal{A}_{\mathcal{I}})  =  s_i \left(\left(\frac{\bchi(\Pi)}{2^{p/2} }\right)\mathcal{A}_{\mathcal{I}}^T\mathcal{A}_{\mathcal{I}}\left(\frac{\bchi(\Pi)}{2^{p/2}}\right)^T\right) =  s_i(\E[\bY_{\mathcal{I}}^{(\Pi)}]\E[\bY_{\mathcal{I}}^{(\Pi)}]^T)  
\end{equation*}
Using the equation above and by assuming \eqref{eq:experimental_design_donor_set_balanced_spectrum} in the set-up of the Lemma, we have 
\begin{equation}
\label{eq:rth_singular_value_donor_set}
    s_r(\E[\bY_{\mathcal{I}}^{(\Pi)}]^T\E[\bY_{\mathcal{I}}^{(\Pi)}]) = 2^p  s_r (\mathcal{A}_{\mathcal{I}}^T\mathcal{A}_{\mathcal{I}}) \geq \frac{c|\mathcal{I}|2^p}{r}
\end{equation}
for a universal constant $C > 0$. Next, using \eqref{eq:experimental_design_donor_set_balanced_spectrum} again, we have that
\begin{equation}
\label{eq:top_singular_value_donor_set}
 s_1(\E[\bY_{\mathcal{I}}^{(\Pi)}]^T\E[\bY_{\mathcal{I}}^{(\Pi)}]) = 2^p s_1 (\mathcal{A}_{\mathcal{I}}^T\mathcal{A}_{\mathcal{I}})  \leq \frac{C|\mathcal{I
 }|2^p}{r}
\end{equation}
Putting both \eqref{eq:rth_singular_value_donor_set} and \eqref{eq:top_singular_value_donor_set} together, we get
\begin{equation}
\label{eq:subspace_inclusion_singular_value_bound}
    \frac{C|\mathcal{I}|2^p}{r} \ge s_1(\E[\bY_{\mathcal{I}}^{(\Pi)}]^T\E[\bY_{\mathcal{I}}^{(\Pi)}]) \ge s_r(\E[\bY_{\mathcal{I}}^{(\Pi)}]^T\E[\bY_{\mathcal{I}}^{(\Pi)}])  \ge  \frac{c|\mathcal{I}|2^p}{r}
\end{equation}
which establishes that $\E[\bY_{\mathcal{I}}^{(\Pi)}]$ has a balanced spectrum. 
Further, observe that for $\pi \in \Pi$ chosen uniformly at random, we have
\begin{equation*}
    \E\left[\E[\bY_{\mathcal{I}}^{(\pi)}]^T\E[\bY_{\mathcal{I}}^{(\pi)}]\right] = \frac{1}{2^p}\sum_{\pi \in \Pi} \E[\bY_{\mathcal{I}}^{(\pi)}]^T\E[\bY_{\mathcal{I}}^{(\pi)}] = \E[\bY_{\mathcal{I}}^{(\Pi)}]^T\E[\bY_{\mathcal{I}}^{(\Pi)}]/2^p
\end{equation*}
where the outer expectation is taken with respect to the randomness in choosing $\pi$ uniformly and independently at random from $\Pi$. Then, we can apply Theorem \ref{thm:matrix_bernstein} with $X = \E[\bY_{\mathcal{I}}^{(\Pi_N)}]$, $m = \lVert \E[\bY_{\mathcal{I}}^{\pi}] \rVert \leq \sqrt{|\mathcal{I}|}$ for any $\pi \in \Pi$, $\Sigma = \E[\bY_{\mathcal{I}}^{(\Pi)}]^T\E[\bY_{\mathcal{I}}^{(\Pi)}]/2^p$, $||\Sigma|| = C|\mathcal{I}|/r$ (obtained from \eqref{eq:top_singular_value_donor_set}), $t = \sqrt{\log(8|\mathcal{I}|/\gamma)/c'}$ to give us that 
\begin{equation}
\label{eq:matrix_bernstein_subspace_inclusion_bound}
     \left \lVert \frac{1}{|\Pi_N|}\E[\bY_{\mathcal{I}}^{(\Pi_N)}]^T\E[\bY_{\mathcal{I}}^{(\Pi_N)}] - \Sigma \right \rVert \leq \sqrt{\frac{|\mathcal{I}|^2\log(8|\mathcal{I}|/\gamma)}{c'r|\Pi_N|}} \vee \frac{|\mathcal{I}|\log(8|\mathcal{I}|/\gamma)}{c''|\Pi_N|}
\end{equation}
holds with probability at least $1 - \gamma/8$.  Hence, we have that the following holds with probability at least $1 - \gamma/8$. 
\begin{equation*}    s_1(\E[\bY_{\mathcal{I}}^{(\Pi_N)}]^T\E[\bY_{\mathcal{I}}^{(\Pi_N)}]/|\Pi_N|) \leq \frac{C|\mathcal{I}|}{r} + \sqrt{\frac{|\mathcal{I}|^2\log(8|\mathcal{I}|/\gamma)}{c'r|\Pi_N|}} \vee \frac{|\mathcal{I}|\log(8|\mathcal{I}|/\gamma)}{c''|\Pi_N|}
\end{equation*}
Choosing $|\Pi_N|\geq Cr\log(|\mathcal{I}|/\gamma)$ for appropriate constant $C > 0$, gives us 
\begin{equation}
\label{eq:experiment_design_top_singular_value_balanced_spectrum_bound}
s_1(\E[\bY_{\mathcal{I}}^{(\Pi_N)}]^T\E[\bY_{\mathcal{I}}^{(\Pi_N)}]/|\Pi_N|) \leq \frac{C|\mathcal{I}|}{r}
\end{equation}

with probability at least $1 - \gamma/8$. Next, we lower bound $s_r$ as follows. Using Theorem \ref{thm:weyl_inequality}, \eqref{eq:subspace_inclusion_singular_value_bound}, and \eqref{eq:matrix_bernstein_subspace_inclusion_bound}, we have 
\begin{align*}
s_{r}(\E[\bY_{\mathcal{I}}^{(\Pi_N)}]^T\E[\bY_{\mathcal{I}}^{(\Pi_N)}]/|\Pi_N|) & \geq s_{r}(\E[\bY_{\mathcal{I}}^{(\Pi)}]^T\E[\bY_{\mathcal{I}}^{(\Pi)}]/2^p|) -   \lVert \frac{1}{|\Pi_N|}\E[\bY_{\mathcal{I}}^{(\Pi_N)}]^T\E[\bY_{\mathcal{I}}^{(\Pi_N)}] - \Sigma \rVert \\
& \geq \frac{c|\mathcal{I}|}{r} - \sqrt{\frac{|\mathcal{I}|^2\log(8|\mathcal{I}|/\gamma)}{c'r|\Pi_N|}} \vee \frac{|\mathcal{I}|\log(8|\mathcal{I}|/\gamma)}{c''|\Pi_N|} \\
\end{align*}
Using $|\Pi_N|\geq Cr\log(|\mathcal{I}|/\gamma)$ for an appropriate constant $C>0$ implies
\begin{equation}
\label{eq:experiment_design_min_singular_value_balanced_spectrum_bound}
s_{r}(\E[\bY_{\mathcal{I}}^{(\Pi_N)}]^T\E[\bY_{\mathcal{I}}^{(\Pi_N)}]/|\Pi_N|) \geq \frac{C'|\mathcal{I}|}{r} 
\end{equation}
with probability at least $ 1- \gamma/8$. Combining \eqref{eq:experiment_design_top_singular_value_balanced_spectrum_bound}, and \eqref{eq:experiment_design_min_singular_value_balanced_spectrum_bound} completes the proof.

\section{Formal Results for Motivating Examples in Section \ref{sec:identification}}
\label{sec:example_proofs}

In this section, we provide the proofs for both of the motivating examples described in Section \ref{sec:identification}. 

\subsection{Proof of Proposition \ref{prop:motivating_example}}
\label{subsec:proof_experimental_motivating_example}

\begin{proof} Let $\mathcal{E}_{H}$, $\mathcal{E}_{L}$ denote the events that horizontal span and linear inclusion holds. We proceed by showing that both $\mathcal{E}_{H}$, $\mathcal{E}_{L}$ hold with high probability. 
\vspace{2mm}

\noindent \emph{Step 1: $\mathcal{E}_H$ high probability bound.} Recall the following notation defined in Section \ref{subsec:experiment_design_lemmas_proof_horizontal_regression}. For a given  unit $u \in \mathcal{I}$, and let $\mathcal{S}_u = \{S \subset [p] ~ | ~ \alpha_{u,S} \neq 0 \}$. That is, $\mathcal{S}_u$ denotes the subset of coordinates where $\balpha_u \in \mathbb{R}^{2^p}$ is non-zero. Note that by Assumption \ref{ass:observation_model},  $|S_u| \leq s$. For any combination $\pi$, let $\bchi_{\mathcal{S}_u}^{\pi} = [\chi^{\pi}_{S} : S \in \mathcal{S}_u] \in \{-1,1\}^{|\mathcal{S}_u|}$ denote the projection of $\bchi^\pi$ to the non-zero coordinates of $\balpha_u$. Let $\bchi_{\mathcal{S}_u}(\Pi_{\mathcal{I}}) = [\bchi_{\mathcal{S}_u}^{\pi}: \pi \in \Pi_{\mathcal{I}}] \in \{-1,1\}^{|\Pi_{\mathcal{I}}| \times |\mathcal{S}_u|}$. Additionally, define $\Tilde{\bchi}_u(\Pi_{\mathcal{I}}) = [\Tilde{\bchi}^{\pi}_u : \pi \in \Pi_{\mathcal{I}}] \in \{-1,1\}^{2^{p}}$ where $\Tilde{\bchi}_u$ is defined in Section \ref{sec:identification}. 
\vspace{2mm}

\noindent Notice that our sampling scheme for assigning combinations to donor units is precisely the same as that described in our experimental design mechanism in Section \ref{sec:experimental_design}.  Using this observation and Equation \eqref{eq:lower_bound_min_singular_value_fourier_characteristic} from the proof of Lemma \ref{lem:experiment_design_horizontal_span_inclusion} gives us that $s_{\text{min}}(\bchi_{\mathcal{S}_u}(\Pi_{\mathcal{I}})) > 0$ with probability at least $1 - 2|\mathcal{S}_u|\cdot\exp(-c|\Pi_{\mathcal{I}}|/4|\mathcal{S}_u|)$ for a given unit $u$. Plugging in our assumption that $|\Pi_{\mathcal{I}}| \geq Cs\log(s|\Pi_{\mathcal{I}}|/\gamma)$ for an appropriate universal constant ensures that 
\begin{equation*}
    \P\left(s_{\text{min}}(\bchi_{\mathcal{S}_u}(\Pi_{\mathcal{I}})) > 0\right) \geq 1 - \gamma/2|\mathcal{I}|
\end{equation*}
\noindent Next, observe that $s_{\text{min}}(\bchi_{\mathcal{S}_u}(\Pi_{\mathcal{I}})) = s_{\text{min}}(\Tilde{\bchi}_{u}(\Pi_{\mathcal{I}}))$. Using this fact shows that horizontal linear span inclusion holds for unit $u$ with probability at least $1 - \gamma/2|\mathcal{I}|$. Taking a union bound over all donor units shows that $\P(\mathcal{E}_H) \geq 1 - \gamma/2$. 
\vspace{2mm}

\noindent \emph{Step 2: $\mathcal{E}_L$ high probability bound.} We begin by defining some notation. Let $N_j$ denote the number of times a unit of type $j$ appears. More formally, we have
\begin{equation*}
    N_j = \sum_{u \in \mathcal{I}} \ \mathbbm{1}[\balpha_{u} = \balpha(j)]
\end{equation*}
Additionally, let $X \sim \text{Bin}(n,p)$ denote a draw from a binomial distribution with parameters $n$ and $p$, where $n$ is the nymber of trials and $p \in [0,1]$ is the success probability of each trial.  Note that $N_j$  is distributed as a Binomial random variable with parameters $|\mathcal{I}|,p_j$ where $p_j$ is the probability of choosing a unit of type $j$. Since each donor unit is drawn independently and uniformly at random, and we assume that there are at least $cN/r$ units of each type, we have that $p_j \geq c/r$. 
\vspace{2mm}

\noindent To show linear span inclusion holds, it suffices to show that $\min_{j \in [r]} N_j > 0$. Note that for a fixed $j \in [r]$, we have
\begin{equation}
\label{eq:motivating_example_zero_type_probability}
    \P(N_j = 0) = (1 - p_j)^{|\mathcal{I}|} \leq (1 - c/r)^{|\mathcal{I}|} \leq \exp(-c|\mathcal{I}|/r)
\end{equation}
where we use the inequality $(1 - x)^a \leq \exp(-ax)$. Next, by Demorgan's law, the union bound and \eqref{eq:motivating_example_zero_type_probability}, we have
\begin{align*}
    \P((\min_{j \in [r]} N_j > 0)^c) & =  \P((\cap^r_{j=1}N_j > 0)^c)  \\
    & =  \P(\cup ^r_{j=1}N_j = 0) \\
    & \leq \sum^r_{j=1} \P(N_j = 0) \\
    & \leq r\exp(-c|\mathcal{I}|/r)
\end{align*}
Plugging in our assumption $|\mathcal{I}| \geq cr\log(r/\gamma)$ into the equation above gives us that 
\begin{equation*}
     \P((\min_{j \in [r]} N_j > 0)^c)  \leq \gamma/2
\end{equation*}
for an appropriately chosen constant $c > 0$. Hence, we have that $\P(\mathcal{E}_L) \geq 1 - \gamma/2$. 
\vspace{2mm}

\noindent \emph{Step 3: Collecting Terms.} Let $\mathcal{E} = \mathcal{E}_H \cap \mathcal{E}_L$. To complete the proof, it suffices to show that $\P(\mathcal{E}) \geq 1 - \gamma$. From steps 1 and 2, we have that $\P(\mathcal{E}_H)$ and $\P(\mathcal{E}_L)$ occur with probability at least $ 1 - \gamma/2$. Hence, we have that
\begin{equation*}
    \P(\mathcal{E}^c) =  \P(\mathcal{E}_H^c \cup \mathcal{E}_L^c) \leq  \P(\mathcal{E}_H^c) + \P(\mathcal{E}_L^c) \leq 1 - \gamma
\end{equation*}
This completes the proof. 
\end{proof}

\subsection{Proofs for Natural Model of Unobserved Confounding}
\label{subsec:proofs_natural_model}

We verify that the under the DGP for the natural model of unobserved confounding described in Section \ref{sec:identification}, units of type $\{\balpha(3),\balpha(4)\}$ do not satisfy horizontal span inclusion. 
For units of these type, we observe combinations $\{\pi_1,\pi_2,\pi_3\}$ with the following binary representations $\{v(\pi_1) = (1,1,1,1,\ldots,1),v(\pi_2) = (1,-1,-1,1,\ldots,1), v(\pi_3)\} = (1,-1,1,1,\ldots,1),$. 
These combinations have restricted Fourier characteristics $\Tilde{\bchi}^\pi$ as follows: $\{\Tilde{\bchi}^{\pi_1} = (1,1,1,0,\ldots,0), \Tilde{\bchi}^{\pi_2} = (1,-1,-1,0,\ldots,0), \Tilde{\bchi}^{\pi_3} = (1,-1,1,0,\ldots,0)\}$. 
It is easy to check that $\Tilde{\bchi}^{\pi} = (1,1,-1,0,\ldots,0)$ does not belong to the span of $\{\Tilde{\bchi}^{\pi_1},\Tilde{\bchi}^{\pi_2},\Tilde{\bchi}^{\pi_3}\}$.
Hence horizontal span inclusion does not hold of type $\{\balpha(3),\balpha(4)\}$.

\allowdisplaybreaks

\section{Asymptotic Normality}
\label{sec:asymptotic_normality_supp}

In this section, we provide proofs for Proposition \ref{prop:horizontal_asymptotic_normality} and Theorem \ref{thm:vertical_regression_normality}.

%




\subsection{Proof of Proposition \ref{prop:horizontal_asymptotic_normality}}
For ease of exposition, we suppress conditioning on $\mathcal{A}$.
Then, from the law of total expectation, this yields, 
\begin{align*}
    &\hat{\E}_{SR}[Y_u^{(\pi)}] - \E[Y_u^{(\pi)}] 
    \\ &= \left( \hat{\E}_{SR}[Y_u^{(\pi)} | \hat{\mathcal{S}}_u = \mathcal{S}_u] -  \E[Y_u^{(\pi)}] \right) \P(\hat{\mathcal{S}}_u = \mathcal{S}_u) + \left( \hat{\E}_{SR}[Y_u^{(\pi)} | \hat{\mathcal{S}}_u \neq \mathcal{S}_u]  - \E[Y_u^{(\pi)}]\right) \P(\hat{\mathcal{S}}_u \neq \mathcal{S}_u) 
    \\ & =  \left( \hat{\E}_{SR}[Y_u^{(\pi)} | \hat{\mathcal{S}}_u = \mathcal{S}_u] -  \E[Y_u^{(\pi)}] \right) \P(\hat{\mathcal{S}}_u = \mathcal{S}_u) +  \hat{\E}_{SR}[Y_u^{(\pi)} | \hat{\mathcal{S}}_u \neq \mathcal{S}_u]  \P(\hat{\mathcal{S}}_u \neq \mathcal{S}_u)  - \E[Y_u^{(\pi)}] \P(\hat{\mathcal{S}}_u \neq \mathcal{S}_u) 
 \end{align*}

\noindent We scale each of three terms in the equation above by $\sqrt{|\Pi_u|}/\sqrt{\sigma^2(\bchi^{\pi})^T\mathbf{K}^{-1}_u\bchi^{\pi}}$, and analyze each of them separately. 

\noindent \emph{Term 1.} 
To proceed, we state the following lemma.
\begin{lemma}
\label{lem:fourier_coefficient_asymptotic_normality}
Let the set-up of Proposition \ref{prop:horizontal_asymptotic_normality} hold. Then, as $|\Pi_u| \rightarrow \infty$, we have, 
\begin{equation}
\label{eq:fourier_coefficient_asymptotic_normality}
     \sqrt{\frac{|\Pi_u|}{\sigma^2(\bchi^{\pi})^T\mathbf{K}^{-1}_u\bchi^{\pi}}}\left( \hat{\E}_{SR}[Y_u^{(\pi)} | \hat{\mathcal{S}}_u = \mathcal{S}_u ] - \E[Y_u^{(\pi)}] \right)\xrightarrow{d} N(0,1).
\end{equation}
\end{lemma}

\noindent Next, from condition (e) of Proposition \ref{prop:horizontal_asymptotic_normality}, as  $|\Pi_u| \rightarrow \infty$, $\P(\hat{\mathcal{S}}_u = \mathcal{S}_u) \rightarrow 1$.
Then, using this observation, substituting \eqref{eq:fourier_coefficient_asymptotic_normality} into term 1, and using Slutsky's theorem, gives the following
\begin{equation}
\label{eq:horizontal_regression_term1_normality}
      \sqrt{\frac{|\Pi_u|}{\sigma^2(\bchi^{\pi})^T\mathbf{K}^{-1}_u\bchi^{\pi}}}\left( \hat{\E}_{SR}[Y_u^{(\pi)} | \hat{\mathcal{S}}_u = \mathcal{S}_u] -  \E[Y_u^{(\pi)}] \right) \P(\hat{\mathcal{S}}_u = \mathcal{S}_u) \xrightarrow{d} N (0,1),
\end{equation}
as $|\Pi_u| \rightarrow \infty$.
\vspace{1mm}

\noindent \emph{Term 2.} We introduce the following lemma for the analysis of term 2. 
\begin{lemma}
\label{lem:largest_eigenvalue_covariance_matrix}
Let the set-up of Proposition \ref{prop:horizontal_asymptotic_normality} hold. Then, we have 
\begin{equation*}
      \sqrt{\frac{|\Pi_u|}{\sigma^2(\bchi^{\pi})^T\mathbf{K}^{-1}_u\bchi^{\pi}}} = O\left(\sqrt{\frac{s|\Pi_u|}{2^p}} \right)
\end{equation*}
    
\end{lemma}

\noindent Substituting the result of Lemma \ref{lem:largest_eigenvalue_covariance_matrix}, and conditions (d) and (e) into the expression for term 2 yields,
\begin{equation}
\label{eq:horizontal_regression_term2_intermediate1}
    \sqrt{\frac{|\Pi_u|}{\sigma^2(\bchi^{\pi})^T\mathbf{K}^{-1}_u\bchi^{\pi}}} \hat{\E}_{SR}[Y_u^{(\pi)} |\hat{\mathcal{S}}_u  \neq \mathcal{S}_u]  \P(\hat{\mathcal{S}}_u \neq \mathcal{S}_u) = o\left(\sqrt{\frac{|\Pi_u|^{1+c_2}}{2^p} }e^{-|\Pi_u|^{c_3}} \hat{\E}_{SR}[Y_u^{(\pi)} | \hat{\mathcal{S}}_u  \neq \mathcal{S}_u] \right)
\end{equation}
     
\noindent We continue by stating the following lemma.

\begin{lemma}
\label{lem:ridge_bound}
Let the set-up of Proposition \ref{prop:horizontal_asymptotic_normality} hold. Then, we have that
\begin{equation*}
   \hat{\E}_{SR}[Y_u^{(\pi)} | \hat{\mathcal{S}}_u  \neq \mathcal{S}_u]= O_p \left(|\Pi_u| \sqrt{2^{p}} \right)
\end{equation*}
\end{lemma}

\noindent Then, substituting the result of Lemma \ref{lem:ridge_bound} into \eqref{eq:horizontal_regression_term2_intermediate1} gives us 
\begin{equation}
\label{eq:horizontal_regression_term2}
    \sqrt{\frac{|\Pi_u|}{\sigma^2(\bchi^{\pi})^T\mathbf{K}^{-1}_u\bchi^{\pi}}} \hat{\E}_{SR}[Y_u^{(\pi)} | \hat{\mathcal{S}}_u  \neq \mathcal{S}_u]  \P(\hat{\mathcal{S}}_u \neq \mathcal{S}_u) = o_p \left(\sqrt{|\Pi_u|^{3+c_2}} e^{-|\Pi_u|^{c_3}} \right) = o_p(1)
\end{equation}

\vspace{1mm}

\noindent \emph{Term 3.} By Lemma \ref{lem:largest_eigenvalue_covariance_matrix}, Assumption \ref{ass:boundedness_potential_outcome}, conditions (d) and (e) of Proposition \ref{prop:horizontal_asymptotic_normality}, we have 
\begin{equation}
\label{eq:horizontal_asymptotic_normalty_term3_bound}
     \sqrt{\frac{|\Pi_u|}{\sigma^2(\bchi^{\pi})^T\mathbf{K}^{-1}_u\bchi^{\pi}}} \E[Y_u^{(\pi)}]\P(\hat{\mathcal{S}}_u \neq \mathcal{S}_u) =  o\left(\sqrt{\frac{|\Pi_u|^{1+c_2}}{2^p}} e^{-|\Pi_u|^{c_3}}\right) = o(1), 
\end{equation}

\vspace{1mm}

\noindent Collecting \eqref{eq:horizontal_regression_term1_normality}, \eqref{eq:horizontal_regression_term2}, and \eqref{eq:horizontal_asymptotic_normalty_term3_bound} gives us the claimed result.

\subsection{Helper Lemmas for Proposition \ref{prop:horizontal_asymptotic_normality}}

\subsubsection{Proof of Lemma \ref{lem:fourier_coefficient_asymptotic_normality}}
\noindent We establish some notation required for the proof. 
Define $\balpha_{\mathcal{S}_u} = [\alpha_{u,S}: S \in \mathcal{S}_u] \in \mathbb{R}^{|\mathcal{S}_u|}$, and $\hat{\balpha}^{SR}_{\mathcal{S}_u} = [\hat{\alpha}^{SR}_{u,S}: S \in \mathcal{S}_u] \in \mathbb{R}^{|\mathcal{S}_u|}$.
Then, we begin our proof by stating the following lemma, and state the delta theorem. 

\begin{lemma} [Theorem 3 of \cite{liu2013asymptotic}]
\label{lem:fourier_coefficient_asymptotic_normality_liu}
Let the set-up of Proposition \ref{prop:horizontal_asymptotic_normality} hold. Then, as $|\Pi_u| \rightarrow \infty$, we have that 
\begin{equation*}
    \sqrt{|\Pi_u|} \left( \hat{\balpha}^{SR}_{\mathcal{S}_u} - \balpha_{\mathcal{S}_u} \right) \xrightarrow{d} N\left(0,\sigma^2\mathbf{K}^{-1}_u\right)
\end{equation*}
\end{lemma}

\begin{theorem} [Delta Theorem \cite{ding2024linear}] 
\label{thm:delta_theorem}
Let $f(\mathbf{z})$ be a function from $\mathbb{R}^p \rightarrow \mathbb{R}$, and $\frac{\partial f(\bz)}{\partial \bz} \in \mathbb{R}^p$ denote the partial derivative of $f$ with respect to $\bz$.
Additionally, suppose $\sqrt{n}(\mathbf{Z}_n - \mathbf{\theta}) \xrightarrow{d} N(0,\Sigma)$ as $n \rightarrow \infty$. 
Then, we have
\begin{equation*}
    \sqrt{n}\left(f(\mathbf{Z}_n) - f(\mathbf{\theta}) \right) \xrightarrow{d} N(0,(\frac{\partial f(\bz)}{\partial \bz})^T \Sigma \frac{\partial f(\bz)}{\partial \bz} )
\end{equation*}
as $n \rightarrow \infty$. 
\end{theorem}

\noindent Applying Lemma \ref{lem:fourier_coefficient_asymptotic_normality}, and Theorem \ref{thm:delta_theorem} (with $f(\cdot) = \langle \cdot, \bchi^\pi \rangle$) gives us 
\begin{equation*}
     \sqrt{|\Pi_u|} \left(\hat{\E}_{SR}[Y_u^{(\pi)} | \hat{\mathcal{S}}_u =\mathcal{S}_u ]  - \E[Y_u^{(\pi)}] \right) =  \sqrt{|\Pi_u|} \left(\langle  \hat{\balpha}^{SR}_{\mathcal{S}_u}, \bchi^{\pi} \rangle - \langle  \balpha_{\mathcal{S}_u}, \bchi^{\pi} \rangle  \right) \xrightarrow{d} N(0, \sigma^2 (\bchi^{\pi})^T\mathbf{K}^{-1}_u \bchi^{\pi} ),
\end{equation*}
as $|\Pi_u| \rightarrow \infty$. 
Scaling both sides of the equation above by $\sqrt{1/\sigma^2 (\bchi^{\pi})^T\mathbf{K}^{-1}_u \bchi^{\pi}}$ finishes the proof. 

\subsubsection{Proof of Lemma \ref{lem:largest_eigenvalue_covariance_matrix}} 

We begin by defining some notation.
For a square matrix $\bX$, let $\lambda_{\min}(\bX)$ and let $\lambda_{\max}(\bX)$ denote the minimum and maximum eigenvalues of $\bX$ respectively. 
Additionally, for a positive-definite invertible matrix $\bX$, recall the fact that $\lambda_{\min}(\bX^{-1}) = 1/\lambda_{\max}(\bX)$.
Finally, for a general matrix $\mathbf{A}$, let $s_{\max}(\mathbf{A})$ denote its largest singular value. 
Then, this gives

\begin{align}
    \sqrt{\frac{|\Pi_u|}{\sigma^2(\bchi^{\pi})^T\mathbf{K}^{-1}_u\bchi^{\pi}}}  & \leq \sqrt{\frac{|\Pi_u|}{\sigma^2 \lambda_{\min}(\mathbf{K}^{-1}_u) (\bchi^{\pi})^T \bchi^{\pi}}}  \nonumber \\
    & = \sqrt{\frac{|\Pi_u| \lambda_{\max}(\mathbf{K}_u) }{\sigma^2 (\bchi^{\pi})^T \bchi^{\pi}}}  \nonumber \\
    & = \sqrt{\frac{|\Pi_u| \lambda_{\max}(\mathbf{K}_u) }{\sigma^2 2^p}} \nonumber \\
    & =  \sqrt{\frac{\lambda_{\max}((\bchi_{\mathcal{S}_u}(\Pi_u))^T \bchi_{\mathcal{S}_u}(\Pi_u)) }{\sigma^2 2^p}} \nonumber \\
    & =  \frac{s_{\max}(\bchi_{\mathcal{S}_u}(\Pi_u))}{\sqrt{\sigma^2 2^p}} ,\label{eq:horizontal_regression_term2_largest_singular_value}
\end{align}
where in the last line we use the fact that for a matrix $\mathbf{A}$, $s_{\max}(\mathbf{A}) = \sqrt{\lambda_{\max}(\mathbf{A}^T\mathbf{A})}$. 
To proceed, recall that $s_{\max}(\mathbf{A}) \leq \lVert \mathbf{A} \rVert_{F}$, where $\lVert \cdot \rVert_{F}$ denote the Frobenius norm.
Since $\bchi_{\mathcal{S}_u}(\Pi_u) \in \{-1,1\}^{|\Pi_u| \times |\mathcal{S}_u|}$, we have $\lVert \bchi_{\mathcal{S}_u}(\Pi_u) \rVert_{F} = \sqrt{|\Pi_u| \times |\mathcal{S}_u|} \leq \sqrt{s|\Pi_u|}$. 
Substituting $s_{\max}(\bchi_{\mathcal{S}_u}(\Pi_u)) \leq \sqrt{s|\Pi_u|}$ into \eqref{eq:horizontal_regression_term2_largest_singular_value} gives us
\begin{equation*}
     \sqrt{\frac{|\Pi_u|}{\sigma^2(\bchi^{\pi})^T\mathbf{K}^{-1}_u\bchi^{\pi}}}  = O\left(\sqrt{\frac{s|\Pi_u|}{2^p}}\right),
\end{equation*}
which is the claimed result. 

\subsubsection{Proof of Lemma \ref{lem:ridge_bound}}
Using Cauchy-Schwarz gives,
\begin{equation*}
    \hat{\E}_{SR}[Y_u^{(\pi)} | \hat{\mathcal{S}}_u \neq \mathcal{S}_u] = \langle \hat{\balpha}_u^{SR}, \bchi^{\pi} \rangle \leq \lVert  \hat{\balpha}_u^{SR} \rVert_2 \lVert \bchi^{\pi} \rVert_2 =  \sqrt{2^p} \lVert  \hat{\balpha}^{SR}_{u} \rVert_2 
\end{equation*}
\noindent Next, we upper bound $\lVert  \hat{\balpha}^{SR}_{u} \rVert_2 $.
Let $\mathbf{U}_{\hat{\mathcal{S}}_u} \mathbf{D}_{\hat{\mathcal{S}}_u} \mathbf{V}^T_{\hat{\mathcal{S}}_u}$ denote the SVD of $\bchi_{\hat{\mathcal{S}}_u}(\Pi_u)$. 
Then, substituting the SVD of $\bchi_{\hat{\mathcal{S}}_u}(\Pi_u)$ into the definition of $\hat{\balpha}^{SR}_{u}$ (see \eqref{eq:Ridge_estimator}), it is easy to obtain 
\begin{equation} \label{eq:ridge_SVD_1}
    \hat{\balpha}^{SR}_{u} = \mathbf{V}_{\hat{\mathcal{S}}_u}\left(\mathbf{D}^2_{\hat{\mathcal{S}}_u} + \frac{1}{|\Pi_u|} \mathbf{I}_{|\hat{\mathcal{S}}_u|}\right)^{-1}\mathbf{D}_{\hat{\mathcal{S}}_u} \mathbf{U}^T_{\hat{\mathcal{S}}_u} \bY_{\Pi_u}
\end{equation}

\noindent To proceed, define the diagonal matrix $\mathbf{D}'_{\hat{\mathcal{S}}_u} =  \left(\mathbf{D}^2_{\hat{\mathcal{S}}_u} + \frac{1}{|\Pi_u|} \mathbf{I}_{|\hat{\mathcal{S}}_u|}\right)^{-1} \mathbf{D}_{\hat{\mathcal{S}}_u} $.
Then, using \eqref{eq:ridge_SVD_1}, we obtain the following upper bound for $\lVert  \hat{\balpha}^{SR}_{u} \rVert^2_2$.
\begin{align}
    \lVert  \hat{\balpha}^{SR}_{u} \rVert^2_2 & = \left(\mathbf{V}_{\hat{\mathcal{S}}_u}  \mathbf{D}'_{\hat{\mathcal{S}}_u} \mathbf{U}^T_{\hat{\mathcal{S}}_u} \bY_{\Pi_u} \right)^T \left(\mathbf{V}_{\hat{\mathcal{S}}_u}  \mathbf{D}'_{\hat{\mathcal{S}}_u} \mathbf{U}^T_{\hat{\mathcal{S}}_u} \bY_{\Pi_u} \right) \nonumber \\
    &  = \bY^T_{\Pi_u} \mathbf{U}_{\hat{\mathcal{S}}_u} \mathbf{D}'_{\hat{\mathcal{S}}_u} \mathbf{V}^T_{\hat{\mathcal{S}}_u} \mathbf{V}_{\hat{\mathcal{S}}_u}  \mathbf{D}'_{\hat{\mathcal{S}}_u} \mathbf{U}^T_{\hat{\mathcal{S}}_u} \bY_{\Pi_u} \nonumber\\
    & = \bY^T_{\Pi_u} \mathbf{U}_{\hat{\mathcal{S}}_u} (\mathbf{D}'_{\hat{\mathcal{S}}_u})^2 \mathbf{U}^T_{\hat{\mathcal{S}}_u} \bY_{\Pi_u} \nonumber \\
    & \leq s_{\max}\left((\mathbf{D}'_{\hat{\mathcal{S}}_u})^2\right) \lVert \mathbf{U}^T_{\hat{\mathcal{S}}_u} \bY_{\Pi_u} \rVert^2_2 \nonumber \\
    & \leq  s_{\max}\left((\mathbf{D}'_{\hat{\mathcal{S}}_u})^2\right) \lVert \mathbf{U}_{\hat{\mathcal{S}}_u}  \rVert^2_2 
 \lVert \bY_{\Pi_u} \rVert^2_2 \nonumber \\
    & \leq s_{\max}\left((\mathbf{D}'_{\hat{\mathcal{S}}_u})^2\right) \lVert \bY_{\Pi_u} \rVert^2_2 \label{eq:ridge_SVD_2}
\end{align}
\noindent where the last inequality follows from the fact that $\mathbf{U}_{\hat{\mathcal{S}}_u}$ is a orthonormal matrix. 

For the rest of the proof, some additional notation is required.
Define $\bepsilon^{\Pi_u} = [\epsilon_n^{\pi}: \pi \in \Pi_u] \in \R^{|\Pi_u|}$.
Let $s_{i}(\bchi_{\hat{\mathcal{S}}_u}({\Pi_u}))$ denote the $i$th singular value of $\bchi_{\hat{\mathcal{S}}_u}(\Pi_u)$ for $i = 1 \ldots |\hat{\mathcal{S}}_u|$.
For simplicity, suppress dependence on $\bchi_{\hat{\mathcal{S}}_u}(\Pi_u)$, and denote  $s_{i}(\bchi_{\hat{\mathcal{S}}_u}({\Pi_u}))$ as $s_i$.

To proceed, observe that the matrix $(\mathbf{D}'_{\hat{\mathcal{S}}_u})^2$ is diagonal with elements $s^2_i/(s^2_i + 1/|\Pi_u|)^2$ for $i = 1 \ldots |\hat{\mathcal{S}}_u|$.
Next, noticing that $(a+b)^2 \geq ab$ for any $a,b \geq 0$, we have $s^2_i/(s^2_i + 1/|\Pi_u|)^2 \leq |\Pi_u|$ for all $i = 1 \ldots |\hat{\mathcal{S}}_u|$.
As a result, 
$s_{\max}\left((\mathbf{D}'_{\hat{\mathcal{S}}_u})^2\right) \leq |\Pi_u|$.
Substituting this inequality into \eqref{eq:ridge_SVD_2}, and simplifying further gives us, 
\begin{align}
    \lVert \hat{\balpha}^{SR}_{u} \rVert^2_2 & \leq |\Pi_u| \lVert \bY_{\Pi_u} \rVert^2_2 \nonumber \\ 
    & =|\Pi_u| \lVert \E[\bY_u^{(\Pi_u)}] + \bepsilon^{\Pi_u} \rVert^2_2 \nonumber \\
    & \leq 2|\Pi_u|\lVert \E[\bY_u^{(\Pi_u)}] \rVert^2_2 + 2|\Pi_u|\lVert \bepsilon^{\Pi_u} \rVert^2_2 \nonumber \\
    & \leq 2|\Pi_u|^2 + 2|\Pi_u| \lVert \bepsilon^{\Pi_u} \rVert^2_2 \label{eq:ridge_SVD_3},
\end{align}

\noindent where the last inequality uses Assumption \ref{ass:boundedness_potential_outcome}. 
Finally, it follows from Theorem 3.1.1 of \cite{vershynin_2018} that $\lVert \bepsilon^{\Pi_u} \rVert^2_2 = O_p (|\Pi_u|)$. 
Substituting this into \eqref{eq:ridge_SVD_3} completes the proof.

%

\subsubsection{Proof of Lemma \ref{lem:fourier_coefficient_asymptotic_normality}}
The proof of this lemma follows immediately from adapting notation of Theorem 3 in \cite{liu2013asymptotic} to this paper. 
In particular, $Y = \bY_{\Pi_u}$, $X = \bchi(\Pi_u)$, $p = 2^p$, $\beta = \balpha_u$, $\Tilde{\beta}_{\text{Select + Ridge}} = \hat{\balpha}_u^{SR}$, $\beta_S = \balpha_{\mathcal{S}_u}$, $\Tilde{\beta}_{\text{Select + Ridge}, \mathcal{S}_u}= \hat{\balpha}^{SR}_{\mathcal{S}_u} $,  $\mu_n = 1/|\Pi_u|$, where the left hand side of each equality is the notation used in \cite{liu2013asymptotic}.

\subsection{Proof of Theorem \ref{thm:vertical_regression_normality}}

In this section, we provide the proof of Theorem \ref{thm:vertical_regression_normality}, i.e., establish asymptotic normality of the vertical regression step of \method. We also recall the following notation, let $\Delta^n_w = \hat{\bw}^n - \Tilde{\bw}^n \in \R^{|\mathcal{I}|}$, and $\Delta_{\mathcal{I}}^{\pi} = \hat{\E}[\bY_{\mathcal{I}}^{(\pi)}] - \E[\bY_{\mathcal{I}}^{(\pi)}] \in \R^{|\mathcal{I}|}$. 

\noindent Using Lemma \ref{lem:w_tilde_transfer_outcomes}, and the notation established above, we have
\begin{align}
     \hat{\E}[Y_n^{(\pi)}] - \E[Y_{n}^{(\pi)}] & = \langle \hat{\E}[\bY_{\mathcal{I}}^{(\pi)}], \hat{\bw}^n \rangle - \langle \E[\bY_{\mathcal{I}}^{(\pi)}], \Tilde{\bw}^n \rangle  \nonumber \\
     & = \langle \Delta_{\mathcal{I}}^{\pi}, \Tilde{\bw}^n \rangle  +   \langle \E[\bY_{\mathcal{I}}^{(\pi)}] ,\Delta^n_w   \rangle  +\langle \Delta^n_w, \Delta_{\mathcal{I}}^{\pi} \rangle. \nonumber
\end{align}
From Assumption \ref{ass:rowspace_inclusion}, it follows that $\E[\bY_{\mathcal{I}}^{(\pi)}] = \mathcal{P}_{V_{\mathcal{I}}^{(\Pi_n)}} \E[\bY_{\mathcal{I}}^{(\pi)}]  $ , where $\bV_{\mathcal{I}}^{(\Pi_n)}$ are the right singular vectors of $\E[\bY_{\mathcal{I}}^{(\Pi_n)}]$. Plugging this into the equation above gives us
\begin{equation}
\label{eq:three_term_asymptotic_normality}
    \hat{\E}[Y_n^{(\pi)}] - \E[Y_{n}^{(\pi)}] = \langle \Delta_{\mathcal{I}}^{\pi}, \Tilde{\bw}^n \rangle  + \langle   \E[\bY_{\mathcal{I}}^{(\pi)}] ,\mathcal{P}_{V_{\mathcal{I}}^{(\Pi_n)}}\Delta^n_w   \rangle  +\langle \Delta^n_w, \Delta_{\mathcal{I}}^{\pi} \rangle. 
\end{equation}
Next, we scale the left-hand side (LHS) of \eqref{eq:three_term_asymptotic_normality}  by $ \left( \sum_{u \in \mathcal{I}} \hspace{0.5mm} \Tilde{\sigma}^2_u \right)^{-1/2}$, and analyze term 1, and terms 2 \& 3 together  on the right-hand side (RHS) of \eqref{eq:three_term_asymptotic_normality} separately. 

\noindent \emph{Term 1.} Expanding the first term gives us
\begin{equation} \label{eq:vertical_asymptotic_normality_term1_intermediate1}
   \frac{1}{\left( \sum_{u \in \mathcal{I}} \hspace{0.5mm} \Tilde{\sigma}^2_u \right)^{1/2}}  \langle \Delta_{\mathcal{I}}^{\pi}, \Tilde{\bw}^n \rangle = \frac{1}{\left( \sum_{u \in \mathcal{I}} \hspace{0.5mm} \Tilde{\sigma}^2_u \right)^{1/2}}  \sum_{u \in \mathcal{I}}  \Tilde{w}^n_u \left(\hat{\E}[Y_u^{(\pi)}] - \E[Y_u^{(\pi)}] \right)
\end{equation}

\noindent From \eqref{eq:asymptotic_normality_condition_2}, as $|\Pi_u| \rightarrow \infty$, we have that 
\begin{equation*}
     \Tilde{w}^n_u \left(\hat{\E}[Y_u^{(\pi)}] - \E[Y_u^{(\pi)}] \right) \xrightarrow{d} N(0,\Tilde{\sigma}^2_u)
\end{equation*}

\noindent Since $\Tilde{w}^n_u \left(\hat{\E}[Y_u^{(\pi)}] - \E[Y_u^{(\pi)}] \right)$ is independent across $u$, as $M \rightarrow \infty$, the equation above implies
\begin{equation*}
    \sum_{u \in \mathcal{I}}
     \Tilde{w}^n_u \left(\hat{\E}[Y_u^{(\pi)}] - \E[Y_u^{(\pi)}] \right) \xrightarrow{d} N(0,\sum_{u \in \mathcal{I}}\Tilde{\sigma}^2_u)
\end{equation*}

\noindent Substituting the equation above into \eqref{eq:vertical_asymptotic_normality_term1_intermediate1} yields the following, 
\begin{equation} \label{eq:vertical_asymptotic_normality_term1}
   \frac{1}{\left( \sum_{u \in \mathcal{I}} \hspace{0.5mm} \Tilde{\sigma}^2_u \right)^{1/2}}  \sum_{u \in \mathcal{I}}  \Tilde{w}^n_u \left(\hat{\E}[Y_u^{(\pi)}] - \E[Y_u^{(\pi)}] \right) \xrightarrow{d} \mathcal{N}(0,1),
\end{equation}
as $M \rightarrow \infty$.

\noindent \emph{Terms 2 and 3.} We scale the right hand side of \eqref{eq:collecting_terms_simplified_log} by $ \left( \sum_{u \in \mathcal{I}} \hspace{0.5mm} \Tilde{\sigma}^2_u \right)^{-1/2}$ which yields the following,
\begin{equation} \label{eq:vertical_asymptotic_normality_term23_intermediate1}
      \frac{1}{\left( \sum_{u \in \mathcal{I}} \hspace{0.5mm} \Tilde{\sigma}^2_u \right)^{1/2}} \left(\langle \E[\bY_{\mathcal{I}}^{(\pi)}] ,\Delta^n_w   \rangle   +\langle \Delta^n_w, \Delta_{\mathcal{I}}^{\pi} \rangle \right) = O_p\left(\frac{\log^{3}(|\Pi_n||\mathcal{I}|)}{\left( \sum_{u \in \mathcal{I}} \hspace{0.5mm} \Tilde{\sigma}^2_u \right)^{1/2}}\left(r^2_n\sqrt{\frac{{s^2p}}{{M}}} +  \frac{r_n}{|\Pi_n|^{1/4}}\right) \right)
\end{equation}

\vspace{1mm}

\noindent To proceed, we state the following lemma 

\begin{lemma} \label{lem:scaled_variance_upper_bound}
Let the set-up of Theorem \ref{thm:vertical_regression_normality} hold. Then, we have
\begin{equation*}
  \left( \sum_{u \in \mathcal{I}}  \Tilde{\sigma}^2_u \right)^{-1/2} = O\left(\frac{1}{\lVert \Tilde{\bw}^n \rVert_2} \right) 
\end{equation*}
  
\end{lemma}

\noindent Substituting the result of Lemma \ref{lem:scaled_variance_upper_bound} into \eqref{eq:vertical_asymptotic_normality_term23_intermediate1}, and recalling \eqref{eq:asymptotic_normality_condition_1} yields
\begin{equation} \label{eq:asymptotic_normality_term23_op_bound}
     \frac{1}{\left( \sum_{u \in \mathcal{I}} \hspace{0.5mm} \Tilde{\sigma}^2_u \right)^{1/2}} \left(\langle \E[\bY_{\mathcal{I}}^{(\pi)}] ,\Delta^n_w   \rangle   +\langle \Delta^n_w, \Delta_{\mathcal{I}}^{\pi} \rangle \right) = o_p(1)
\end{equation}

\noindent Collecting \eqref{eq:vertical_asymptotic_normality_term1}, and \eqref{eq:asymptotic_normality_term23_op_bound} gives us the result.

\subsection{Proof of Lemma \ref{lem:scaled_variance_upper_bound}}
\label{subsec:scale_variance_proof}
For a donor unit $u \in \mathcal{I}$, we have, 
\begin{align}
    \Tilde{\sigma}^{2}_u & = \frac{(\bchi^{\pi})^T \mathbf{K}^{-1}_u \bchi^{\pi} \left( \Tilde{w}^n_u 
    \right)^2 }{|\Pi_u|}  \nonumber \\
    & \geq \frac{\lambda_{\min}(\mathbf{K}^{-1}_u)(\bchi^{\pi})^T \bchi^{\pi} \left( \Tilde{w}^n_u 
    \right)^2}{|\Pi_u|} \nonumber \\
    & = \frac{\lambda_{\min}(\mathbf{K}^{-1}_u)2^p \left( \Tilde{w}^n_u 
    \right)^2}{|\Pi_u|} \nonumber \\
    & = \frac{2^p \left( \Tilde{w}^n_u 
    \right)^2 }{\lambda_{\max}(\mathbf{K}_u)|\Pi_u|} \nonumber \\
    & \geq \frac{\left( \Tilde{w}^n_u 
    \right)^2}{\lambda_{\max}(\mathbf{K}_u)} \label{eq:sigma_tilde_largest_eigenvalue},
\end{align}
where in the last line we use the fact that $|\Pi_u| \leq 2^p$. 
To proceed, we upper bound $\lambda_{\max}(\mathbf{K}_u)$.
To do so, we define some notation. 
Let $\mathbf{I}_m \in \mathbb{R}^{m \times m}$ denote the identity matrix of dimension $m$.
Additionally, recall the following fact: for a matrix $\mathbf{A} \in \mathbb{R}^{m \times m}$, we have that $\lVert \mathbf{A} \rVert_2 \leq m \lVert \mathbf{A} \rVert_{\infty}$.
Using this and Assumption \ref{ass:incoherence}, we have that
\begin{align*}
    \lVert \mathbf{K}_u - \mathbf{I}_{|\mathcal{S}_u|} \rVert_2 & = \lVert \frac{(\bchi_{\mathcal{S}_u}(\Pi_u))^T\bchi_{\mathcal{S}_u}(\Pi_u)}{|\Pi_u|} - \mathbf{I}_{|\mathcal{S}_u|} \rVert_2 \\ 
    & \leq s \lVert \frac{(\bchi_{\mathcal{S}_u}(\Pi_u))^T\bchi_{\mathcal{S}_u}(\Pi_u)}{|\Pi_u|} - \mathbf{I}_{|\mathcal{S}_u|} \rVert_\infty \\
    & \leq  s \lVert \frac{(\bchi(\Pi_u))^T\bchi(\Pi_u)}{|\Pi_u|} - \mathbf{I}_{2^p} \rVert_\infty \leq C
\end{align*}
\noindent for a positive constant $C > 0$.
Next, using the equation above and Weyl's inequality \cite{wainwright2019high}, we have that 
\begin{equation*}
    \lambda_{\max}(\mathbf{K}_u)  \leq \lVert \mathbf{K}_u - \mathbf{I}_{|\mathcal{S}_u|}  \rVert_2 +  \lambda_{\max}(\mathbf{I}_{|\mathcal{S}_u|}) \leq 1 + C
\end{equation*}

\noindent Substituting the result of Lemma \ref{lem:largest_eigenvalue_covariance_matrix} into \eqref{eq:sigma_tilde_largest_eigenvalue} gives us $\Tilde{\sigma}^{2}_u \geq \left( \Tilde{w}^n_u \right)^2/(1 + C)$. 
Hence, we have $\left( \sum_{u \in \mathcal{I}} \hspace{0.5mm} \Tilde{\sigma}^2_u \right)^{-1/2} = O(1/\lVert \Tilde{\bw}^n \rVert_2)$, which is the claimed result. 
\section{Extension to Permutations}
\label{sec:permutations}
In this section, we show how we can extend our causal framework to estimate potential outcomes under different permutations of $p$ items, i.e., rankings. 
As discussed earlier, learning to rank has been widely studied and is of great practical interest in a number of important practical applications such as personalizing results on search engines or matching markets. 
To show how \method~can be adapted to permutations, we first show how  functions over permutations can be re-expressed as Boolean functions.
As a result, we establish that functions over permutations admit a Fourier representation as shown in Section \ref{sec:model}.
Then, we adapt our generating model established for combinations in Section \ref{sec:model} to permutations.  
Next, we discuss how our theoretical guarantees can be adapted to this setting as well. 

\subsection{Fourier Expansion of Functions of Permutations}
\label{subsec:fourier_series_permutations}
We show how we can represent functions of permutations as Boolean functions. 
We had discussed how permutations can be represented as binary vectors in Section \ref{sec:model}, but repeat it here for completeness.

\noindent \textbf{Binary Representation of Permutations.} 
Let $\tau: [p] \rightarrow [p]$ denote a permutation on a set of $p$ items such that $\tau(i)$ denotes the rank of item $i \in [p]$.  
There are a total of $p!$ different rankings, and we denote the set of permutations by $\mathbb{S}_p$.
Every permutation $\tau \in \mathbb{S}_p$ induces a binary representation $\mathbf{v}(\tau) \in \{-1,1\}^{\binom{p}{2}}$, which can be constructed as follows.
For an item $i$, define  $\mathbf{v}^{i}(\tau) \in \{-1,1\}^{p-i}$ as follows: $\mathbf{v}_j^{i}(\tau)= \mathbbm{1}\{ \tau(i) > \tau(j)\} - \mathbbm{1}\{ \tau(i) < \tau(j)\}$ for items $1 \leq i  <  j \leq p$.
That is, each coordinate of $\mathbf{v}^{i}(\tau)$ indicates whether items $i$ and $j$ have been swapped for items $j > i$. 
Then, $\mathbf{v}(\tau) = [\mathbf{v}^{i}(\tau): i \in [p]] \in \{-1,1\}^{\binom{p}{2}}$.
For example, with $p = 4$, the permutation $\tau([1,2,3,4]) = [1,3,4,2]$ has the binary representation $\mathbf{v}(\tau) = (-1,-1,-1,1,1,-1)$.

\noindent \textbf{Fourier Expansion of Functions of Permutations.}
Since a permutation $\tau$ can be expressed as a binary vector, any function $f: \mathbb{S}_p \rightarrow \mathbb{R}$ can be thought of as a Boolean function. 
Then, given the discussion on Fourier expansions of Boolean functions in Section \ref{sec:notation}, we have that any function $f: \mathbb{S}_p \rightarrow \mathbb{R}$ admits the following Fourier decomposition:
$f(\tau) = \sum_{S \subset [\binom{p}{2}]} \alpha_S \chi_{S}(\mathbf{v}(\tau)) \coloneqq \langle  \balpha_f, \bchi^{\tau} \rangle$, where $\balpha_f = [\alpha_S]_{S \in [\binom{p}{2}]} \in \mathbb{R}^{\binom{p}{2}}$, and $\bchi^{\tau} = [\chi_S(\mathbf{v}(\tau)]_{S \in [\binom{p}{2}]} \in \{-1,1\}^{\binom{p}{2}}$ for $\tau \in \mathbb{S}_p$.
\subsection{Model}
\label{subsec:model_permutations}

In this section we show how we can adapt our notation for observed  and potential outcomes as well as our model to the permutation setting.

\noindent \textbf{Notation, and Outcomes.} Let $Y_n^{(\tau)}$ denote the potential outcome for a unit $n$ under permutation $\tau$, and $Y_{n\tau} \in \{\mathbb{R} \cup \star \}$ denote an observed value, where $\star$ denotes an unobserved entry. 
For a given unit $n$ and subset $G \subseteq \mathbb{S}_p$, let $\bY_{G,n} = [Y_{n\tau_i} : \tau_i \in G] \in \{\mathbb{R} \cup \star \}^{|G|}$ represent the vector of observed outcomes. 
Similarly, let $\bY_n^{(G)} = [Y^{(\tau_i)}_{n} : \tau_i \in G] \in \mathbb{R}^{|G|}$ represent the vector of potential outcomes.
Denote $\bchi(G) = [\bchi^{\tau_i} : \tau_i \in G] \in \{-1, 1\}^{|G| \times 2^{\binom{p}{2}}}$.
Let $\mathcal{D} \subset [N] \times [\mathbb{S}_p],$ refer to the subset of unit-combination pairs we do observe, i.e., 
\begin{equation}\label{eq:SUTVA_permutation}
Y_{n\tau} = 
\begin{cases}
Y^{(\tau)}_n, & \text{if $(n, \tau) \in \mathcal{D}$}\\
\star, & \text{otherwise}.
\end{cases}
\end{equation}
Note that \eqref{eq:SUTVA_permutation} implies stable unit treatment value assignment (SUTVA)  holds. 
Further, for a unit $n \in [N]$, denote the subset of permutations we observe them under as $G_n \subseteq \mathbb{S}_p$. 
Our target causal parameter is the potential outcome $\E[Y_n^{(\tau)}]$ for all $N
\times p!$ unit-permutation pairs. 

\noindent \textbf{Model and DGP.} We propose a similar model to the one discussed for combinations in Section \ref{sec:model}. That is, we model $Y_n^{(\tau)} = \langle \balpha_n, \bchi^{\tau} \rangle + \epsilon_n^{\tau}$, where we assume sparsity ($||\balpha||_0 = s \leq 2^{\binom{p}{2}}$), low-rank structure ($\text{rank}(\mathcal{A}) = r \in \{\min\{N,2^{\binom{p}{2}}\}\}$), and $\E[\epsilon_n^{\tau} ~|~ \mathcal{A}] = 0$. 
Additionally, the DGP for permutations is identical to that for combinations discussed in Section \ref{sec:model}.

\subsection{Synthetic Permutations Estimator}
\label{subsec:synth_combo_permutations}

We introduce the Synthetic Permutations Estimator, which is identical to \method~estimator, where we first perform horizontal regression for the donor set, and then transfer these outcomes to non-donor units via PCR. 
Specifically, we detail the two steps of the Synthetic Permutations estimator as follows. 

\emph{Step 1: Horizontal Regression.} For every donor unit $u$, we run a Lasso regression (or fit a ML model of choice) on the outcomes $\bY_{u,G_u} = [Y_{u,\tau_u}: \tau_u \in G_u] \in \mathbb{R}^{|G_u|}$ against the Fourier characteristics $\bchi(G_u)  \in \mathbb{R}^{|G_u| \times 2^{\binom{p}{2}}}$. 
For any unit $n$, we denote $\bY_{n,G_n}$ as $\bY_{G_n}$
Then, we solve the following convex program with penalty parameter $\lambda_u$: 
\begin{align}\label{eq:Lasso_estimator_permutation}
 \hat{\balpha}_u=  \argmin_{\balpha} \ \frac{1}{|G_u|}\lVert \bY_{G_u} - \bchi(G_u)\balpha \rVert^2_2 + \lambda_u \lVert \balpha \rVert_1
\end{align}
For any donor unit-permutation pair $(u,\tau)$, let $\hat{\E}[Y_u^{(\tau)}] = \langle  \hat{\balpha}_u, \bchi^{\tau} \rangle$ denote the estimate of the potential outcome $\E[Y_u^{(\tau)}]$.

\emph{Step 2: Vertical Regression.} Before we detail the second step for permutations, we first define some notation: for $G \subseteq \mathbb{S}_p$, let the vector of estimated potential outcomes $\hat{\E}[\bY^{(G)}_{u}] = [ \hat{\E}[Y_u^{(\tau)}]: \tau \in G] \in \R^{|G|}$. 
Additionally, let $\hat{\E}[\bY^{(G)}_{\mathcal{I}}] = [\hat{\E}[\bY^{(G)}_{u}]: u \in \mathcal{I}] \in \mathbb{R}^{|G| \times |\mathcal{I}|}$.

\vspace{2mm}
\emph{Step 2(a): Principal Component Regression.} Perform a singular value decomposition (SVD) of $\hat{\E}[\bY^{(G_n)}_{\mathcal{I}}]$ to get $\hat{\E}[\bY^{(G_n)}_{\mathcal{I}}] = \sum^{\min(|G_n|,|\mathcal{I}|)}_{l = 1} \hat{s_l}\hat{\bmu}_{l}\hat{\bnu}^T_{l}$. Using a hyper-parameter $\kappa_n \leq \min(|G|,|\mathcal{I}|)$, compute $\hat{\bw}^{n} \in \mathbb{R}^{|\mathcal{I}|}$ as follows:
\begin{equation}
\label{eq:pcr_linear_model_def_permutation}
    \hat{\bw}^{n} = \left(\sum^{\kappa_n}_{l = 1} \hat{s_l}^{-1}\hat{\bnu}_{l}\hat{\bmu}^T_{l}\right)\bY_{G_n} 
\end{equation}

\emph{Step 2(b): Estimation.} Using $\hat{\bw}^{n} = [\hat{w}_u^n : u \in \mathcal{I}]$, we have the following estimate for any intervention $\pi \in \Pi$
\begin{equation}
\label{eq:potential_outcome_estimate_vertical_regression_permutation}
     \hat{\E}[Y_n^{(\tau)}] = \sum_{u \in \mathcal{I}} \hat{w}_u^{n} \hat{\E}[Y_u^{(\tau)}]
\end{equation}

\subsection{Theoretical Results}
We discuss how we can adapt our key theoretical results, in particular identification, and finite-sample consistency of Synthetic Permutations.
In particular, all of our key theoretical results for combinations are easily adapted under the appropriate change in notation. 
We present the following table that summarizes the changes in the notation from combinations to permutations.

\begin{table}[H]
\centering
\begin{tabular}{||c c||} 
 \hline
 Combinations & Permutations \\ [1ex] 
 \hline
 $\pi$ & $\tau$ \\ 
 \hline
  $\Pi$ & $\mathbb{S}_p$  \\ [1ex]
 \hline
 $\Pi_n$ & $G_n$ \\  [1ex]
 \hline
  $\E[Y_n^{(\pi)}]$ &  $\E[Y_n^{(\tau)}]$\\
 \hline
 $Y_n^{(\pi)}$ &  $Y_n^{(\tau)}$  \\ [1ex]
 \hline 
 $Y_{n\pi}$ & $Y_{n\tau}$ \\
 \hline 
 $Y_n^{(\Pi)}$ & $Y_n^{(G)}$ \\
 \hline 
$Y_{\Pi_n}$ & $Y_{G_n}$ \\
\hline 
$p$ & $\binom{p}{2}$ \\
 \hline 
 \end{tabular}
 \vspace{1mm}
\caption{Notational substitutions between combinations and permutations.}
\label{tab:notational_changes_permutation_to_combination}
\end{table}

\noindent Given this summary of notational changes, we now present our identification, and finite-sample consistency results. 

\subsubsection{Identification}
\label{subsec:identification_permutation}

\begin{theorem}
\label{thm:identification_permutation}

Let  Assumptions \ref{ass:observation_model},  \ref{ass:selection_on_fourier}, \ref{ass:donor_set_identification} hold where we make the notational changes presented in Table \ref{tab:notational_changes_permutation_to_combination}. Then, the following hold.

\noindent (a) Donor units: For $u \in \mathcal{I}$, and $\tau \in \mathbb{S}_p$, 
$
 \E[Y^{(\tau)}_{u} ~ | ~ \mathcal{A}]  =  \sum_{\tau_u \in G_{u}} \beta_{\tau_{u}}^{\tau}
 \E[Y_{u,\tau_{u}} \ | \  \mathcal{A}, \ \mathcal{D}].
$

\noindent
(b) Non-donor units: For $n \in [N] \setminus \mathcal{I}$, and $\tau \notin G_n$,
$
\E[Y^{(\tau)}_{n} ~ | ~ \mathcal{A}] = \sum_{u \in \mathcal{I},\tau_u \in G_u}w_{u}^{n}  \beta_{\tau_{u}}^{\tau}  \E[Y_{u,\tau_u} \ | \  \mathcal{A}, \ \mathcal{D}].
$

\end{theorem} 

\noindent Theorem \ref{thm:identification_permutation} establishes  identification for permutations of items for both the donor set and non-donor units.
Our identification strategy for permutations is identical to the combination setting, where we first identify outcomes of donor units, and then transfer them to non-donor units. 
As with combinations, our result allows for identification even though most permutations are observed for no units. 

\subsubsection{Finite-Sample Consistency and Asymptotic Normality }
\label{subsec:finite_sample_permutations}

First, we present our finite-sample consistency result for Synthetic Permutations when applied to permutations. 

\begin{theorem} [Finite Sample Consistency for Permutations]
\label{thm:potential_outcome_convergence_rate_permutations}
Conditioned on $\mathcal{A}$, let Assumptions \ref{ass:observation_model}--\ref{ass:donor_set_identification}, and \ref{ass:boundedness_potential_outcome}--\ref{ass:incoherence} hold with the notational changes presented in Table \ref{tab:notational_changes_permutation_to_combination}. Then, the following statements hold. 
\begin{itemize}
    \item [(a)] For a given donor unit-permutation pair $(u,\tau)$, let the Lasso regularization parameter satisfy $\lambda_u = \Omega(\frac{p^2}{\sqrt{|G_u|}})$. Then, we have that,
        $$
            |\hat{\E}[Y_u^{(\pi)}] - \E[Y_{u}^{(\pi)}]| = \Tilde{O}_p\left(\frac{sp}{\sqrt{|G_u|}} \right).
        $$
    \item [(b)] Additionally, let Assumptions \ref{ass:balanced_spectrum}, and \ref{ass:rowspace_inclusion} hold under the appropriate notational changes. For the given unit-permutation pair  $(n,\tau)$ where $n \in [N] \setminus \mathcal{I}$, let $\kappa_n = \text{rank}(\E[\bY_{\mathcal{I}}^{(G_n)}]) \coloneqq r_{n}$. Then, provided that  $ \min_{u \in \mathcal{I}} |G_u| \coloneqq M = \omega(r_ns^2p^2)$, we have that,
        $$
            \left |\hat{\E}[Y_n^{(\pi)}] - \E[Y_{n}^{(\pi)}]\right| = \Tilde{O_p}\left(  \frac{r^{2}_nsp}{\sqrt{M}} + \frac{r_n}{|G_n|^{1/4}}\right).
        $$
\end{itemize}
\end{theorem}

\noindent Theorem \ref{thm:potential_outcome_convergence_rate_permutations}
establishes finite-sample consistency of Synthetic Permutations when applied to permutations.
The proof follows in an identical fashion to the proof of Theorem \ref{thm:potential_outcome_convergence_rate} under the approriate notational changes. 
Following the argument in Section \ref{subsec:sample_complexity_synth_combo}, the sample complexity of Synthetic Permutations to estimate all $N \times p!$ outcomes with accuracy $O(\delta)$ scales as $O(\text{poly}(r/\delta) \times (N + s^2p^2))$.
In comparison, the sample complexity of Lasso and Matrix completion algorithms scales as $O(N\times s^2p^2/\delta^2)$ and $O(\text{poly}(r/\delta) \times (N + p!))$ respectively.
As compared to the result established for combinations, the sample complexity for permutations is worse by a factor of $p$. 
This additional factor of $p$ arises because the binary representation $\bv(\tau) \in \mathbb{R}^{\binom{p}{2}}$ for a permutation $\tau$ as compared to $\bv(\pi) \in \mathbb{R}^{p}$ for a combination $\pi$. 
Since the error of the Lasso is proportional to the dimension of the binary encoding, the estimation error for donor units is inflated by a factor of $p$ for permutations as compared to combinations. 
As a result, the sample complexity for permutations worse by a factor of $p$. 


Next, we present an informal discussion of the sample complexity of Synthetic Permutations with CART when applied to rankings. 
Under analogous assumptions to Theorem \ref{thm:CART_convergence_rate} (i.e., $\E[Y_n^{(\tau)}]$ is a $k$-Junta), $\hat{\E}[Y_u^{(\tau)}] - \E[Y_u^{(\tau)}] = \Tilde{O}_p\left(s/\sqrt{|G_u|} \right)$
for every donor unit $u$.
That is, like combinations, the estimation error for the horizontal regression with CART is independent of $p$.
Then, following the arguments in Sections \ref{subsec:CART_finite_sample} and \ref{subsec:sample_complexity_synth_combo}, the sample complexity of CART  scales as  $O(\text{poly}(r/\delta) \times (N + s^2))$. 
This efficiency is a result of CART's ability to exploit the $k$-Junta structure of $\E[Y_n^{(\tau)}]$.
\vspace{1mm}

%
%



\end{document}